\documentclass[aps,showpacs,pre,amsmath,amsfonts,amssymb,superscriptaddress,nofootinbib,notitlepage,a4paper]{revtex4}

\usepackage{graphicx}
\usepackage{psfrag}
\usepackage[naturalnames]{hyperref}

\newcommand{\beq}{\begin{equation}} \newcommand{\eeq}{\end{equation}}
\newcommand{\bes}{\begin{split}} \newcommand{\ees}{\end{split}} 
\newcommand{\bea}{\begin{eqnarray}} \newcommand{\eea}{\end{eqnarray}}
\def\I{\mathbb{I}}
\def\ach{\text{ach}\,}
\def\acos{\text{acos}\,}
\def\Image{\text{Image}\,}
\def\sign{\text{sign}\,}
\def\s{\sigma}
\def\us{\underline{\sigma}}

\def\hS{{\widehat{S}}}
\def\hSx{{\widehat{S}^x}}
\def\hSy{{\widehat{S}^y}}
\def\hSz{{\widehat{S}^z}}

\def\hsx{{\widehat{\sigma}^x}}

\def\hsz{{\widehat{\sigma}^z}}
\def\hsix{{\widehat{\sigma}_i^x}}
\def\hsiy{{\widehat{\sigma}_i^y}}
\def\hsiz{{\widehat{\sigma}_i^z}}
\def\hun{\widehat{1}}
\def\hH{\widehat{H}}
\def\hHK{\widehat{H}^{(K)}}
\def\hHzero{\widehat{H}^{(0)}}
\def\hHKm1{\widehat{H}^{(K-1)}}
\def\NNK{{{\cal N}^N_K}}
\def\D{{\cal D}}
\def\MNK{{{\cal M}^N_K}}
\def\MNzero{{{\cal M}^N_0}}
\def\fhN{{\lfloor\frac{N}{2}\rfloor}}
\def\chN{{\lceil\frac{N}{2}\rceil}}
\def\hJ{\widehat{J}}
\def\hHb{\widehat{H}_\bullet}

\def\hHi{\widehat{H}_{\rm i}}
\def\hHf{\widehat{H}_{\rm f}}
\def\hm{\widehat{m}}
\def\hmx{\widehat{m}^x}
\def\hmy{\widehat{m}^y}
\def\hmz{\widehat{m}^z}

\def\hbeta{\widehat{\beta}}
\def\la{\langle}
\def\ra{\rangle}
\def\laz{\phantom{}_z\langle}
\def\raz{\rangle_z}
\def\lax{\phantom{}_x\langle}
\def\rax{\rangle_x}
\def\eres{e_{\textrm{res}}}
\def\hefin{\hat{e}_{\textrm{fin}}}
\def\efin{e_{\textrm{fin}}}
\def\etriv{e_{\textrm{triv}}}
\def\egain{e_{\textrm{gain}}}
\def\U{\mathcal{U}}
\def\tU{\widetilde{\mathcal{U}}}
\def\Ns{{N_{\rm s}}}
\def\dd{{\rm d}}
\def\betasp{\beta_{\rm sp}}
\def\qsp{q_{\rm sp}}
\def\Gammasp{\Gamma_{\rm sp}}
\def\ssp{s_{\rm sp}}
\def\usp{u_{\rm sp}}
\def\esp{e_{\rm sp}}
\def\betac{\beta_{\rm c}}
\def\Tsp{T_{\rm sp}}
\def\Tc{{T_{\rm c}}}
\def\ec{{e_{\rm c}}}
\def\mc{{m_{\rm c}}}

\def\epm{e_{\rm pm}}
\def\efm{e_{\rm fm}}
\def\epfm{e'_{\rm fm}}
\def\msp{m_{\rm sp}}
\def\mpsp{m'_{\rm sp}}
\def\egs{e_{\rm gs}}
\def\egsk{e_{\rm gs}^{(k)}}
\def\mi{m_{\rm i}}
\def\ei{e_{\rm i}}
\def\gammapm{\gamma_{\rm pm}}
\def\gammafm{\gamma_{\rm fm}}
\def\hgammap{\widehat{\gamma}_p}
\def\gb{\gamma_\bullet}
\def\cH{{\cal H}}
\def\numf{\nu_{\rm mf}}
\def\dc{d_{\rm c}}
\def\F{\mathcal{F}}
\def\G{\mathcal{G}}
\def\eiso{e_{\rm iso}}
\def\sturn{s_{\rm turn}}
\def\uturn{u_{\rm turn}}

\def\tq{\widetilde{q}}
\def\tgamma{\widetilde{\gamma}}
\def\qmin{{q_{\rm min}}}
\def\qmax{{q_{\rm max}}}

\begin{document}

\title{On quantum mean-field models and their quantum annealing}

\author{Victor Bapst}

\author{Guilhem Semerjian}
\affiliation{LPTENS, Unit\'e Mixte de Recherche (UMR 8549) du CNRS et
  de l'ENS, associ\'ee \`a l'UPMC Univ Paris 06, 24 Rue Lhomond, 75231
  Paris Cedex 05, France.}

\begin{abstract}
This paper deals with fully-connected mean-field models of quantum spins
with $p$-body ferromagnetic interactions and a transverse field. 
For $p=2$ this corresponds to the quantum Curie-Weiss model (a special case of 
the Lipkin-Meshkov-Glick model) 
which exhibits a second-order phase transition, while for $p>2$ the 
transition is first order. We provide a refined analytical description
both of the static and of the dynamic properties of these models. In particular
we obtain analytically the exponential rate of decay of the gap at the 
first-order transition. We also study the slow annealing from the pure 
transverse field to the pure ferromagnet (and vice versa) and discuss the 
effect of the first-order transition and of the spinodal limit of 
metastability on the residual excitation energy, both for finite and 
exponentially divergent annealing times. In the quantum computation
perspective this quantity would assess the efficiency of the quantum adiabatic
procedure as an approximation algorithm.

\end{abstract}

\maketitle


\section{Introduction}

Finding the minimum of a cost function defined on a discrete configuration
space is the central task of combinatorial optimization. Depending on the
problem considered (i.e. the shape of the cost function), there exists or
not fast (running in polynomial time with respect to the number of variables) 
algorithms for classical computers that performs the minimization,
as classified by the computational complexity theory~\cite{GareyJohnson}.

In more physical terms this problem corresponds to finding the groundstate
of an Hamiltonian (cost function) depending on discrete degrees of freedom
(spins). This analogy has suggested an optimization algorithm named
simulated annealing~\cite{sa}, that proceeds through a stochastic 
exploration of the phase space, according to transition rules obeying 
the detailed
balance condition for a positive temperature, which is slowly reduced from
a very high value down to zero. In this way energy barriers can be 
jumped over through thermal fluctuations, the final state of the system is
the equilibrium at zero temperature, hence concentrated in the sought-for 
minimum of the cost function. Provided with a quantum computer, that is a
device that obeys the laws of quantum mechanics at the level of its computing 
units, one can follow a similar idea
but with quantum fluctuations replacing the thermal ones; this strategy
is known as quantum annealing~\cite{qa_first,qa}, or quantum adiabatic 
algorithm~\cite{qaa}, see~\cite{qa_review_santoro,qa_book_das_chakrabarti}
for reviews. The control 
parameter that replaces the temperature allows to tune the relative strength
of the potential energy (the cost function) and of the ``kinetic energy''
(for instance a transverse field for spins 1/2). The system is initially 
prepared in the groundstate of the latter, then evolves according to 
Schr\"odinger's equation with an Hamiltonian that slowly interpolates 
between the kinetic and the potential energy. If this interpolation is 
sufficiently slow the system remains at all times in the instantaneous 
groundstate of the Hamiltonian, and in particular at the end of
the evolution it is found in the desired minimum of the cost function.

To assess the efficiency of these algorithms one has to specify how slow
the evolution of the control parameter has to be in order that the final
state indeed corresponds to the groundstate. In the quantum setting, which
shall be the focus of this article, this condition is provided by the
quantum adiabatic theorem~\cite{messiah76} which, roughly speaking, states
that the interpolation time has to be larger that the inverse square of
the minimal energy gap between the instantaneous groundstate and the first
excited state encountered along the interpolation. It thus appears that
quantum phase transitions~\cite{sachdev}, where the gap closes in the 
thermodynamic limit,
constitute the bottleneck for the efficiency of the quantum annealing.
First-order phase transitions, at which the gap is typically exponentially 
small in the system size, are in this respect worse than second-order 
transitions for which, at least in non-disordered systems, the gap is only 
polynomially small.

Random instances of combinatorial optimization problems provide useful
benchmark ensembles of cost functions on which to test various 
algorithms~\cite{random_sat}.
They have been the object of an intense research activity at the crossroad
between computer science, mathematics and theoretical 
physics~\cite{book_Marc_Andrea}. 
Several phase transitions have been unveiled that affect the typical number 
and organization of their ground and excited 
states~\cite{BiMoWe00,MePaZe02,pnas07}. 
More recently the tools that allowed
the description of these transitions have been extended to take into account
the additional effect of a transverse field on the corrugated random cost 
function~\cite{cavity_first,cavity_ours}. 
First order phase transitions as a function of the interpolating
transverse field have been observed in some models~\cite{qXOR,young10}; 
this did not come as a surprise as it is a recurrent feature of mean-field 
quantum disordered systems that have
been extensively studied~\cite{Goldschmidt90,NiRi98,BiCu01,jorg08}.

Even if a first order transition in a given model means that the corresponding 
combinatorial optimization problem will only be solved exactly in a time 
exponentially large in the system size, many questions remain open at this 
point. First, one should try to compute the exponential rate of growth of
the adiabatic time. Second, and maybe more importantly, one should 
investigate what is the final energy of an evolution that is too fast to
respect the adiabaticity criterion (a question reminiscent of the Kibble-Zurek 
mechanism, see~\cite{review_kz} for a recent review). Besides its intrisic
physical relevance, this point is also deeply related to important
issues in computational complexity theory, namely hardness of approximation
results~\cite{book_vazirani}. 
Indeed, for some combinatorial optimization problems (MAX-3-SAT for instance,
or even MAX-3-XORSAT whose decision version is in P) 
it is not only difficult to compute the exact value of the minimum cost 
function, but even providing an approximate answer that is asymptotically more
precise than taking the value of the function at a random point in the
configuration space is also a difficult problem~\cite{hastad01}. 
Hence a fast non-adiabatic
evolution has a computational interest if one can find a good compromise 
between the evolution time and the residual energy. In the classical case 
hardness of approximation results are often obtained via the PCP 
theorem~\cite{pcp}. In the quantum complexity litterature
a quantum analog of the PCP theorem has been conjectured 
in~\cite{qpcp_conjecture}. For recent works on the approximation algorithms in 
the quantum complexity setting we refer the readers 
to~\cite{qpcp_hastings,qapprox_kempe}.

In this paper we shall investigate the annealing on non-adiabatic timescales
for a class of ferromagnetic, non disordered, mean-field models of the 
fully-connected type, with $p$-spin interactions. These can of 
course not be considered as difficult optimization problems. However, despite
their simplicity that allows for an analytical resolution, they exhibit
some of the features expected also in more realistic optimization problems,
and therefore constitute useful toy-models to study. The 
statics~\cite{botet83,ViMoDu04,ribeiro06,ribeiro08,jorg10,FiDuVi11,SeNi12}
and the dynamics, both for quantum annealings~\cite{santoro08,solinas08,itin09_1,itin09_2}
and for quantum quenches~\cite{ViPaAs04,DaSeSeCh06,sciolla11}, 
of this kind of models have been largely studied.
From a technical point of view these models are relatively simple because
their mean-field character allows for a semi-classical treatment,
the small parameter in this limit being the inverse of the size of the system 
(instead of $\hbar$ in usual semi-classical computations). In most of
previous works this semi-classical limit has been achieved through the
introduction of spin coherent-states~\cite{ribeiro08,RiPa09}, or instantonic 
computations~\cite{jorg10}. Here our treatment will have more of a WKB 
flavour, with the magnetization playing the role of a particle coordinate. 
Moreover most of these works dealed with second-order phase transitions, at 
the exception of~\cite{jorg10,FiDuVi11}, which studied the statics of models 
with first-order transitions. The annealing dynamics of such models with 
first-order phase transition was not investigated before, to the best of our 
knowledge.

Let us now explain the structure of the paper. In Sec.~\ref{sec_def_and_thermo}
we introduce the definition of the models (\ref{sec_def}) and present
their thermodynamic behavior and phase diagram (\ref{sec_thermo}).
Section~\ref{sec_statics} contains our investigations on their static 
properties, at a refined level with respect to the thermodynamic
quantities. As explained in a first part (Sec.~\ref{subsubsec_spin_sectors}) 
of this section, their
mean-field character induces strong symmetries that allow to decompose
their spectrum in various disconnected sectors. We provide an ordering
theorem between different sectors in Sec.~\ref{sec_ordering}.
In Sec.~\ref{sec_salient} we discuss the qualitative features of the
spectrum of the model, and point to the following parts of the text where
they are quantitatively derived. The main technical result is established
in Sec.~\ref{sec_stat_sc}, where we show how to determine the eigenvectors, 
at the leading level in the thermodynamic limit.
This is then applied to the computation of various quantities: 
the density of states inside one sector (Sec.~\ref{subsubsec_dos}), 
the finite gap between levels away from transitions 
(Sec.~\ref{sec_finite_gaps}), and the exponentially small gaps in 
Sec.~\ref{sec_small_gaps}. The latter part is divided according to the
location of the quasi-degenerate levels in the spectrum, 
we consider in particular the exponentially small gap between the groundstate
and the first excited state at a first-order 
transition in Sec.~\ref{subsubsec_gap_first_order}, and the 
exponentially small gap between the two ferromagnetic phases for even $p$
in Sec.~\ref{subsubsec_gap_ferro}. The dynamics of the models is studied in 
Sec.~\ref{sec_dynamics}. After a precise definition of the annealing
procedure in Sec.~\ref{sec_dynamics_definitions} we recall the basic 
mechanism of the Landau-Zener model in Sec.~\ref{sec_dynamics_LZ} and
discuss the behavior of the dynamics that it suggests, in view of the
properties of the spectrum derived previously. The actual results
are presented in Sec.~\ref{subsec_exptimes} (resp. Sec.~\ref{subsec_csttimes}) 
for annealing on exponentially large (resp. finite) timescales. A simplified
model, introduced in Sec.~\ref{subsubsec_def_full_matrix}, is also studied
for comparison. We finally draw our conclusions in Sec.~\ref{sec_conclu}.
Some technical details are deferred to a series of Appendices.

\section{Definition and thermodynamic properties 
of the models}
\label{sec_def_and_thermo}

\subsection{Definition}
\label{sec_def}

We shall consider Hamiltonians of interacting spins 1/2, acting on the
Hilbert space spanned by 
\hbox{$\{|\us \ra | \us=(\s_1,\dots,\s_N)\in\{-1,+1\}^N \}$}.
We denote $\hsix$, $\hsiy$ and $\hsiz$ the Pauli matrices acting on the $i$-th
spin, and recall that in this basis, $\hsiz |\us \ra = \s_i |\us \ra $, 
$\hsix |\us \ra = |\us^{(i)} \ra $, where $\us^{(i)}$ is the configuration
obtained from $\us$ by flipping the $i$-th spin. The transverse and 
longitudinal magnetization per spin operators are defined as follows~:
\beq 
\hmx = \frac{1}{N} \sum_{i=1}^N \hsix  \ , \qquad 
\hmz = \frac{1}{N} \sum_{i=1}^N \hsiz \ .
\eeq
The Hamiltonian of the fully-connected $p$-spin ferromagnet is usually defined 
as $-N (\hmz)^p - \Gamma N \hmx$,
i.e. with $p$-body interactions along the $z$ axis, and a transverse field
$\Gamma$ along the $x$ axis. The dependency in $N$ is chosen to ensure the
extensivity of the model in the thermodynamic limit. For future convenience
we shall trade $\Gamma$ for a parameter $s\in[0,1]$ and define
\beq
\hH(s) = - N s (\hmz)^p - N (1-s) \hmx \ .
\label{eq_def_model}
\eeq
Up to a change of the energy scale these two definitions are equivalent, with
the correspondance $\Gamma = \frac{1-s}{s}$. The two limits $s=0$ and $s=1$
corresponds to a pure transverse field and pure ferromagnetic interactions
along $z$, respectively. The mean-field character of the model arises from
the form of the interacting term, which depends on the total magnetization 
only. The $p=2$ case corresponds to the quantum Curie-Weiss model, which can
also be viewed as the anisotropic version of the Lipkin-Meshkov-Glick (LMG) 
model~\cite{LMG,botet83,ribeiro08} (the general LMG model contains pair-wise 
interactions in the $y$ and $z$ directions). The case $p\ge 3$ was 
investigated in~\cite{jorg10}, and generalized in~\cite{FiDuVi11} to a model 
where both $\hmz$ and $\hmx$ are raised to arbitrary powers, and 
in~\cite{SeNi12} with the addition of antiferromagnetic pairwise interactions.
The methods and results developed in this paper for the model of 
Eq.~(\ref{eq_def_model}) are easily extended to these generalizations, as
sketched in the conclusions.

\subsection{Thermodynamic properties}
\label{sec_thermo}

We shall first briefly explain how to compute the free-energy density of 
this model, in the thermodynamic limit, and discuss its phase diagram. 
Similar derivations can be found in~\cite{cavity_ours,jorg10,SeNi12}.
For a rigorous treatment of such models we refer the 
reader to~\cite{QCW_rigorous}. The partition function at inverse 
temperature $\beta$ can be obtained by mapping the quantum problem to 
a classical one with one additional imaginary time direction.
Using the Suzuki-Trotter formula to disentangle the two non-commuting
terms in the Hamiltonian, and inserting representations of the
identity between each of the $\Ns$ Suzuki-Trotter slices one indeed
obtains:
\beq \label{eq_trotter_1} 
Z(\beta,s) \equiv \textrm{Tr} \, e^{- \beta \hH(s)}  
= \lim_{\Ns \rightarrow \infty} \underset{\us(1),\dots,\us(\Ns)}{\sum}
\prod_{\alpha=1}^{\Ns} \la \us(\alpha) | 
e^{\frac{\beta}{\Ns} s N (\hmz)^p} e^{\frac{\beta}{\Ns}  (1-s)N \hmx} 
|\us(\alpha+1) \ra  \ .
\eeq
In this expression $\us(1),\dots,\us(\Ns)$ are $\Ns$ Ising spin 
configurations, with periodic boundary conditions $\us(\Ns +1) = \us(1)$.
As $\hmz$ is diagonal in the basis chosen one obtains
\beq
Z(\beta,s) = \lim_{\Ns \rightarrow \infty} 
\underset{\us(1),\dots,\us(\Ns)}{\sum}
\prod_{\alpha=1}^{\Ns} 
e^{\frac{\beta}{\Ns} s N \left(\frac{1}{N}\sum_{i=1}^N \s_i(\alpha)\right)^p}
\la \us(\alpha) | 
e^{\frac{\beta}{\Ns}  (1-s)N \hmx} 
|\us(\alpha+1) \ra  \ .
\eeq
Thanks to the mean-field character of the model one can reduce the problem
to a single-site one by defining 
\hbox{$m(\alpha) = \frac{1}{N}\sum_{i=1}^N \s_i(\alpha)$} 
and imposing this definition,
for each $\alpha$, by an exponential representation of the Dirac distribution
with conjugate parameter $\lambda(\alpha)$:
\bea 
Z &=& \lim_{\Ns \rightarrow \infty} 
\int \prod_{\alpha=1}^{N_s} \frac{\dd m(\alpha) \dd \lambda(\alpha)}
{2 \pi \Ns/(\beta N)} \, 
e^{\frac{\beta N}{N_s} \sum_{\alpha=1}^{N_s} 
( s m(\alpha)^p -  \lambda(\alpha) m(\alpha))}
\underset{\us(1),\dots,\us(\Ns)}{\sum}
\prod_{\alpha=1}^{\Ns} 
\la \us(\alpha) | 
e^{\frac{\beta}{\Ns} \sum_{i=1}^N [\lambda(\alpha) \hsiz +  (1-s) \hsix] } 
|\us(\alpha+1) \ra \nonumber \\
&=&\lim_{\Ns \rightarrow \infty} 
\int \prod_{\alpha=1}^{N_s} 
\frac{\dd m(\alpha) \dd \lambda(\alpha)}{2 \pi \Ns/(\beta N)} \, 
\exp\left[ N \left( \frac{\beta}{N_s} \sum_{\alpha=1}^{N_s} 
( s \, m(\alpha)^p -  \lambda(\alpha) m(\alpha))+ 
 \ln \, \textrm{Tr} \prod_{\alpha=1}^{N_s} 
e^{\frac{\beta}{\Ns} (\lambda(\alpha) \hsz +  (1-s) \hsx)} 
\right) \right] \ .
\label{eq_trotter_2}  
\eea
Making the natural assumption that the dominant contribution comes
from values of $m(\alpha)$ and $\lambda(\alpha)$ that are constant
in imaginary time and equal to $m,\lambda$ respectively, and
evaluating the integral via the saddle-point method yields
\beq
f(\beta,s) \equiv \lim_{N \to \infty} - \frac{1}{\beta N} \ln  Z(\beta,s)
= \inf_m \underset{\lambda}{\text{ext}} \left[ - s \, m^p + \lambda \, m - \frac{1}{\beta}
\ln 2 \cosh(\beta \sqrt{\lambda^2 + (1-s)^2}) \right] \ .
\label{eq_f_variational}
\eeq
The stationarity conditions for this function of $(m,\lambda)$ are
\beq
\lambda = p \, s \, m^{p-1} \ , \qquad 
m = \frac{\lambda}{\sqrt{\lambda^2 + (1-s)^2}} 
\tanh (\beta\sqrt{\lambda^2 + (1-s)^2}  ) \ .
\label{eq_stationarity}
\eeq
Note that an alternative derivation of this result consists in making
a mean-field approximation $(\hmz)^p \to \la \hmz \ra^p + p \la \hmz \ra^{p-1} 
(\hmz - \la \hmz \ra)$ in
the Hamiltonian and setting self-consistently the average value in the 
single-spin problem thus obtained~\cite{BrMuTh66,WiViVeDu12}.

The various observables can be expressed in terms of the relevant 
critical point $(m_*(\beta,s),\lambda_*(\beta,s))$, in particular the 
longitudinal and transverse magnetization per spin read respectively
\beq
\la \hmz \ra = m_*(\beta,s) \ , \qquad 
\la \hmx \ra = \frac{1-s}{\sqrt{\lambda_*(\beta,s)^2 + (1-s)^2}} 
\tanh (\beta\sqrt{\lambda_*(\beta,s)^2 + (1-s)^2}  ) = 
\frac{1-s}{s}\frac{1}{p} m_*(\beta,s)^{2-p} \ .
\label{eq_thermo_mzmx}
\eeq
These expressions are easily obtained by adding to the Hamiltonian 
appropriate fields conjugated to the observables and by deriving the 
variational free-energy with respect to these additional fields; the
last expression of Eq.~(\ref{eq_thermo_mzmx}) is only valid under the 
assumption $m_*(\beta,s) \neq 0$.

\begin{figure}
\centerline{
\includegraphics[width = 5.5cm]{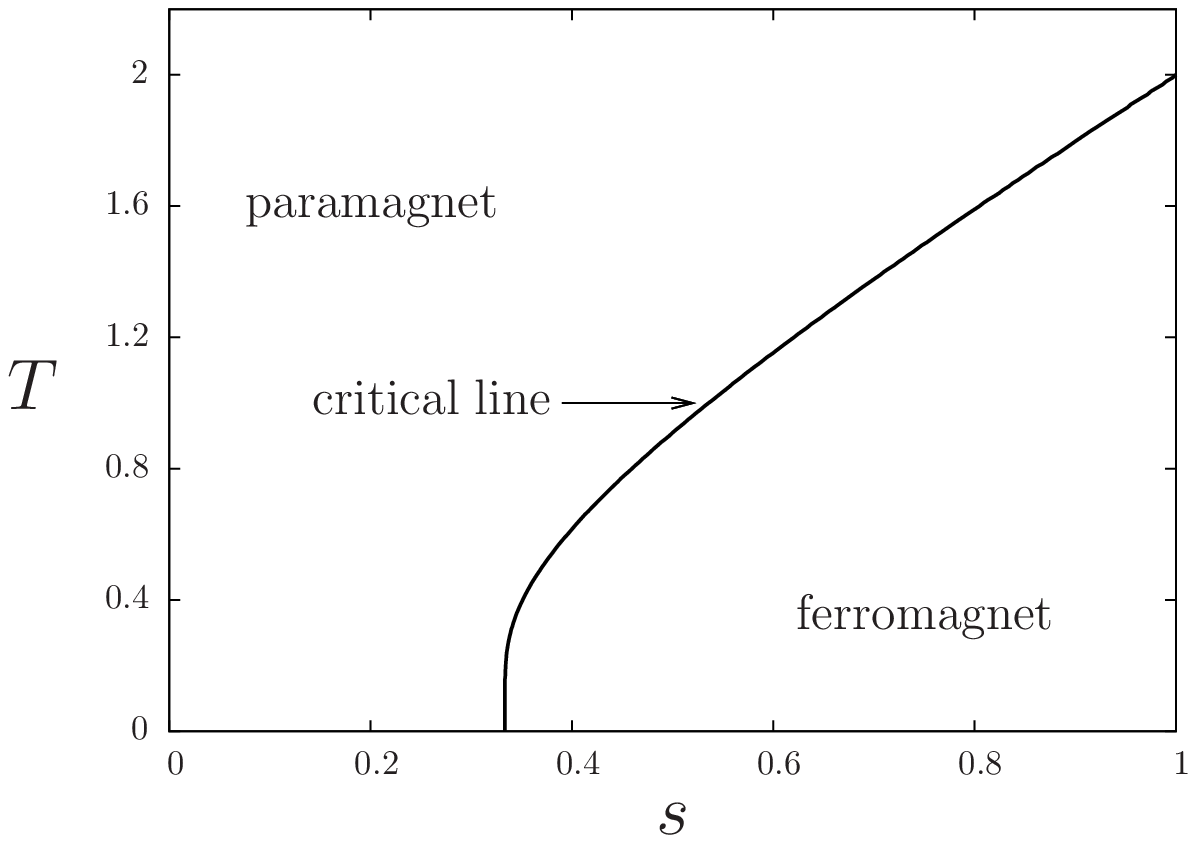}\hspace{6mm}
\includegraphics[width = 5.2cm]{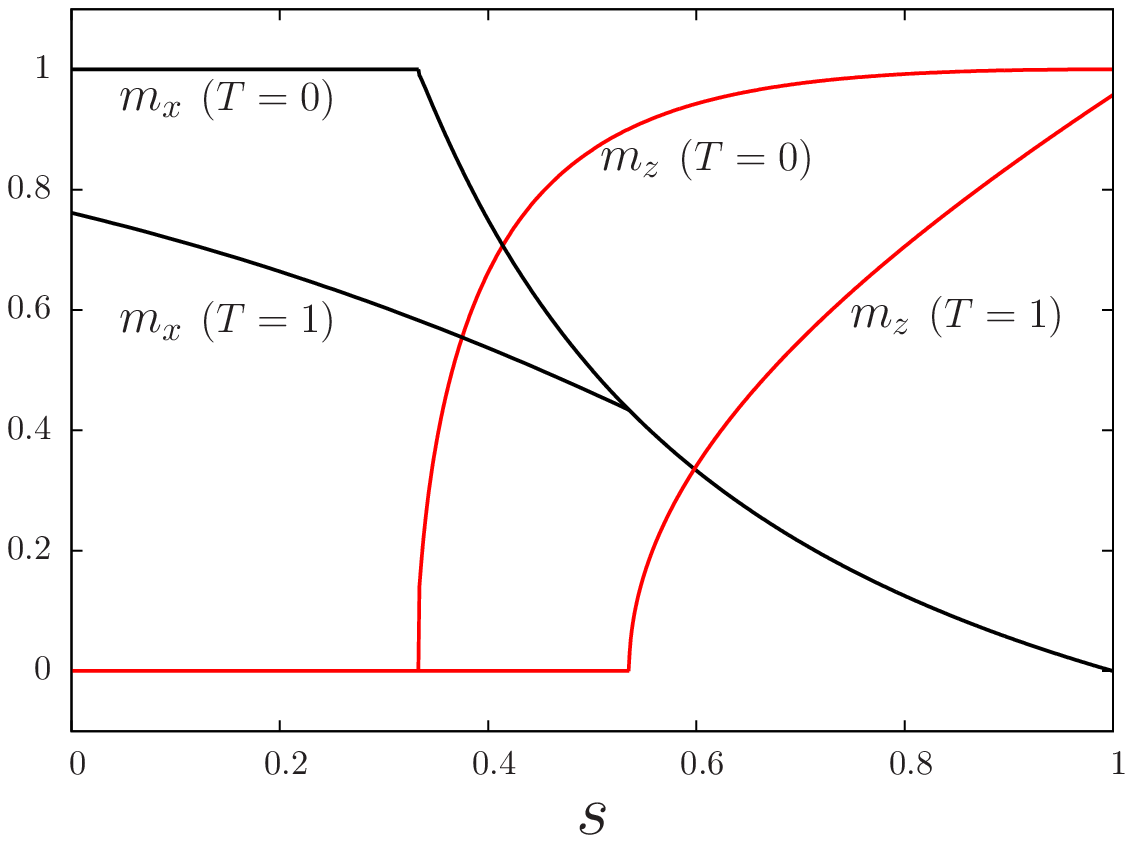}\hspace{6mm}
\includegraphics[width = 5.5cm]{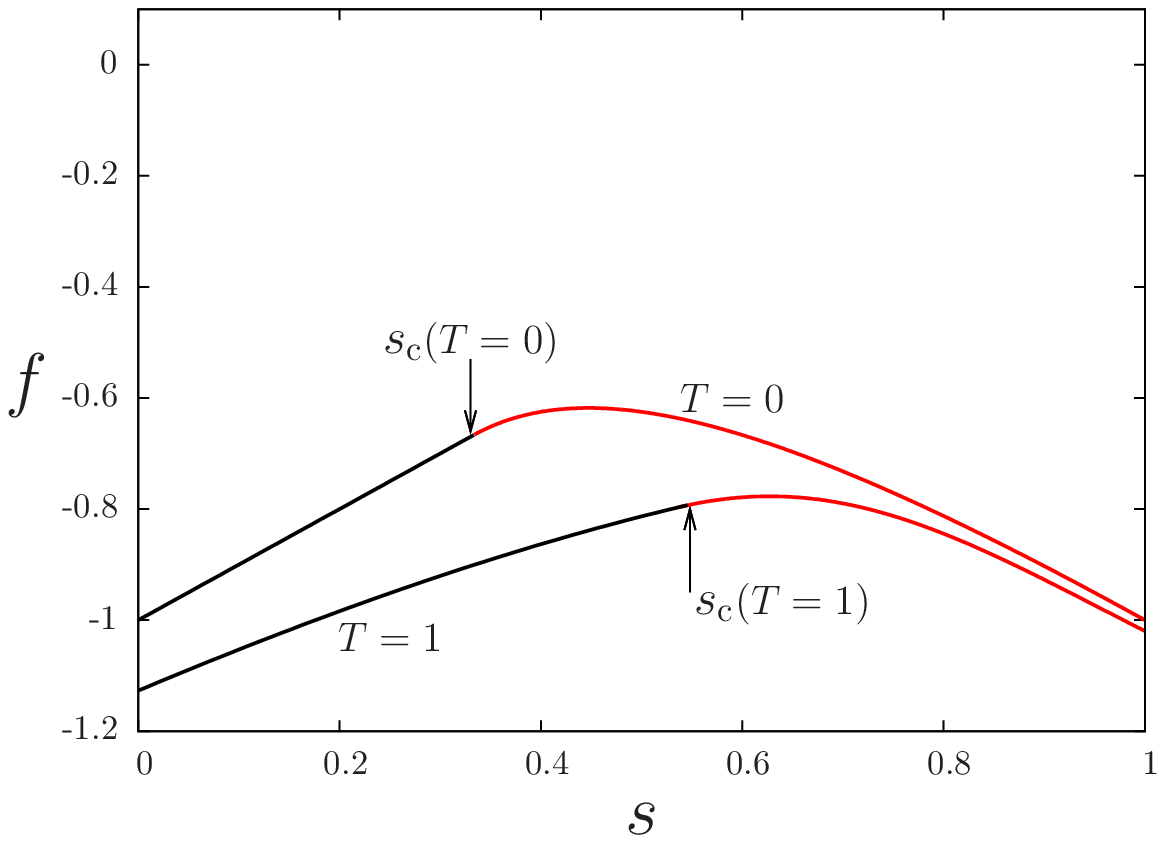}}
\caption{Thermodynamic properties of the $p=2$ model.
Left panel: phase diagram, the line indicates the value $\Tc(s)$
of the critical temperature for the second-order transition between the 
paramagnetic and the ferromagnetic phases. 
Center panel: longitudinal ($m_z$, red lines) 
and transverse ($m_x$, black lines) magnetizations as 
a function of $s$, for $T=1$ and $T=0$; in the ferromagnetic phase $m_x$
is independent of $T$, as apparent from the last expression of 
Eq.~(\ref{eq_thermo_mzmx}) with $p=2$.
Right panel: free-energy density as a function of $s$ for $T=1$, and
ground-state energy as a function of $s$; the arrows indicate the values
$s_{\rm c}(T)$ of the transition between the paramagnet (black lines) and
the ferromagnet (red lines).
}
\label{fig_thermo_p2}
\end{figure}

Let us first discuss the solution of these equations for $p=2$, and present
the associated phase diagram. The point $(m,\lambda)=(0,0)$ is always
a solution of Eq.~(\ref{eq_stationarity}). There is however a line in the
$(s,\beta)$ plane separating a paramagnetic phase (at low values of $\beta,s$,
i.e. high values of the temperature and transverse field) where it corresponds
to the global minimum of the function in~(\ref{eq_f_variational}), from
a ferromagnetic phase where it becomes a local maximum. In the latter phase
there appears two global minima related by the symmetry operation 
$(m,\lambda) \to (-m,-\lambda)$. The spontaneous longitudinal magnetization
$m_*(\beta,s)>0$ grows continuously from 0 at the border of the ferromagnetic
phase, with the usual mean-field exponent $\beta=1/2$. The phase transition
is thus of second order, the free-energy and its first derivatives being
continuous at the transition. These properties are illustrated in 
Fig.~\ref{fig_thermo_p2}.

\begin{figure}[h]
\centerline{
\includegraphics[width = 5.5cm]{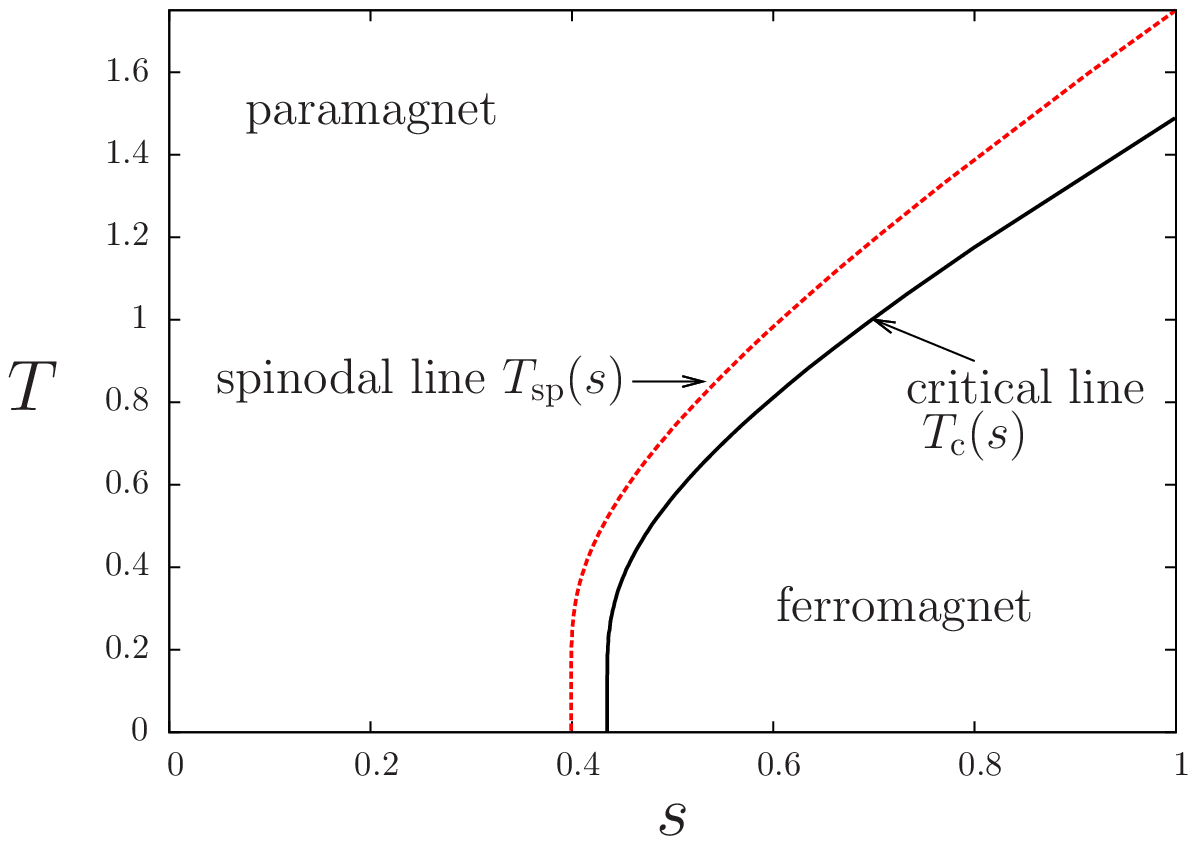}\hspace{6mm}
\includegraphics[width = 5.2cm]{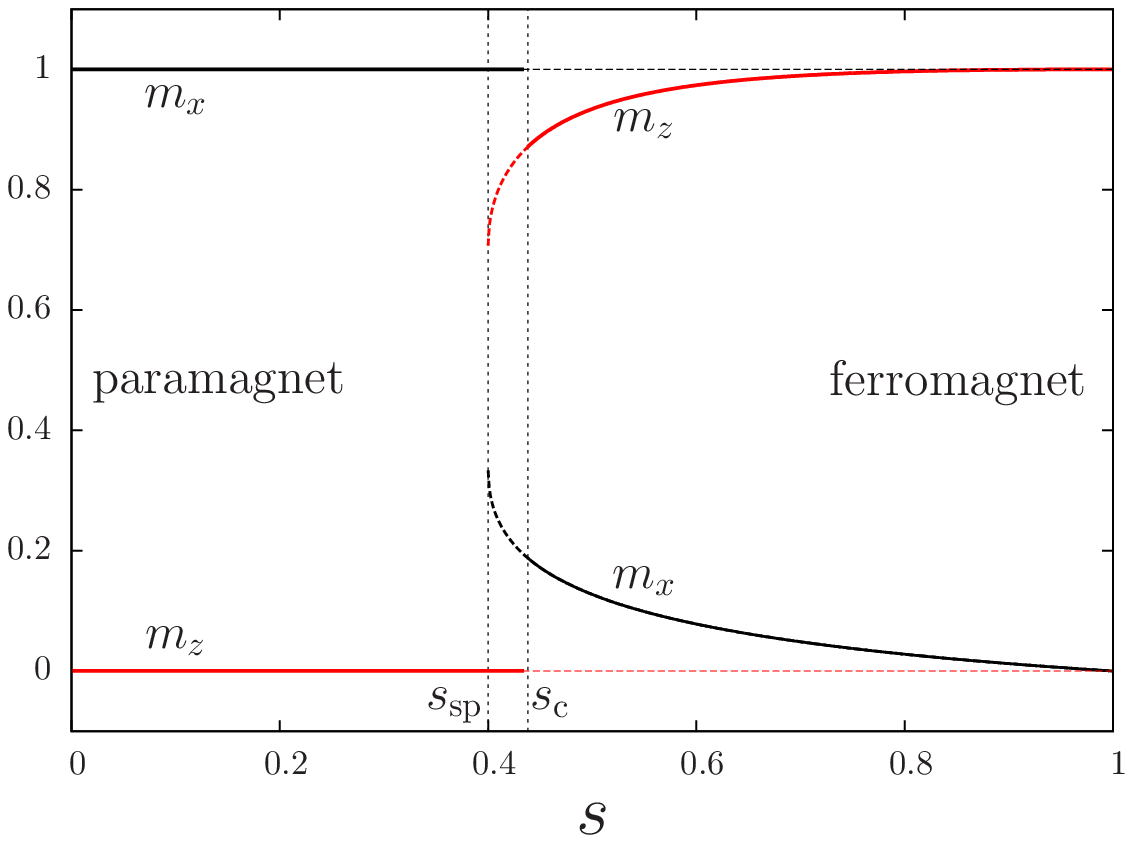}\hspace{6mm}
\includegraphics[width = 5.5cm]{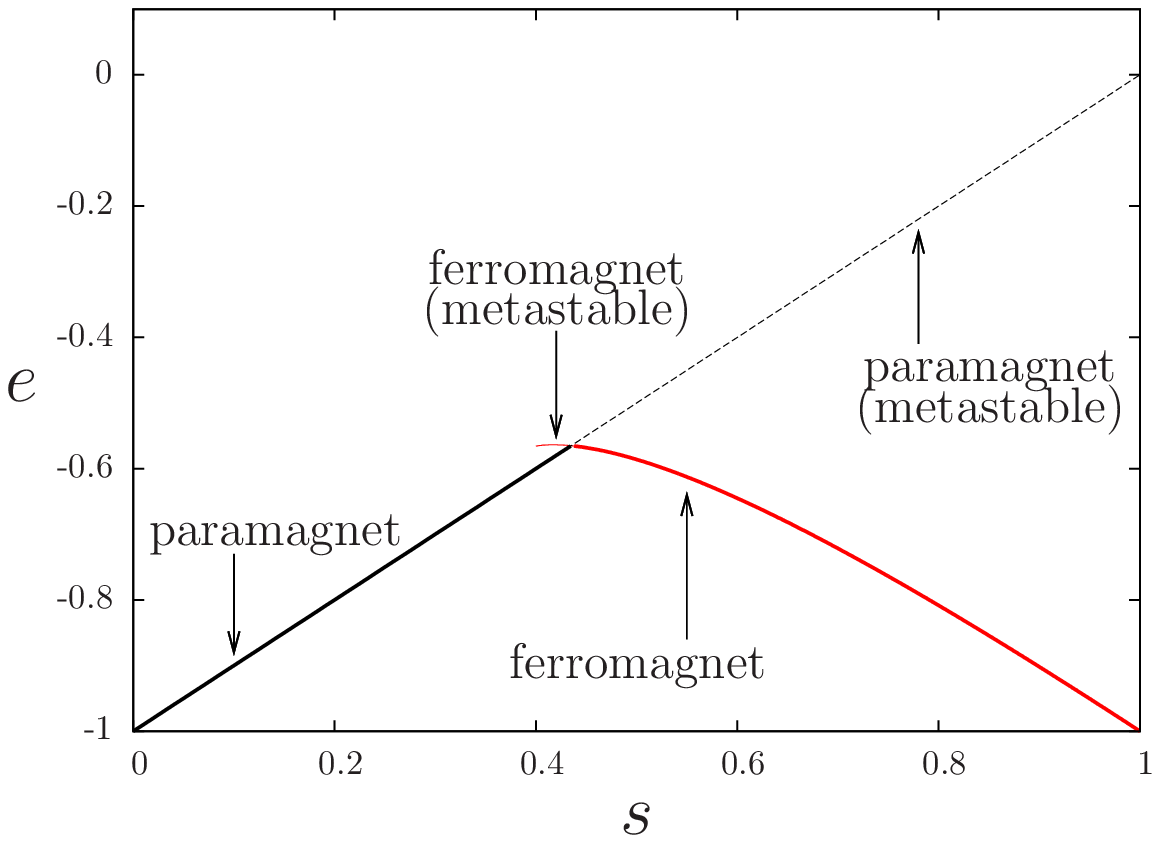}}
\caption{Thermodynamic properties of the $p=3$ model (all cases with $p \ge 3$
are qualitatively identical). Left panel: phase diagram, the solid (black) line
stands for the first-order transition line $\Tc(s)$, the dashed (red) line
being the spinodal curve $\Tsp(s)$ for the limit of existence of the 
ferromagnetic phase.
Center panel:  longitudinal ($m_z$, red line) and transverse ($m_x$, black line)
magnetizations as a function of $s$, at $T=0$. Solid thick part of the curves 
correspond to the thermodynamically relevant phase (paramagnetic for
$s<s_{\rm c}$, ferromagnetic for $s>s_{\rm c}$), dashed light ones to the 
metastable ones ($s>s_{\rm c}$ for the metastable paramagnet, 
$s\in[\ssp,s_{\rm c}]$ for the metastable ferromagnet). Note the square-root 
singularity at $\ssp$ for the magnetizations of the ferromagnetic
phase. 
Right panel: groundstate energy density as a function of $s$; solid thick and
dashed light have the same meaning as in the center panel.}
\label{fig_thermo_p3}
\end{figure}

Consider now the case $p \ge 3$. The paramagnetic solution $(m,\lambda)=(0,0)$
of Eq.~(\ref{eq_stationarity}) is then a local minimum (with respect to $m$) 
of the function in~(\ref{eq_f_variational}) for all values of $(\beta,s)$. 
For low values of $\beta,s$ this is the only minimum 
of~(\ref{eq_f_variational}). Beyond a line $\betasp(s)$ (or equivalently 
$\ssp(\beta)$), another local minimum appears discontinuously in 
$m_*(\beta,s)>0$ (if $p\ge 4$ is even there is also a symmetric one in 
$-m_*(\beta,s)$). At its appearance
this non-trivial local minimum corresponds to an higher free-energy density
than the paramagnetic one. It is only for strictly larger values of
$\beta,s$ than their free-energy density becomes equal, on the line 
$\betac(s)>\betasp(s)$ (or $s_{\rm c}(\beta)>\ssp(\beta)$). The model thus
exhibits a first-order phase transition along the line $\betac(s)$, 
associated to a discontinuity in the first derivatives of the free-energy 
density, which implies in particular a discontinuity of the magnetizations. 
The line $\betasp(s)$ is the
spinodal of the ferromagnetic phase, i.e. the limit of its existence as
a metastable local minimum of the free-energy. Note that the paramagnetic
phase is always locally stable, there is thus no spinodal line for this
phase. The features of this first-order transition are illustrated on 
Fig.~\ref{fig_thermo_p3}; to anticipate the discussion of the rest of the
paper the results displayed there are at zero temperature, yet they would be
qualitatively identical at any positive temperature below the transition
temperature of the classical model (at $s=1$).

In the remaining of this section we shall collect for future use some more 
explicit formulas valid in the zero-temperature limit, which will 
be the most useful case in the following of the paper. The groundstate energy 
density is obtained from~(\ref{eq_f_variational}) as
\beq
\egs(s) = \lim_{\beta \to \infty} f(\beta,s) =  
\inf_{m} \underset{\lambda}{\text{ext}} \left[ - s \, m^p + \lambda \, m 
- \sqrt{\lambda^2 + (1-s)^2} \right] \ .
\eeq
One can solve explicitly the stationarity condition with respect to $\lambda$,
which yields $\lambda=(1-s) m /\sqrt{1-m^2}$ and thus
\beq
\egs(s)=\inf_m \left[ - s \, m^p - (1-s) \sqrt{1-m^2} \right] \ .
\label{eq_egs_thermo}
\eeq
The energy corresponding to the paramagnetic state $m=0$ is $\epm(s)=-(1-s)$.
For $p\ge 3$ the ferromagnetic phase exists when $s\in[\ssp,1]$, where $\ssp$ 
is the zero-temperature limit of the spinodal line. We shall denote $m_*(s)>0$
the non-trivial solution of the stationarity equations corresponding to a local
minimum for $s\in[\ssp,1]$, and 
\hbox{$\efm(s)=-s \, m_*(s)^p - (1-s)\sqrt{1-m_*(s)^2}$} 
the corresponding energy. We also define $\mi(s)$ and $\ei(s)$ as the 
magnetization and energy of the local maximum (unstable phase of intermediate
magnetization) of the function in (\ref{eq_egs_thermo}).
These magnetizations are the solutions, ordered with $0<\mi(s)<m_*(s)$, of the 
equation
\beq
m = p \frac{s}{1-s} m^{p-1} \sqrt{1-m^2} \ .
\label{eq_mag_thermo}
\eeq
The average magnetizations are then given by
\beq
\la \hmz \ra = m_*(s) \ , \qquad 
\la \hmx \ra = \sqrt{1-m_*(s)^2} \ .
\eeq

\begin{figure}
\centerline{\includegraphics[width=5.5cm]{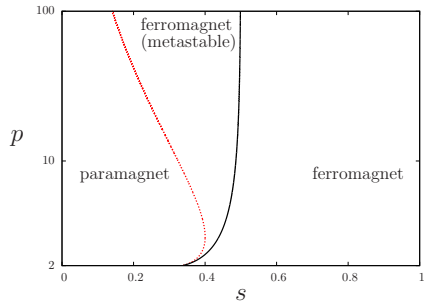}}
\caption{
Spinodal line $\ssp(p)$ and phase transition line $s_{\rm c}(p)$ in the 
$(s,p)$ plane, in logarithmic scale on the $p$ axis. 
For $p$ large, $s_{\rm c} \rightarrow 1/2$ whereas 
$\ssp \rightarrow 0$ (the domain of existence of the metastable ferromagnetic 
phase grows with growing $p$).}
\label{fig_thermo_largep}
\end{figure}

At the spinodal point we call
$\msp=m_*(\ssp)=\mi(\ssp)$ and $\esp=\efm(\ssp)=\ei(\ssp)$ the longitudinal 
magnetization and
the energy, while the first-order transition happens for $s_{\rm c}$ such
that $\epm(s_{\rm c})=\efm(s_{\rm c})=\ec$, with $\mc=m_*(s_{\rm c})$.
Explicit formulas can be given for these quantities. Consider first the
spinodal point. $m_*(s)$ is the solution of an implicit equation of the
form $m=g(m,s)$, with the function $g$ defined by the r.h.s. of 
Eq.~(\ref{eq_mag_thermo}). The spinodal corresponds to a bifurcation of
this implicit equation, in consequence $(\msp,\ssp)$ are solutions of
$\msp=g(\msp,\ssp)$ and 
$1=\left. \frac{\partial g}{\partial m} \right|_{(\msp,\ssp)}$. Solving this 
system yields
\beq 
\msp=\sqrt{1-\frac{1}{p-1}} \ , \qquad 
\ssp = \frac{1}{1+ p \frac{(p-2)^{(p-2)/2}}{(p-1)^{(p-1)/2}}}
\ .
\label{eq_thermo_spin} 
\eeq
At the first-order transition point, the equation $\mc=g(\mc,s_{\rm c})$
is supplemented by the condition $\epm(s_{\rm c})=\efm(s_{\rm c})$, which leads
to
\beq 
\mc = \sqrt{\frac{p(p-2)}{(p-1)^2}} \ , \qquad
s_{\rm c} = \frac{1}{1+\frac{p}{p-1} \left( \frac{p(p-2)}{(p-1)^2}\right)^{\frac{p-2}{2}}} \ , \qquad 
\ec = -(1-s_{\rm c}) \ . 
\label{eq_critical_parameters}
\eeq
The dependency on $p$ of the spinodal and critical point parameters $\ssp$ and
$s_{\rm c}$ are plotted on Fig.~\ref{fig_thermo_largep}. From the above explicit
expressions one can in particular work out the large $p$ asymptotics, that
read
\beq
\msp=1-\frac{1}{2p} + O(p^{-2}) \ , \quad \ssp = \frac{e^{1/2}}{\sqrt{p}} +
O(p^{-3/2}) \ , \quad \mc = 1-\frac{1}{2p^2} + O(p^{-3}) \ , \quad
s_{\rm c} = \frac{1}{2} - \frac{1}{8p} + O(p^{-2})  \ .
\label{eq_critical_parameters_largep}
\eeq

\section{Detailed description of the spectrum}
\label{sec_statics}

We shall now turn to a refined description of the statics of the model,
beyond the computation of the thermodynamic limit of its free-energy density.
This detailed study of the eigenvalues and eigenvectors of $\hH(s)$ will be 
crucial for the understanding of the annealing dynamics presented in 
Sec.~\ref{sec_dynamics}. This section is organized as follows. 
In Sec.~\ref{subsubsec_spin_sectors} we exploit the symmetries of the model
to decompose its Hilbert space into several disconnected sectors. 
In Sec.~\ref{sec_ordering} we prove a result on the relative ordering of the 
eigenvalues between different symmetry sectors, and provide a finer 
conjecture motivated by numerical evidences. 
The qualitative features of the spectrum inside one symmetry sector are
discussed in Sec.~\ref{sec_salient}, and the rest of the section is
devoted to the quantitative derivation of these properties.
In Sec.~\ref{sec_stat_sc} we obtain 
the solution of the eigenvalue equation inside one sector, at the leading
exponential level (in a semi-classical fashion). This main technical result
is then exploited to obtain the density of states inside each sector 
(in Sec.~\ref{subsubsec_dos}), the finite gaps between eigenvalues 
(in Sec.~\ref{sec_finite_gaps}) and the exponentially small gaps
(in Sec.~\ref{sec_small_gaps}), in particular at the first order transition
of models with $p \ge 3$ (see Sec.~\ref{subsubsec_gap_first_order}), and in
the ferromagnetic phases for even $p$ (cf. Sec.~\ref{subsubsec_gap_ferro}).

\subsection{Decomposition of the Hilbert space in spin sectors}
\label{subsubsec_spin_sectors}

The diagonalization, be it numerical or analytical, of a quantum Hamiltonian
is in general a very difficult task because of the exponential growth of
the dimension of the Hilbert space with the size of the system. For the
fully-connected mean-field models under study this difficulty is greatly 
reduced thanks to their highly symmetric structure: as a matter of fact
the Hamiltonian is invariant under the permutation of any pair of spin
indices. 

Let us first briefly explain how to exploit this symmetry in an abstract and 
general way. 
The Hilbert space $\cH$ of an $N$-component system is the tensorial 
product $\cH = V^{\otimes N}$ of the space $V$ of each component. The theory
of representation~\cite{FultonHarris} asserts that such a tensor product
can be decomposed as a direct sum of vector spaces, classified according to
their symmetry properties with respect to permutations. More precisely,
in order to construct $V^{\otimes N}$ one has to sum over the Young diagrams 
with $N$ boxes and no more than $d$ rows, where $d$ is the dimension of $V$.
Each of these diagrams gives rise to a Young symmetrizer, i.e. an operator
on $V^{\otimes N}$ that, roughly speaking, completely symmetrizes along each row
and antisymmetrizes along each column of the diagram. The tensor product
$V^{\otimes N}$ can then be written as the direct sum of the images of the Young
symmetrizers; the degeneracies in this sum, as well as the dimensions of these
images, can be computed from the shape of the diagram. This decomposition can
be useful only if the Hamiltonian itself respect such permutation symmetries, 
as it becomes block-diagonal once written in this basis.

This general theory that we only sketched above greatly simplifies in our 
case, and its consequences can be understood with more physical arguments.
The important point is that the dimension $d$ of the base space of a
spin $1/2$ is only $2$ here. In consequence the Young diagrams have at most two
rows, and the sum over the diagrams reduces to a sum over the number 
$K=0,1,\dots,\fhN$ of elements in the second row. This number counts the
pair of spins over which the antisymmetrization procedure is accomplished.
It can be given a more intuitive interpretation as follows. The operators
$\hS^\alpha = \frac{N}{2} \hm^\alpha$ with $\alpha=x,y,z$
obey the commutation rules of an angular momentum; the total spin operator 
$\hS^2 = (\hSx)^2 + (\hSy)^2+(\hSz)^2$ has thus eigenvalues of the
form $S(S+1)$ with $S$ integer or half-integer. It turns out that the images
of a Young symmetrizer with a given value of $K$ are eigenspaces of $\hS^2$,
with total spin $S=\frac{N}{2}-K$. In particular the states of maximal
spin $N/2$ correspond to fully symmetric states. More generally, the results
of the abstract construction can be recovered by using recursively the 
standard rules for the addition of angular momenta.

Let us now summarize these results and write explicit formulas for the
matrix elements of the Hamiltonian. There are 
\beq
\NNK= \binom{N}{K} \frac{N+1-2K}{N+1-K} = \binom{N}{K} - \binom{N}{K-1} 
\label{eq_value_NNK}
\eeq
distinct eigenspaces of $\hS^2$ with spin $S=N/2-K$ (as could be expected
the fully symmetric space $K=0$ is unique). Each of them has dimension
$2S+1=N+1-2 K$; using the second expression of $\NNK$ an easy computation 
allows 
to check that the total dimension of the Hilbert space is indeed
\beq
\sum_{K=0}^\fhN \NNK (N+1-2K) = 2^N \ .
\eeq
The Hamiltonian $\hH(s)$ is stable with respect to this decomposition; moreover
its action on one of these subspaces depends only on the value of $K$, not
on the choice of one of the $\NNK$ degenerate sectors.
We shall denote $\hHK(s)$ the restriction of $\hH(s)$ to one of the 
subspaces of spin $N/2-K$, or equivalently view $\hHK(s)$ as a square matrix 
of order 
$N+1-2 K$. We will also use the notation $k=K/N$.
The subspace on which $\hHK(s)$ acts is spanned by the basis of
eigenvectors of $\hmz$, written $| m ; K \raz$ with the $N+1-2 K$ possible
values of $m$: $\MNK=\{-1+2 k, -1 + 2 k + 2/N,\dots,1-2 k -2/N,1-2k\}$.
The action of $\hmx$ on this vector amounts to increase or decrease the
value of $m$ by its minimal amount $2/N$, i.e.
\beq
\laz m ; K |\hmx| m';K \raz =  
\frac{1}{2} \sqrt{(1-2k+\max(m,m'))(1-2k-\min(m,m'))} \quad \text{for}
\ |m-m'|=2/N \ .
\label{eq_mx_on_mz}
\eeq
The matrix representing $\hHK(s)$ in this basis
has thus a symmetric tridiagonal form, with matrix elements:
\bea
\laz m;K | \hHK(s) | m;K \raz &=& - N \, s \, m^p \ , 
\label{eq_hHK_zbasis_d} \\ 
\laz m;K |\hHK(s)| m';K \raz &=& 
- N \frac{1-s}{2} \sqrt{(1-2k+\max(m,m'))(1-2k-\min(m,m'))} \quad \text{for}
\ |m-m'|=2/N \ .
\label{eq_hHK_zbasis_offd}
\eea
One can also define a second basis spanned by the eigenvectors 
$| m ; K \rax$ of $\hmx$. The expression of $\hmz$ in this basis is nothing
but Eq.~(\ref{eq_mx_on_mz}) with the interversion of the indices $z$ and $x$.
In this basis the matrix representation of $\hHK(s)$ has a diagonal part
corresponding to the action of the transverse field; the interaction term
$(\hmz)^p$ has a band diagonal form, with non-zero matrix elements between
eigenvectors $| m ; K \rax$ and $| m' ; K \rax$ when $\frac{N}{2}(m-m')\in 
\{p,p-2,\dots,-p+2,-p \}$. We shall give and use their explicit form, in the
large $N$ limit, in Sec.~\ref{sec_stat_sc}.

These symmetry considerations thus allow to reduce the complexity of the
full diagonalization of the Hamiltonian from a matrix problem of size $2^N$
to $\fhN +1$ matrices of sizes at most $N+1$. This great simplification
will be used in the following both for numerical and analytical computations.

The above reduction is valid for any model symmetric under all permutations
of spins, irrespectively of the precise form of the interactions. The 
Hamiltonian of Eq.~(\ref{eq_def_model}) exhibit additional symmetries:
\begin{itemize}
\item for odd values of $p$ the spectrum is invariant under the 
transformation $E \to -E$. Consider indeed an eigenvector $|\psi \ra$ of 
$\hHK(s)$, with eigenvalue $E$, written as 
$|\psi \ra = \underset{m \in \MNK}{\sum} c_m | m;K \raz$. Then the vector 
\hbox{
$|\psi' \ra= \underset{m \in \MNK}{\sum} c_{-m} (-1)^{\frac{N}{2} m} | m;K \raz$}
is an eigenvector of $\hHK(s)$, with eigenvalue $-E$.
\item for even values of $p$ the Hamiltonian is symmetric under global 
longitudinal magnetization reversal, and this implies that
each $\hHK(s)$ can be further decomposed in a
block diagonal form, with two blocks of sizes $\fhN +1 - K$ and $\chN - K$.
One can justify this statement in two ways. When acting on a basis vector
$| m ; K \rax$, the operator $(\hmz)^p$, with $p$ even, produces
a vector whose decomposition over $| m' ; K \rax$ is non-zero only for 
magnetizations $m'$ such that $N/2(m-m')$ is even:
non-zero off-diagonal matrix elements in the $x$ basis
are only found at an even distance from 
the main diagonal, hence in that basis the parity of the number of spin flips 
with respect to the fully polarized vector in the $x$ direction is conserved 
by the Hamiltonian.

In addition, the matrix representing $\hHK(s)$ in the 
$| m ; K \raz$ basis commutes with the matrix with 1 on the anti-diagonal 
(from bottom left to top right), that represents the reversal of the 
magnetization along the $z$ axis. The eigenvectors of $\hHK(s)$ can thus
be divided between those that are symmetric or antisymmetric under this
transformation.

\end{itemize}

\subsection{Ordering properties of the spectra}
\label{sec_ordering}

\begin{figure}
\centerline{
\includegraphics[width=8.3cm]{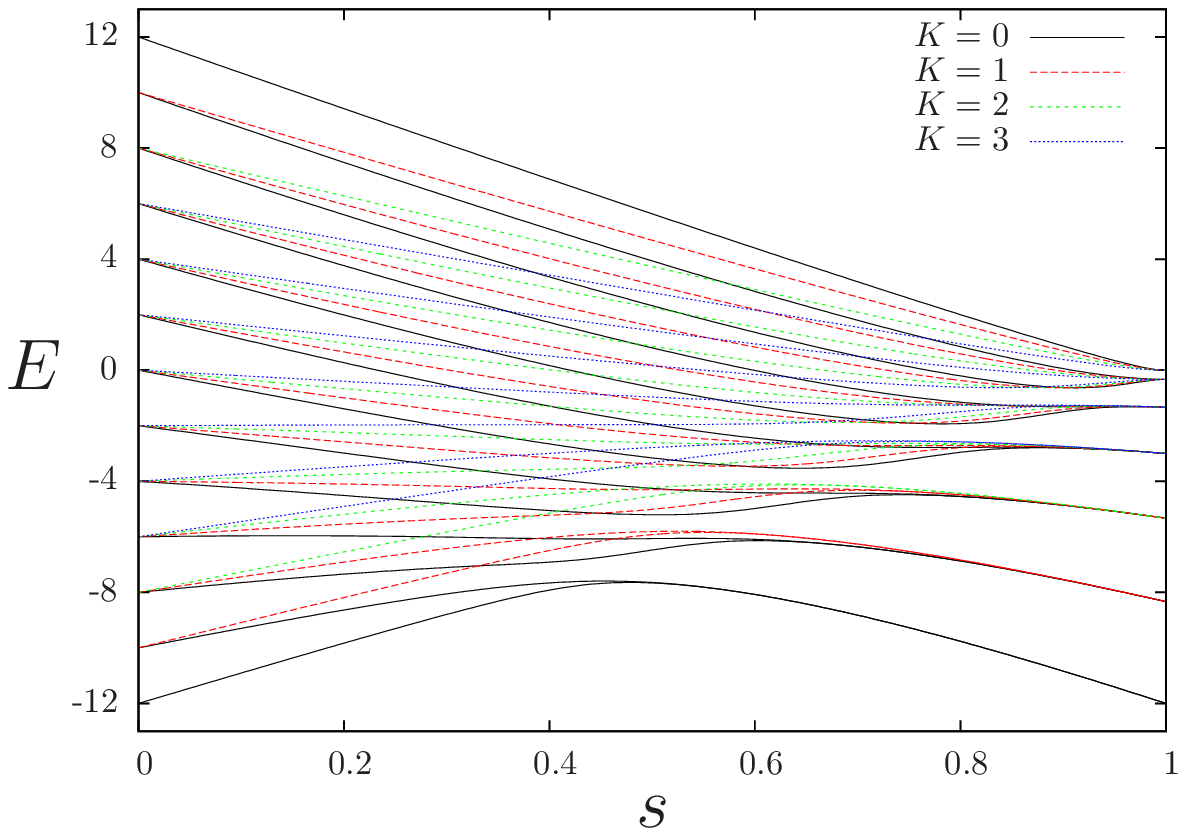} \hspace{6mm}
\includegraphics[width=8.3cm]{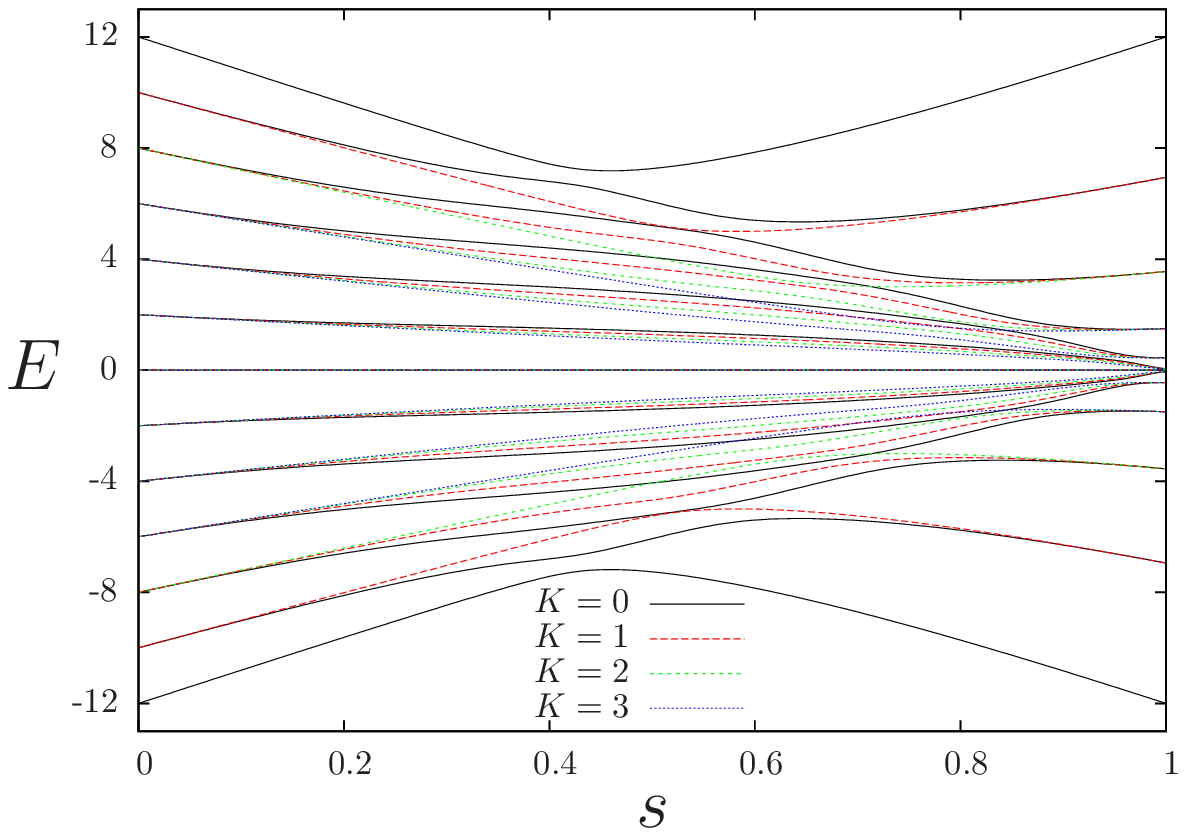}
}
\caption{The eigenvalues $E_i^{(K)}(s)$ of the Hamiltonian $\hH(s)$, 
for $N=12$, $p=2$ (left) and $N=12$, $p=3$ (right). The different values
of $K$ are distinguished by different colors and line styles.}
\label{fig_spectrum_spin_sectors}
\end{figure}

For each value of $K$ the restriction $\hHK(s)$ of $\hH(s)$ to a subspace of
spin $N/2-K$ has $N+1-2K$ real eigenvalues, that we shall denote
$E_0^{(K)}(s) \le E_1^{(K)} (s)\le \dots \le E_{N-2K}^{(K)}(s)$. We leave implicit
the dependency of these quantities on $N$ and $p$, that are understood to
be fixed in this whole subsection. By construction the 
Hamiltonian has no matrix elements between sectors of the Hilbert space
corresponding to different values of $K$; one could thus a priori think
that the spectrum $\{E_i^{(K)}(s)\}$ has no relationship to $\{E_i^{(K')}(s)\}$ for
$K \neq K'$. A quick look at the numerical results displayed in 
Fig.~\ref{fig_spectrum_spin_sectors} reveals on the contrary that
there are strong ordering rules between the energy levels of different spin 
sectors, reminiscent of the Lieb-Mattis theorem for the antiferromagnetic
Heisenberg model~\cite{LiMa62} (see also~\cite{NaSpSt04} for a more recent
treatment of the ferromagnetic case). As a first step we shall prove 
that the groundstates of each sector are strictly ordered according to the 
spin of the sector, i.e. that
\beq
E_0^{(0)}(s) < E_0^{(1)}(s) < \dots E_0^{(\fhN)}(s) \ \ \forall s \in[0,1] \ ,
\label{eq_ordering_gs}
\eeq
in such a way that the global groundstate of $\hH(s)$ lies in the fully 
symmetric subspace $K=0$, of maximal spin $N/2$.

The proof goes as follows. Let us denote $| \psi \ra$ the eigenvector of
$\hHK$ corresponding to its groundstate eigenvalue $E_0^{(K)}$, for some value
of $K>0$. We decompose this vector on the basis in which $\hmz$ is diagonal,
\beq
|\psi \ra = \sum_{m \in \MNK} c_m |m;K\raz \ .
\eeq
We saw above that in this basis $\hHK$ is a tri-diagonal matrix whose 
off-diagonal elements are all positive (cf. Eq.~(\ref{eq_hHK_zbasis_offd})).
The Perron-Frobenius theorem thus ensures that
the coefficients $c_m$ can be chosen to be all strictly positive.
Let us now define a vector $|\psi'\ra$ belonging to the space on which
$\hHKm1$ acts, according to
\beq
|\psi' \ra = \sum_{m \in \MNK} c_m |m;K-1\raz \ .
\eeq
In a column representation this amounts to supplement $|\psi \ra$ with two
null rows corresponding to the two values $m=\pm (1-2k+2/N)$. This vector
being normalized, the variational principle asserts that 
$E_0^{(K-1)} \le \la \psi'| \hHKm1 | \psi' \ra$. One finds easily that
\bea
\la \psi'| \hHKm1 | \psi' \ra = E_0^{(K)} - N (1-s) 
\sum_{m \in \MNK \setminus 1-2k} c_m c_{m+\frac{2}{N}} 
&&\left[ \sqrt{\left(1-2k+m+\frac{4}{N}\right)\left(1-2k-m+\frac{2}{N}\right)} 
\right. \nonumber \\ & & \left. - 
\sqrt{\left(1-2k+m+\frac{2}{N}\right)\left(1-2k-m\right)} \right] 
\ .
\eea
The difference of the square roots being strictly positive, as well as the 
product $c_m c_{m+\frac{2}{N}}$, one thus obtains
\hbox{$E_0^{(K-1)}(s) < E_0^{(K)} (s)$} 
as long as $s<1$. On the other hand for $s=1$ the
matrices are diagonal and the groundstate is obviously 
$E_0^{(K)}(s=1)=-N(1-2k)^p$, which also obeys the strict inequality 
$E_0^{(K-1)}(s=1) < E_0^{(K)} (s=1)$. This completes the proof of 
Eq.~(\ref{eq_ordering_gs}).

A closer look at the plots in Fig.~\ref{fig_spectrum_spin_sectors} suggests
that not only the groundstates are ordered between one sector and another,
but also that excited states are interleaved in a regular way. For instance
the first excited state of one sector, $E_1^{(K)}$, seems to always have
a lower energy than the groundstate of the following sector, $E_0^{(K+1)}$. 
More generally,
we propose the following conjecture based on this numerical investigation:
for all $n \leq N/2$ if $p$ is odd, $n \leq N$ if $p$ is even, one has
\beq
E_n^{(0)}(s) < E_{n-1}^{(1)}(s) < \dots < E_0^{(n)}(s) \ \ 
\forall s \in (0,1) \ .
\eeq
In particular, if this statement is true, the first excited state of the
Hamiltonian $\hH(s)$ in the full Hilbert space is always in the sector of 
maximal spin. The consistency of this conjecture for $s$ close to $0$ and $1$
can easily be checked by perturbative expansions.

\subsection{The salient features of the spectrum}
\label{sec_salient}
In this section we describe qualitatively the main features of the spectrum
of eigenstates in the symmetric sector of maximal spin (all sectors behaving
in a similar way), that are apparent by visual inspection of the
figures. We emphasize the connections with the thermodynamic computations of
Sec.~\ref{sec_thermo}, and also point to the following parts of the article 
where these properties are derived quantitatively.

\begin{figure}
\centerline{
\includegraphics[width=8.3cm]{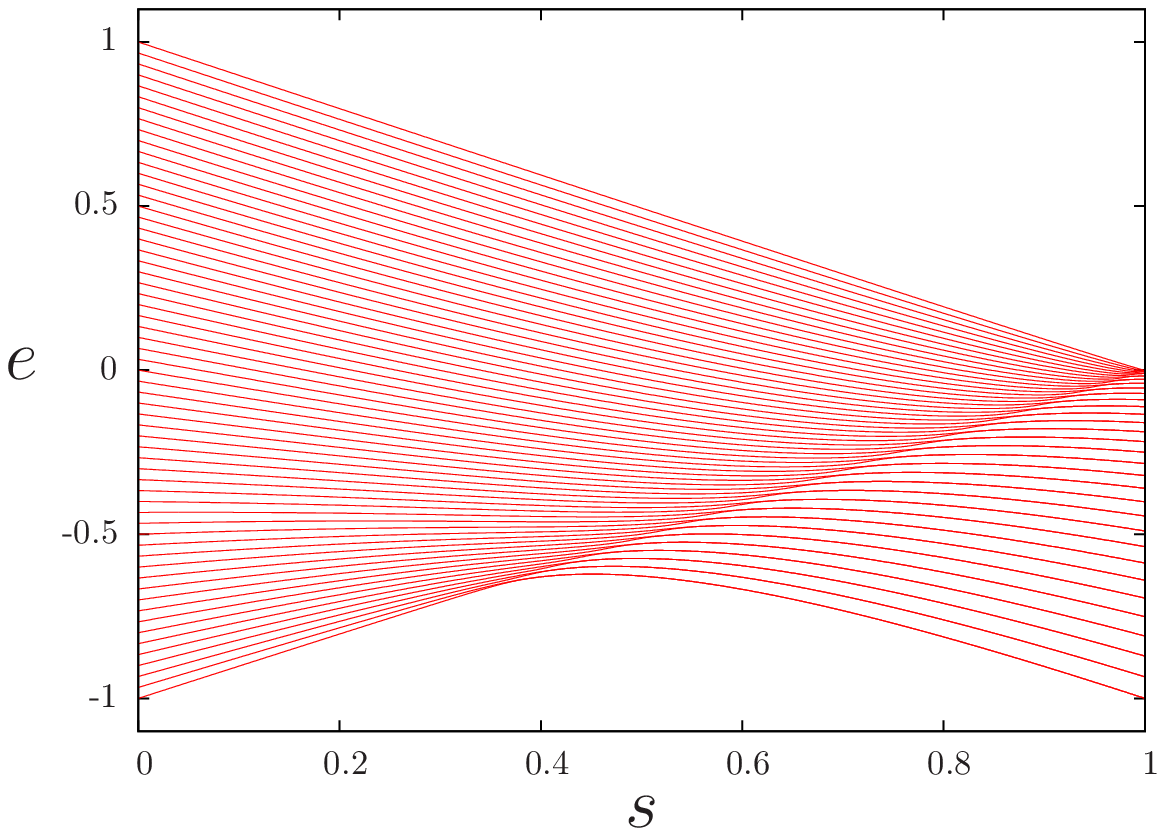} \hspace{6mm}
\includegraphics[width=8.3cm]{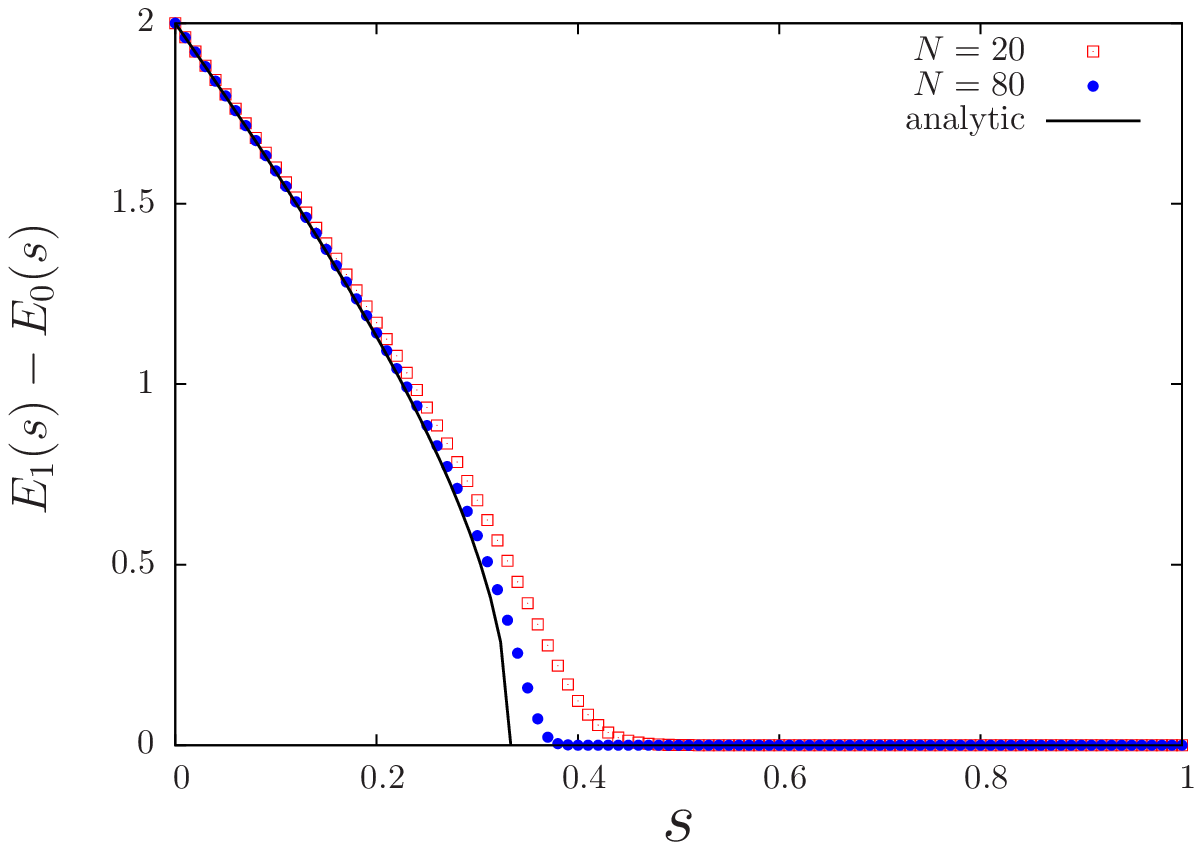}
}
\caption{Left panel: the spectrum of the symmetric sector of the 
$p=2$ model for $N=60$, obtained by numerical diagonalization.
Right panel: the gap between the groundstate and the first excited
state for $p=2$. The solid line is the result of an analytical computation,
see Eq.~(\ref{eq_gap_finite_p2}), the symbols have been obtained by numerical 
diagonalization.}
\label{fig_salient_p2}
\end{figure}

Let us begin with the $p=2$ case, for which the zero-temperature limit of the
thermodynamic computation
predicts a second-order phase transition at $s_{\rm c}=1/3$. The complete
spectrum of the symmetric sector is plotted on the left panel of 
Fig.~\ref{fig_salient_p2} for $N=60$. One observes indeed a good agreement
with the shape of the groundstate energy predicted previously and plotted
in the right panel of Fig.~\ref{fig_thermo_p2}. Looking more carefully at
the two states of lowest energy, one sees that the gap between them is of
order 1 (in extensive energy $E=N e$) in the paramagnetic phase (i.e. for 
$s<s_{\rm c}$), but exponentially small in $N$ in the ferromagnetic phase and
indistinguishable in this figure. This exponentially small splitting is
the consequence of the existence of the two magnetizations $\pm m_*(s)$ 
minimizing the thermodynamic groundstate energy (\ref{eq_egs_thermo}). 
On the right panel of Fig.~\ref{fig_salient_p2} we display the gap between
the lowest states for two finite values of $N$, along with the analytical
prediction of Eq.~(\ref{eq_gap_finite_p2}) for its limit when $N\to \infty$,
that we shall obtain in Sec.~\ref{subsubsec_dos}. The rate of the exponential
splitting in the ferromagnetic phase will be derived in 
Sec.~\ref{subsubsec_gap_ferro}, see Eq.~(\ref{eq_beta_p}) and right panel
of Fig.~\ref{fig_gaps_ferro}. The square-root vanishing of the gap when
$s \to s_{\rm c}^-$ and the behavior of the rate of exponential splitting when
$s \to s_{\rm c}^+$ leads, with a finite-size scaling assumption explained in
Sec.~\ref{subsubsec_gap_ferro}, to a polynomial
closing of the gap as $N^{-1/3}$ in the critical regime $s \approx s_{\rm c}$.
This behaviour has been first predicted on the basis of the scaling analysis
in~\cite{botet83}; the lifting of the degeneracy between
ferromagnetic states was also studied in~\cite{schulman76}. 
Note finally that the quasi-degeneracy of ferromagnetic states also occurs
for excited eigenvalues: on the left panel of Fig.~\ref{fig_salient_p2}
one gets the impression that there are twice as less states on the right of
a diagonal line $e=-(1-s)$ that on the left. This visual impression shall
be confirmed in Sec.~\ref{subsubsec_gap_excited}.

\begin{figure}
\centerline{
\includegraphics[width=8.3cm]{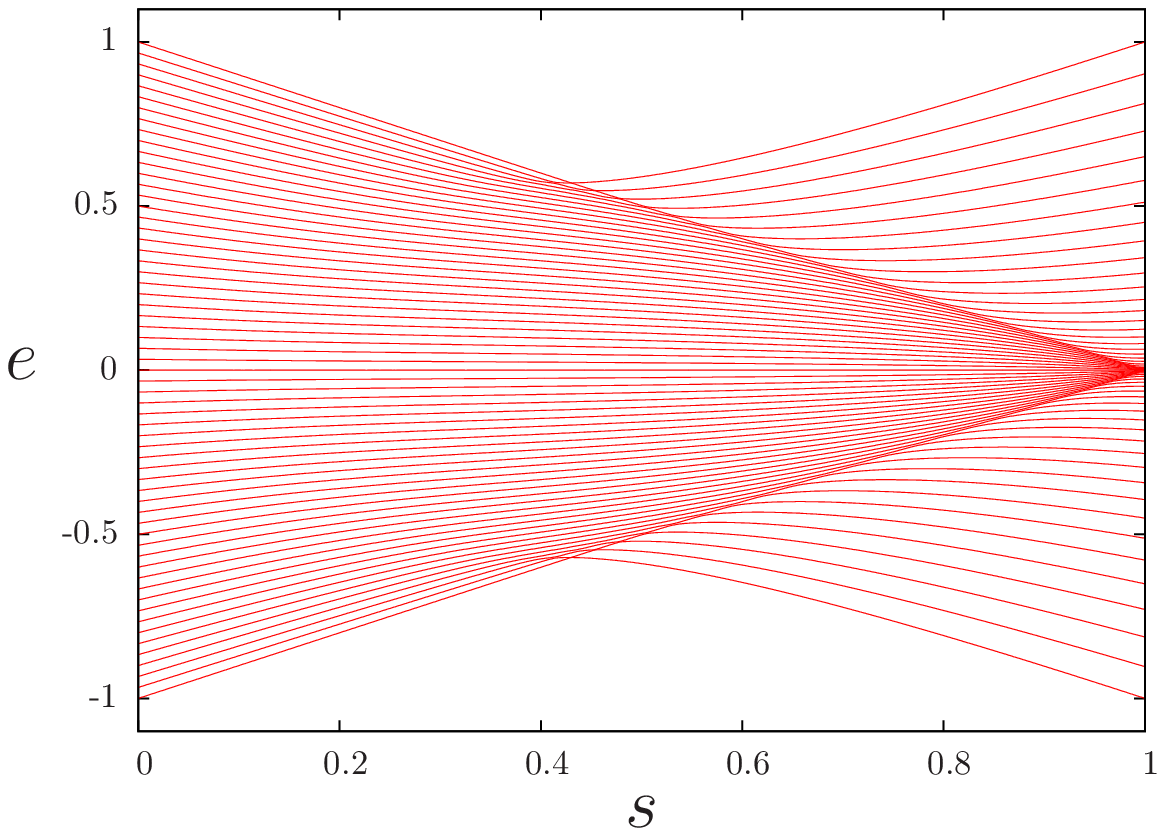} \hspace{6mm}
\includegraphics[width=8.3cm]{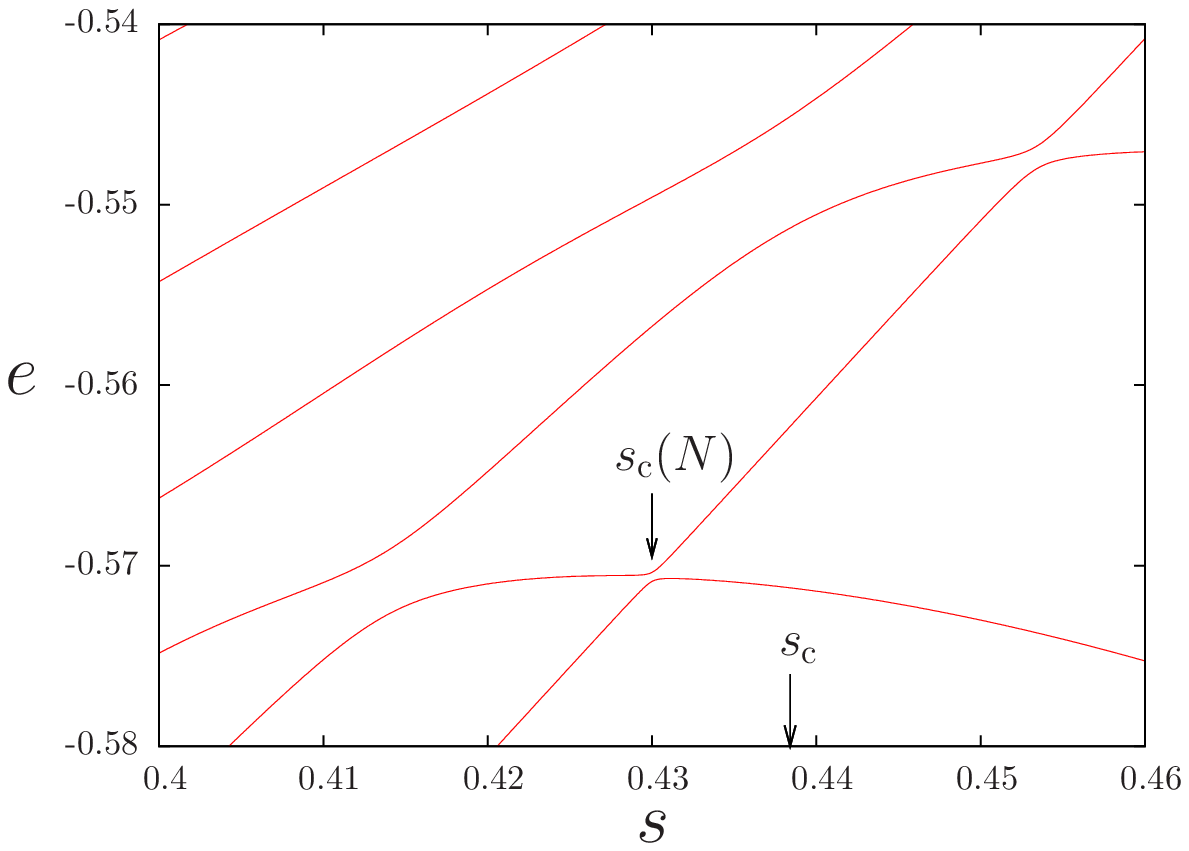}
}
\caption{Left panel: the spectrum of the symmetric sector of the 
$p=3$ model for $N=60$, obtained by numerical diagonalization.
Right panel: a zoom on the lowest energy part of the spectrum around
the avoided crossing between the lowest energy eigenstates at $s_{\rm c}(N)$.
In the thermodynamic limit the first-order transition occurs at
$s=s_{\rm c}$.}
\label{fig_salient_p3_sc}
\end{figure}

\begin{figure}
\centerline{
\includegraphics[width=8.3cm]{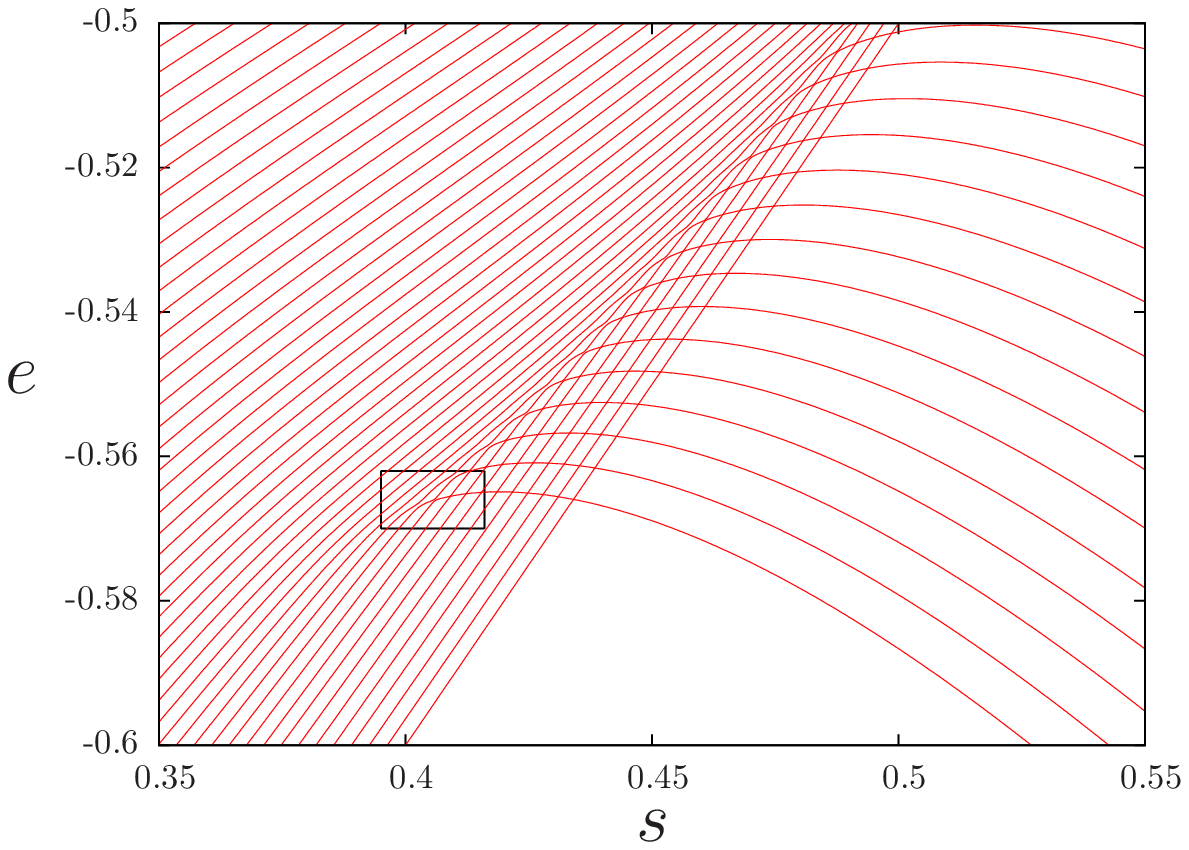} \hspace{6mm}
\includegraphics[width=8.3cm]{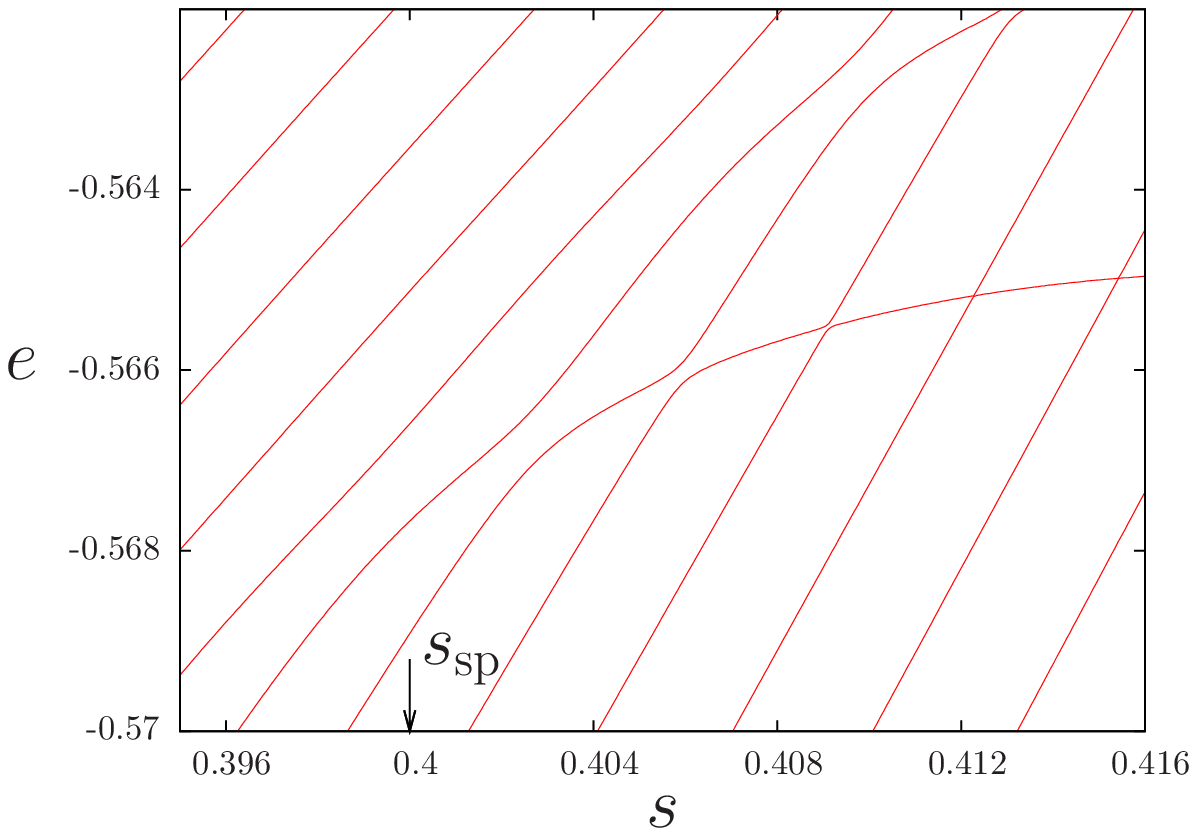}
}
\caption{Left panel: a part of the spectrum for $p=3$, with $N=320$, in
the neighborhood of the spinodal point of the ferromagnetic phase.
Right panel: a zoom on the area of the black box of the left panel.
All the crossings are actually avoided, but some of the gaps are too small 
to be distinguished on the picture.}
\label{fig_salient_p3_ssp}
\end{figure}

Let us now turn to the $p=3$ case, which has a first-order transition at
$s_{\rm c}$. The spectrum of its maximal spin sector is displayed for $N=60$
on the left panel of Fig.~\ref{fig_salient_p3_sc}. The slope of the
groundstate energy is discontinuous at $s_{\rm c}$, as in the thermodynamic
computation (compare with the right panel of Fig.~\ref{fig_thermo_p3}).
At variance with the $p=2$ case the gap between the groundstate and the
first excited state remains of order 1 until one gets very close to 
$s_{\rm c}$; this is best seen on the right panel of 
Fig.~\ref{fig_salient_p3_sc}, which displays a blow up of the lowest energy
states around $s_{\rm c}$. The minimal gap (reached in $s_{\rm c}(N)$ which goes
to $s_{\rm c}$ in the large $N$ limit) is indeed exponentially small
at the first order transition; its exponential rate of decrease, which
has been determined numerically and via an instantonic computation 
in~\cite{jorg10}, will be computed in Sec.~\ref{subsubsec_gap_first_order}
and given as an explicit analytic formula in Eq.~(\ref{eq_result_alpha_p}).
The level repulsion between the two lowest eigenstates is at work only in
a small neighborhood of their point of avoided crossing, and it is 
tempting to infer from this plot that the first excited eigenvector
for $s \gtrsim s_{\rm c}(N)$ is the continuation of the groundstate
eigenvector of $s \lesssim s_{\rm c}(N)$ (and leads thus to the same 
thermodynamic observables), and vice versa. This pattern of avoided crossing 
looks actually very familiar, and can be observed
in many examples involving the eigenvalues of an operator depending on
an external parameter, $s$ here. Let us recall the fundamental reason
behind the universality of such a pattern, and the justification of the
continuation intuition (a detailed discussion of this point can be found 
in~\cite{schulman76}). The matrix elements of the
operator $\hH(s)$ are analytic functions of $s$. As a consequence the
eigenvalues $E_i^{(K)}(s)$ (in any symmetry sector $K$) are the roots
of an algebraic (characteristic) equation of order $N+1-2K$, whose 
coefficients are analytic functions of $s$. Then one can prove (see 
for instance theorem XII.2 in~\cite{reedsimon}) that the $E_i^{(K)}(s)$ are 
analytic functions of $s$, with at most algebraic branchpoint singularities
at the (a priori complex) values of $s$ where the roots are degenerate. An 
avoided crossing is thus due to the two eigenvalues being strictly equal when 
an infinitesimally small imaginary part is added to $s$. By performing the
interpolation between $s \lesssim s_{\rm c}(N)$ and $s \gtrsim s_{\rm c}(N)$
via a detour in the complex plane avoiding the branchpoint singularity one can
thus define in a precise way the first excited eigenvector on one
side of the avoided crossing as the analytic continuation of the groundstate 
on the other side. The thermodynamic calculations of Sec.~\ref{sec_thermo}
suggest that the paramagnetic state of energy $e=-(1-s)$ is metastable
for all values $s\in[s_{\rm c},1]$, and indeed one observes on the plots
of Fig.~\ref{fig_salient_p3_sc} and \ref{fig_salient_p3_ssp} a continuation
of this eigenstate across a series of avoided crossings. On the contrary
the ferromagnetic state which corresponds to the groundstate for 
$s \ge s_{\rm c}$ only exists down to a spinodal point $\ssp$, beyond which the
analytic continuation cannot be performed anymore. This is illustrated
on the two panels of Fig.~\ref{fig_salient_p3_ssp}. The exponential
rate of closing of the gaps encountered along the continuation of the
paramagnetic and ferromagnetic state shall be computed analytically in
Sec.~\ref{subsubsec_gap_metastable}, along with a determination of the
area in the $(s,e)$ plane where avoided crossings do occur (see 
Sec.~\ref{subsubsec_gap_excited}).

All odd values of $p \ge 3$ yield behaviors similar to the $p=3$ case.
The models with an even value of $p \ge 4$ exhibit both the first-order
phenomenology of the $p=3$ case and the exponentially small splitting
between the two ferromagnetic states allowed by the spin-flip symmetry.
This shall be further discussed in Sec.~\ref{subsubsec_gap_ferro} and 
Sec.~\ref{subsubsec_gap_excited}.

\subsection{The semi-classical solution of the eigenvalue equation}
\label{sec_stat_sc}

We shall now determine the structure of the eigenvectors of $\hH(s)$,
in the thermodynamic limit $N\to \infty$, with a calculation formally similar
to the WKB semi-classical treatment of quantum mechanics, the size of the
system $N$ playing the role of $\hbar^{-1}$. Similar semi-classical analysis
have been performed for mean-field spin models 
in~\cite{ulyanov92,ribeiro08,RiPa09}
using a spin-coherent-state representation; at variance with these works 
we shall use here the eigenbasis of $\hmz$ or $\hmx$.

As explained above this computation amounts to diagonalize the matrices of 
order $N+1-2K$
representing the restriction $\hHK(s)$ to a sector of spin $N/2-K$. For
simplicity, and because this will be the most useful case in the following,
we shall concentrate here on the fully symmetric $K=0$ situation. The 
generalization to higher values of $K$ is straightforward and sketched
in Sec.~\ref{subsubsec_dos}. Let us look for an eigenvector $|\phi(s,e)\ra$ of
$\hHzero(s)$ with eigenvalue $N e$, written in the $z$-diagonal basis as
\beq
|\phi(s,e)\ra = \sum_{m \in \MNzero} \phi(m,s,e) \,  |m ;0 \raz \ .
\eeq
Using the expression of the matrix elements given in 
Eqs.~(\ref{eq_hHK_zbasis_d},\ref{eq_hHK_zbasis_offd}), one obtains the 
equation obeyed by the coefficients $\phi(m,s,e)$:
\bea
e \, \phi(m,s,e) = - s  \, m^p \, \phi(m,s,e) 
&- & \frac{(1-s)}{2} \sqrt{1-m^2 + \frac{2}{N}(1-m)} \
\phi\left(m+\frac{2}{N},s,e\right) \nonumber \\ 
&-& \frac{(1-s)}{2} \sqrt{1-m^2 + \frac{2}{N}(1+m)} \
\phi\left(m-\frac{2}{N},s,e\right) \ .
\label{eq_eigenvalue_sigmaz} 
\eea
To deal with the $N \to \infty$ limit we shall make the following Ansatz
on the behaviour of the eigenvector components: 
$\phi(m,s,e) = e^{-N \varphi(m,s,e)}$, with $\varphi(m,s,e)$ an a priori smooth
complex function. Then $\phi(m+2/N,s,e)+\phi(m-2/N,s,e) = 
2 \phi(m,s,e) \cosh \left(2 \varphi'(m,s,e) + O(1/N) \right)$, 
where $'$ denotes the derivative with respect to $m$, and 
(\ref{eq_eigenvalue_sigmaz}) can be rewritten, at the leading order, as:
\beq 
e= - s \, m^p - (1-s) \sqrt{1-m^2} \, \cosh \left (2 \varphi'(m,s,e) \right) \ .
\label{eq_largedev_sigmaz} 
\eeq
Inverting this relation yields
\beq 
\varphi'(m,s,e) = 
\frac{1}{2} \arg\cosh 
\left( - \frac{e+s \, m^p}{(1-s) \sqrt{1-m^2}} \right) \ ,
\label{eq_varphiprime_mz} 
\eeq
the analog of the eikonal equation in the semi-classical one-dimensional
quantum mechanics context.
Several points have to be precised for this equation to unambiguously 
determine the eigenvector rate function $\varphi(m,s,e)$. First of all,
not all values of $e$ should correspond to an authorized eigenvalue of the
Hamiltonian. Then the value of $\varphi$ has to be fixed at one point $m$ to
reconstruct $\varphi$ from its derivative. Finally $\arg \cosh$ is a 
multi-valued function, hence one should precise which of its branches to use.

These ambiguities are actually solved by imposing that $\phi$ is 
normalizable in the large $N$ limit, and that $\varphi'$ is continuous in
$m$ (the coefficients of the eigenvalue equation (\ref{eq_eigenvalue_sigmaz})
being smooth in $m$). By fixing
for instance the norm of $\phi$ to be of order 1, the first requirement imposes 
that $\underset{m}{\inf} [ \Re \, \varphi(m,s,e)] = 0$ 
(we denote $\Re \, z$ and $\Im \, z$
the real and imaginary part of $z$). To precise the meaning of the
$\arg\cosh$ function let us first define the functions $\ach(t)$ as the
reciprocal of $\cosh$ that maps the interval $t \in [1,\infty)$ to 
$[0,\infty)$, and $\acos(t)$ the reciprocal of $\cos$ that maps $t\in[-1,1]$
to $[0,\pi]$. Then, as the argument of the $\arg\cosh$ in 
Eq.~(\ref{eq_varphiprime_mz}) is always real, it is enough to define
\beq
\arg\cosh t = \begin{cases} i \pi \pm \ach(-t) & \text{if}\ t \le -1 \\
i \, \acos(t) & \text{if}\ t \in [-1,1] \\
\pm \ach(t) & \text{if}\ t \ge 1
\end{cases}
\ .
\label{eq_def_branches}
\eeq
There are two branchpoints in $\pm 1$, where the function thus defined is
continuous independently of the choice of the sign of its real part.

Let us now derive from these considerations the authorized value of the
eigenvalue (per spin) $e$. Notice first that the argument of $\arg\cosh$ in 
Eq.~(\ref{eq_varphiprime_mz}) diverges in $m \to \pm 1$, hence it cannot be
confined to $[-1,1]$ for all values of $m \in [-1,1]$. There remains two 
cases to consider: either the argument crosses at least once one of the two
branch-points $\pm 1$, or it remains larger (in absolute value) than 1 for 
all $m$. In the latter case one cannot change branch and the sign of the real 
part of $\varphi'$ is constant on $m \in [-1,1]$ (otherwise $\varphi'$ is not
continuous), hence one cannot fulfill the condition 
$\underset{m}{\inf} [ \Re \, \varphi(m,s,e)] = 0$ in a non-trivial way. A
moment of thought reveals that on the contrary in the former case one can
construct a normalizable eigenvector. This implies that the range of authorized
values for $e$ is
\beq
\underset{m\in[-1,1]}{\Image} [- s \, m^p - (1-s) \sqrt{1-m^2}] \cup
\underset{m\in[-1,1]}{\Image} [- s \, m^p + (1-s) \sqrt{1-m^2}] \ .
\label{eq_e_authorized}
\eeq
In particular the groundstate energy, for a given value of $s$, is obtained
from this reasoning as
\beq
\egs(s)=\inf_{m \in [-1,1]} \left[ - s \, m^p - (1-s) \sqrt{1-m^2} \right] \ ,
\eeq
in perfect agreement with the thermodynamic computation of 
Sec.~\ref{sec_thermo}, see Eq.~(\ref{eq_egs_thermo}).

In the following sections \ref{subsubsec_dos}, \ref{subsubsec_gap_first_order}
and \ref{subsubsec_gap_ferro} we shall show explicitly, in various cases,
how to choose the correct branches of the $\arg\cosh$ function when crossing
a branchpoint and how to determine the rate function $\varphi(m,s,e)$ by 
integration of Eq.~(\ref{eq_varphiprime_mz}). A particularly important
issue will be the occurence of multiple valid eigenvectors corresponding,
at the leading order, to the same eigenvalue $e$. Before that
we shall present a very simple example to check the above computation, and
an alternative formulation in another basis.

A simple consistency check of Eq.~(\ref{eq_varphiprime_mz}) can be performed
for $s=0$, i.e. in a pure transverse field. In that case it is easy to see
that for all $N$ the groundstate has energy $-N$, with the eigenvector
\beq
|\phi(s=0,e=-1) \ra = 
\frac{1}{2^{N/2}} \sum_{m\in \MNzero} \sqrt{\binom{N}{N\frac{1+m}{2}}}  
| m ; 0\raz \ ,
\eeq
corresponding to all spins aligned in the $x$ direction. These values
of $\phi(m,s=0,e=-1)$ solve exactly Eq.~(\ref{eq_eigenvalue_sigmaz}); with 
the help of the Stirling formula one obtains the value of $\varphi$ in the 
$N \to \infty$ limit,
\beq
\varphi_0(m) \equiv 
\varphi(m,s=0,e=-1) = \frac{1+m}{4} \ln(1+m) + \frac{1-m}{4} \ln(1-m) \ .
\label{eq_varphi_check}
\eeq
Let us now check that the computation presented above gives back this
result. We have from Eq.~(\ref{eq_varphiprime_mz})
\beq
\varphi'(m,s=0,e=-1) = 
\frac{1}{2} \arg\cosh\left( \frac{1}{\sqrt{1-m^2}} \right) \ .
\eeq
The argument of the $\arg\cosh$ reaches the branchpoint 1 only in $m=0$.
To enforce the condition $\underset{m}{\inf} [ \Re \, \varphi(m,s,e)] = 0$
one has to choose the branches as
\beq
\varphi'(m,s=0,e=-1) = \sign(m) \frac{1}{2} 
\ach \left( \frac{1}{\sqrt{1-m^2}} \right) \ ,
\eeq
hence upon integration with the boundary condition $\varphi(m=0,s=0,e=-1)=0$,
\beq 
\varphi(m,s=0,e=-1) = \frac{1}{2}  
\int_{0}^m \sign(m) \, \ach \left( \frac{1}{\sqrt{1-m'^2}} \right) \dd m' 
= \frac{1}{2} \int_{0}^m \arg \tanh(m') \dd m'
= \frac{1}{4} \int_{0}^m \ln\left(\frac{1+m'}{1-m'} \right) \dd m' \ ,
\eeq
in agreement with the direct computation yielding (\ref{eq_varphi_check}).

We shall finally present a similar computation of the eigenvectors of
$\hHzero(s)$, but using now the $x$ basis, namely we write
\beq
|\phi(s,e)\ra = \sum_{m \in \MNzero} \phi_x(m,s,e) \,  |m ;0 \rax \ .
\eeq
The coefficients $\phi_x$ obey the following equation (the equivalent of
Eq.~(\ref{eq_eigenvalue_sigmaz}) in the $z$ basis):
\beq 
e \, \phi_x(m,s,e) = - (1-s)  \, m \, \phi_x(m,s,e) 
- s \left ( \frac{1-m^2}{4} \right)^{p/2} 
\sum_{u=0}^{p} \binom{p}{u} \, \phi_x \left(m + \frac{2}{N}(p-2u),s,e \right) 
\ ,
\eeq
in which we have dropped some irrelevant terms of order $1/N$. 
As above we look for a solution of this equation under the form 
$\phi_x(m,s,e)=e^{-N \varphi_x(m,s,e)}$, and find that the leading behaviour of
$\varphi_x$ is ruled by the equation
\bea
e&=&-(1-s) m - s \left ( \frac{1-m^2}{4} \right)^{p/2} 
\sum_{u=0}^{p} \binom{p}{u} (e^{2\varphi_x'(m,s,e)})^{p-u} 
(e^{-2\varphi_x'(m,s,e)})^u \\
&=& - (1-s) m - s (1-m^2)^{p/2} 
\cosh(2\varphi_x'(m,s,e))^p \ .
\label{eq_largedev_sigmax}
\eea
This yields finally the equivalent of Eq.~(\ref{eq_varphiprime_mz}):
\beq
\varphi_x'(m,s,e) = \frac{1}{2} \arg\cosh \left (
\left(-\frac{e + (1-s) m}{s(1-m^2)^\frac{p}{2}} \right)^{1/p}\right) \ .
\label{eq_varphiprime_mx}
\eeq
This equation suffers from the same kind of ambiguities as 
Eq.~(\ref{eq_varphiprime_mz}), the $p$-th root and the $\arg\cosh$ function
being multi-valued. However these ambiguities can also be solved with
exactly the same reasoning as the one following Eq.~(\ref{eq_varphiprime_mz}).
In most of the paper we shall use the $z$-basis computation; the use of
the $x$-basis will however reveal useful in Sec.~\ref{sec_sc_dyn_topara}.

\subsection{The computation of the density of states 
inside one symmetry sector}
\label{subsubsec_dos}

We shall now present the first application of the above computation
of the eigenvectors, that will give an explicit formula for the
integrated density of eigenvalues (similar results for the LMG model can be 
found in~\cite{ribeiro08}). Let us first define this notion
precisely, and emphasize its difference with another, maybe more usual, 
related concept. The full Hilbert
space of the model~(\ref{eq_def_model}) is $2^N$ dimensional, and its 
``density of states'' can be defined as the microcanonical entropy $\s(s,e)$, 
such that $e^{N\s(s,e)}\dd e$ gives, at the leading order, the number of 
eigenvalues of (\ref{eq_def_model}) in the interval $[Ne,N(e+\dd e)]$.
This quantity is obtained from the free-energy density (\ref{eq_f_variational})
via a Legendre transform between $e$ and $\beta$. In this section we shall
however investigate a finer quantity, namely the density of eigenvalues for
the restriction $\hHK$ of the Hamiltonian to one symmetry sector (which has
$N+1-2K$ eigenvalues). Consider for instance the fully-symmetric sector, and 
define
\beq
\D_0(s,e) = \underset{N\to \infty}{\lim} 
\frac{1}{N+1} |\{j| E^{(0)}_j(s) \le N e \}|
\eeq
as the integrated density of states inside that sector. We shall see at the 
end of this section that the knowledge of the integrated density of states of 
all sectors $K=0,\dots,\fhN$ provides a much more detailed information on the
system than the microcanonical entropy.

There is actually a simple relation between $\D_0(e,s)$ and the leading 
order computation of the eigenvectors of the previous Subsection, based
on the following observation: $\hHK$, expressed in the $z$-basis, is a
symmetric tridiagonal matrix with all elements next to the diagonal of the 
same sign (negative). This implies that the ordering in energies of its
eigenvalues corresponds to the number of nodes of the associated eigenstates,
exactly for the same reasons as the $n$-th excited eigenstate of a 
one-dimensional quantum particle described by the Schr\"odinger equation has
precisely $n$ zeroes. A proof for the discrete case can be adapted from the
usual reasonings in the Schr\"odinger case (see~\cite{derrida93} for a similar 
derivation in another context), and shows that the groundstate of $\hHK$ is
a Perron-Frobenius vector whose elements can be taken all positive, while its
first excited state presents exactly one ``domain wall'' between two sets of 
values of $m$ where the eigenvector is positive/negative, and so on and so 
forth. As we defined $\phi(m)=e^{-N \varphi(m)}$, the fact that excited eigenstates
exhibit alternating signs translates into $\varphi$ acquiring an imaginary part.
More precisely, each ``domain wall'' between opposite signs for $\phi(m)$
corresponds to an increase of its phase by $\pm \pi$. One can thus count the
number of nodal points of $\phi$ by integrating the imaginary part of
$\varphi'$ on $m\in[-1,1]$, and deduce from it the number of eigenvalues
that have lower energies. This reasoning yields the following formula,
\beq 
\D_0(s,e) = \frac{1}{\pi} \int_{-1}^1 \dd m \, \Im \, \varphi'(m,s,e) \ ,
\label{eq_dos_first}
\eeq
as we have chosen in Eq.~(\ref{eq_def_branches}) a branch of $\arg\cosh$ with 
positive imaginary part. Using the value (\ref{eq_varphiprime_mz}) for
the derivative of $\varphi$ one obtains a completely explicit formula
for the integrated density of states in the maximal spin sector,
\beq 
\begin{split}
\D_0(s,e) = \frac{1}{2 \pi} \int_{-1}^1  \dd m \, 
\left[ \acos\left(- \frac{e+s \, m^p}{(1-s) \sqrt{1-m^2}} \right)
\I\left(  - \frac{e+s \, m^p}{(1-s) \sqrt{1-m^2}} \in [-1,1]\right) 
\right. \\ \left.
+ \pi \, \I\left(  - \frac{e+s \, m^p}{(1-s) \sqrt{1-m^2}} \le -1\right)
\right] \ ,
\end{split}
\label{eq_dos}
\eeq
where we defined $\I(A)$ to be $1$ if $A$ is true, $0$ otherwise.
Deriving this expression with respect to $e$ one can equivalently obtain
an expression for the density of states,
\beq 
\rho_0(s,e) = \frac{1}{2 \pi} \int_{-1}^1  \dd m \, 
\left[ \frac{1}{\sqrt{(1-s)^2(1-m^2) - (e+s \, m^p)^2}}
\I\left(  - \frac{e+s \, m^p}{(1-s) \sqrt{1-m^2}} \in [-1,1]\right) 
\right] \ .
\label{eq_rho}
\eeq

\begin{figure}
\centerline{
\includegraphics[width=8.3cm]{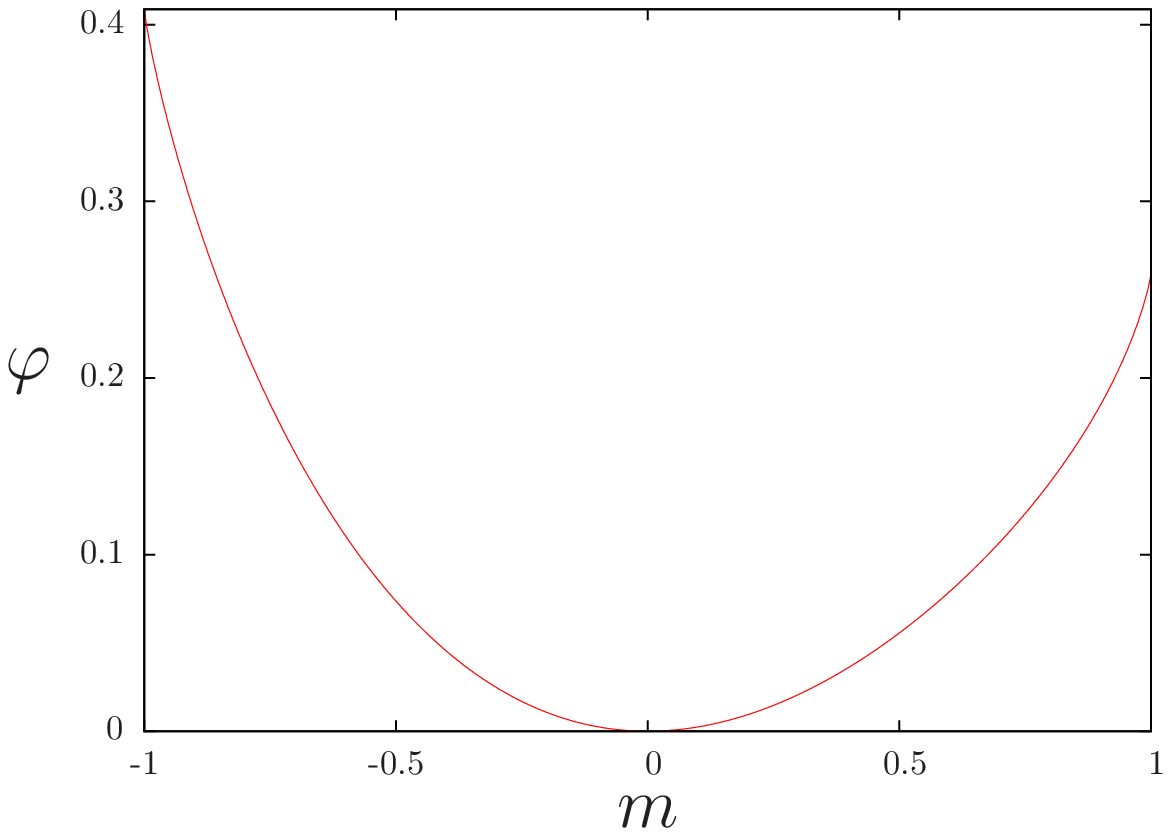} \hspace{6mm}
\includegraphics[width=8.3cm]{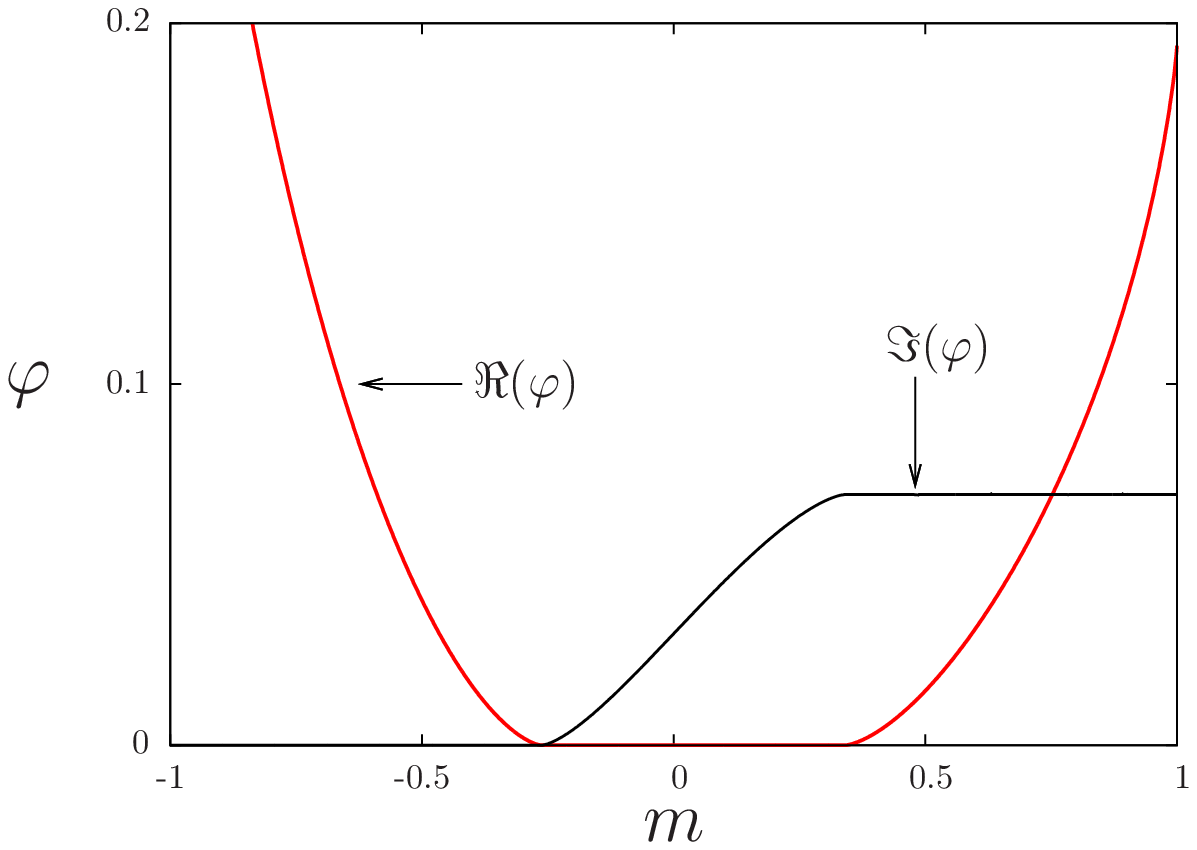}
}
\caption{Plots of the eigenstate function $\varphi$, for $p=3$ and $s=0.3$.
Left panel: for the groundstate energy, $e=\egs$, $\varphi$ is purely real. 
Right panel, for a slightly larger energy, $e=\egs+0.1$, $\varphi$ acquires an 
imaginary part.}
\label{fig_varphi_example1}
\end{figure}

\begin{figure}
\centerline{
\includegraphics[width=8.3cm]{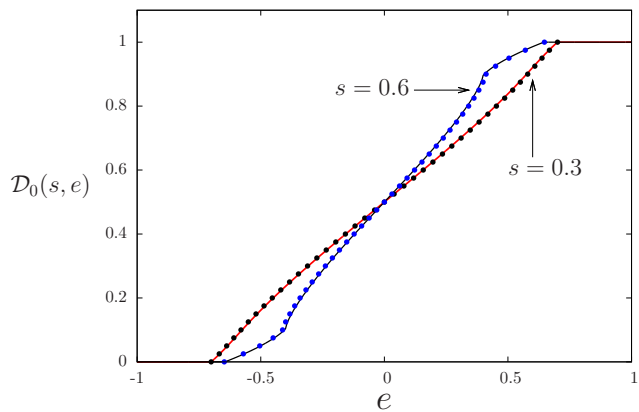}
}
\caption{Plots of the integrated density of states $\D_0$ for $p=3$, $s=0.3$ 
and $s=0.6$. The solid lines are our analytical predictions from 
Eq.~(\ref{eq_dos}); the symbols are the results of numerical diagonalization
of systems with $N=40$.}
\label{fig_ids}
\end{figure}

We give in Fig.~\ref{fig_varphi_example1} some examples of the construction
of $\varphi$, for $p=3$ and $s=0.3$, i.e. in the paramagnetic phase.
For the ground-state energy $e=-(1-s)$ the argument of the $\arg\cosh$ function 
in Eq.~(\ref{eq_varphiprime_mz}) is always $\ge 1$, with a single point of 
equality in $m=0$. As a consequence the corresponding solution for $\varphi$
is everywhere real, see left panel of the figure. On the contrary for a 
slightly higher value of the energy ($e=\egs+0.1$ on the figure) the branchpoint
$1$ is crossed at two values of $m$, hence the imaginary part of $\varphi$
grows on this interval, on which the real part vanishes identically. 
We also present on Fig.~\ref{fig_ids} the curves for the integrated density
of states $\D_0$ for $p=3$ and two values of $s$, $0.3$ and $0.6$ (in the 
latter case one observes a singularity at the crossing of the energy of the
metastable paramagnetic phase). The agreement with the density of states
obtained by numerical diagonalization is very good already for small values
of $N$ ($N=40$ on the figure).

Let us briefly mention here one application of this computation that shall be
useful in the analysis of the annealing dynamics. From the integrated
density of states $\D_0(s,e)$ one can define ``iso-integrated density lines''
$\eiso(s)$ by imposing that $\D_0(s,\eiso(s))$ remains constant when $s$ is
varied. These lines correspond to the thermodynamic limit of the energy density
of some (excited) eigenvalues, as long as no level crossings occurs.

We shall now explain how to generalize the computation of the integrated
density of states to sectors of arbitrary spin $N/2 - K$. The matrix of size 
$N+1-2K$ representing the restriction $\hHK$
of the Hamiltonian to this sector has matrix elements given in 
Eqs.~(\ref{eq_hHK_zbasis_d},\ref{eq_hHK_zbasis_offd}). Denoting $k=K/N$,
one can look for eigenstates of $\hHK$ under the form $e^{-N \varphi(m,s,e)}$,
where the longitudinal magnetization $m$ is now restricted to $[-1+2k,1-2k]$,
and the function $\varphi$ is solution of a generalization of
Eq.~(\ref{eq_varphiprime_mz}), namely
\beq 
\varphi'(m,s,e) = 
\frac{1}{2} \arg\cosh 
\left( - \frac{e+s \, m^p}{(1-s) \sqrt{(1-2k)^2-m^2}} \right) \ .
\eeq
The rest of the computation follows strictly the reasoning made for the 
symmetric ($K=0$) sector. In particular the groundstate energy density for
a sector with $k=K/N$ reads in the thermodynamic limit
\beq
\egsk(s)=\inf_{m \in [-1+2k,1-2k]} \left[ 
- s \, m^p - (1-s) \sqrt{(1-2k)^2-m^2} \right] \ ,
\label{eq_egsk}
\eeq
and the density of states in that sector is
\beq 
\begin{split}
\D_k(s,e) = \frac{1}{2 \pi} \int_{-1+2k}^{1-2k}  \dd m \, 
\left[ \acos\left(- \frac{e+s \, m^p}{(1-s) \sqrt{(1-2k)^2-m^2}} \right)
\I\left(  - \frac{e+s \, m^p}{(1-s) \sqrt{(1-2k)^2-m^2}} \in [-1,1]\right) 
\right. \\ \left.
+ \pi \, \I\left(  - \frac{e+s \, m^p}{(1-s) \sqrt{(1-2k)^2-m^2}} \le -1\right)
\right] \ .
\end{split}
\eeq

We shall finaly compare the amount of information contained in the
densities of states $\D_k(s,e)$ on one hand, and the microcanonical entropy
$\s(s,e)$ on the other. The latter being the Legendre transform of the
free-energy, we shall equivalently discuss this quantity. Using
the decomposition of the Hilbert space into symmetry sectors the partition 
function can be written as
\beq
Z = \sum_{K=0}^{\fhN} \NNK \sum_{j=0}^{N-2K} e^{-\beta E_j^{(K)}} \ ,
\eeq
where $\NNK$ gives the number of degenerate representations of spin $N/2-K$
(see Eq.~(\ref{eq_value_NNK})), and $E_j^{(K)}$ is the $j$-th eigenvalue of
$\hHK$. In the thermodynamic limit the degeneracy $\NNK$ grows exponentially
with $N$; on the contrary the sum over the eigenstates of one sector contains
only a linear number of terms, and is thus dominated at the leading 
exponential order by the greatest of these terms, i.e. the groundstate
energy of the corresponding sector. This leads to the following
expression for the free-energy density,
\beq
f(\beta,s)=\inf_{k\in[0,1/2]} 
\left[\egsk(s) - \frac{1}{\beta}(-k \ln k - (1-k)\ln(1-k)) \right] \ .
\eeq
A short computation based on the expression of $\egsk$ given in (\ref{eq_egsk})
reveals the agreement between this expression and the one obtained in
Sec.~\ref{sec_thermo} (see Eq.~(\ref{eq_f_variational})). What we want to
stress here is that the only ``microscopic'' (i.e. at the level of 
eigenstates) input of the computation is the energy density of the groundstate
in each sector. The microcanonical entropy is thus entirely dominated by the
effect of the degeneracy $\NNK$ of the various spin sectors, and is completely
insensitive to their internal structure beyond their groundstate energy
density.

\subsection{The computation of finite gaps}
\label{sec_finite_gaps}

One can estimate the energy gaps between eigenvalues from the density of 
states obtained in Eq.~(\ref{eq_rho}): in the interval $[e,e+\dd e]$ of 
(intensive) energy one finds $N \rho(s,e) \dd e$ eigenvalues. Assuming these
levels to be equispaced, the gap between two successive eigenvalues is, in 
extensive energy, $1/\rho(s,e)$~\cite{ribeiro08}. This computation can be 
performed in any part of the energy spectrum; for simplicity we shall only
state some results, obtained by combining this observation with the
explicit expressions of the density of states~(\ref{eq_rho}) and of its
integrated form~(\ref{eq_dos}), in the most relevant regions of the spectrum.

For $p=2$, i.e. in the Curie-Weiss model,
one obtains in the paramagnetic phase ($s< s_{\rm c}$) for the gap
between the groundstate and the first excited state:
\beq
\lim_{N \to \infty} [E^{(0)}_1(s) - E^{(0)}_0(s)] = 
\frac{1}{\rho(s,\egs(s))}= 2 \sqrt{3} \sqrt{(1-s)(s_{\rm c}-s)} \ ,
\label{eq_gap_finite_p2}
\eeq
as found in~\cite{botet83}, and plotted on the right panel of 
Fig.~\ref{fig_salient_p2}. Note the square-root closing of the gap
at the second-order transition. The same computation performed in the 
ferromagnetic phase ($s > s_{\rm c}$) yields
\beq
\frac{1}{\rho(s,\egs(s))}= \sqrt{3} \sqrt{(1+s)(s-s_{\rm c})} \ .
\eeq
This should however not be interpreted as the gap between the first two 
eigenvalues, but rather as half the gap between the groundstate and the second
excited state. Indeed the level splitting between the two lowest states
is exponentially small in $N$ (as will be computed in 
Sec.~\ref{subsubsec_gap_ferro}) and the density of states does not distinguish 
them. In other words the hypothesis of equi-spacing of eigenvalues is strongly
broken in this situation.

Consider now the case $p>2$. In the paramagnetic phase ($s\le s_{\rm c}$) 
one finds
\beq
\lim_{N \to \infty} [E^{(0)}_1(s) - E^{(0)}_0(s)] = 
\frac{1}{\rho(s,\egs(s))} = 2 (1-s) 
\eeq
for the gap between the two lowest levels, which is the same result as would
have been obtained if the Hamiltonian contained only the transverse field
term. Note also that the gap thus computed remains positive at the
first-order transition; the exponentially small gap (to be determined
in Sec.~\ref{subsubsec_gap_first_order}) cannot be detected by the
density of states. The computation of the density of states $\rho_0$ at the 
groundstate energy can similarly be performed in the ferromagnetic phase 
(i.e. for $s\ge s_{\rm c}$). For odd values of $p$ one finds
\beq
\underset{N\to \infty}{\lim} [E^{(0)}_1(s) - E^{(0)}_0(s)] = 
\frac{1}{\rho(s,\egs(s))} = 2 p s m_*(s)^{p-2}\sqrt{(p-1) m_*(s)^2 - (p-2)} \ ,
\eeq
which is positive at $s_{\rm c}$. For even values of $p$ this computation, as
explained above in the case $p=2$, gives an information only on the gap
between the groundstate and the second excited state. After a short computation
one obtains a similar formula,
\beq
\underset{N\to \infty}{\lim} [E^{(0)}_2(s) - E^{(0)}_0(s)] = 
\frac{2}{\rho(s,\egs(s))} = 2 p s m_*(s)^{p-2}\sqrt{(p-1) m_*(s)^2 - (p-2)} \ ;
\eeq
one could expect to find an additional factor $2$ in this expression with 
respect to the odd $p$ case, however this factor compensates because of the 
contributions of the two minima in $\pm m_*(s)$ in the density of states $\D_0$.

For $p>2$ a richer behaviour is displayed in the neighborhood of the 
spinodal point of coordinates $(\ssp,\esp)$. In particular in the limit
$s\to \ssp^-$ one finds after some computations a scaling behaviour for
the integrated density of states, of the form
\beq
\D_0(\ssp - \delta s, \esp+\delta e) - \D_0(\ssp,\esp) \sim \delta s \,
\G((\delta e + \epfm(\ssp) \delta s ) \delta s^{-6/5}) \ ,
\label{eq_scaling_D0}
\eeq
where $\epfm(\ssp) = - \msp^p + \sqrt{1-\msp^2}$ is the 
derivative of the energy of the ferromagnetic metastable state at the 
spinodal. The scaling function $\G(z)$ is monotonously increasing,
behaves as $|z|^{5/6}$ for $|z| \to \infty$, and vanishes in one point we shall
denote $z_0$. The iso-integrated density line that goes through the spinodal 
point behaves thus as $\eiso(\ssp - \delta s) \sim - \epfm(\ssp) \delta s 
+ z_0 \delta s^{6/5}$. Moreover in the scaling regime the density of states can 
be obtained by deriving the above relation, namely
\beq
\rho_0(\ssp - \delta s, \esp+\delta e) \sim \delta s^{-1/5} \,
\G'((\delta e + \epfm(\ssp) \delta s ) \delta s^{-6/5}) \ .
\eeq
In consequence the finite gap between the eigenstate level that reaches the
spinodal point and the first excited state above it closes when $s \to \ssp^-$
as $\delta s^{1/5} / \G'(z_0)$. This fifth root is to be contrasted with
the square root singularity for the groundstate of $p=2$. Moreover we should
warn the reader that the correction $z_0 \delta s^{6/5}$ in the expansion
of the eigenstate energy is crucial: in $z=0$ the scaling function $\G$ is 
finite but has no derivative, and the expansion of $\rho_0$ without taking
into account the correction leads to
$\rho_0(\ssp - \delta s, \esp-\epfm(\ssp) \delta s) \propto \delta s^{-1/4}$,
which modifies the exponent from $1/5$ to $1/4$.

\subsection{The computation of  exponentially small gaps}
\label{sec_small_gaps}

As discussed qualitatively in Sec.~\ref{sec_salient} the energy gaps between
two successive levels are, in some regions of the plane $(s,e)$ depending
on $p$, exponentially small in the size $N$ of the system. This section
is devoted to the computation of the exponential rate of closing of those
gaps. It is divided in four parts; we shall first investigate the 
avoided crossing between the groundstate and the first excited state at
the first-order transition of the models with $p\ge 3$ 
(in Sec.~\ref{subsubsec_gap_first_order}), then compute the exponentially 
small splitting between the two lowest levels in the ferromagnetic phase
of even $p$ models (in Sec.~\ref{subsubsec_gap_ferro}). The next two 
subsections will be devoted to exponentially small gaps between excited 
states; in Sec.~\ref{subsubsec_gap_excited} we shall determine the
values of $(s,e)$ where these avoided crossings do occur, and in 
Sec.~\ref{subsubsec_gap_metastable} we will concentrate on the avoided
crossings encountered by the metastable continuations of the paramagnetic
and ferromagnetic groundstates.

From a technical point of view the common pattern behind the appearance
of an exponentially small gap is the existence of two valid solutions of the
semi-classical eigenvalue equation (\ref{eq_varphiprime_mz}) for the
same value of $e$. This approximate degeneracy is lifted at the exponential
order, the splitting between the two levels being proportional to the 
exponentially small scalar product between the two quasi-eigenvectors 
computed at leading order. This can be shown by writing the eigenvalue equation
in the two dimensional Hilbert space spanned by the two quasi-eigenvectors.
In more physical terms this corresponds to the semi-classical 
approximation of quantum mechanics for a double-well potential, in
which the two lowest energy levels have a gap exponentially small in $1/\hbar$. 

\subsubsection{The exponentially small gap at the first order transition for $p \ge 3$}
\label{subsubsec_gap_first_order}

The first case we shall consider is the exponentially small gap between the 
groundstate and the first excited state at the first-order transition for 
$p\ge 3$. We first concentrate on the odd $p$ case for simplicity, the 
modifications to be made when $p$ is even are discussed afterwards.

\begin{figure}
\centerline{
\includegraphics[width=8.3cm]{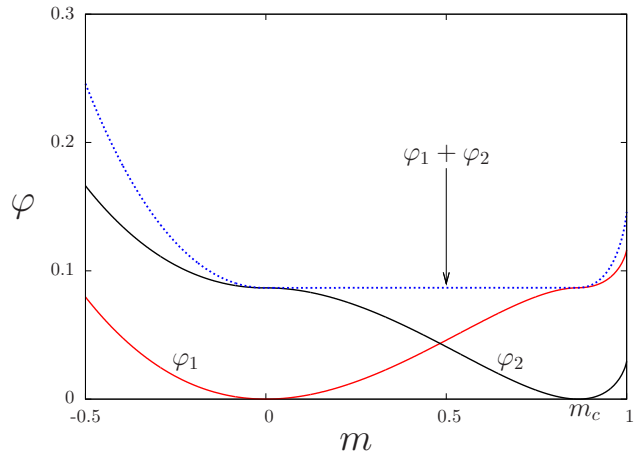}
}
\caption{The two possible eigenstate functions $\varphi_{1,2}(m,s_{\rm c},\ec)$
at the first order transition point of the $p=3$ model.}
\label{fig_varphi_p3_critical}
\end{figure}

From the thermodynamic considerations of Sec.~\ref{sec_thermo} we showed that
this transition happens at a ($p$-dependent) value of $s$ denoted $s_{\rm c}$,
where the infimum in the definition (\ref{eq_egs_thermo}) of the groundstate 
energy is reached for two distinct values of $m$, i.e. in $m=0$ and $m=\mc>0$ 
(see Eq.~(\ref{eq_critical_parameters}) for the values of $s_{\rm c}$ and $\mc$).
The paramagnetic and ferromagnetic phases have thus the same energy
$\ec$. In terms of the eigenstate computation, this observation translates
into the fact that the argument of the $\arg\cosh$ in the expression of
$\varphi'(m,s_{\rm c},\ec)$ given by (\ref{eq_varphiprime_mz}) is $\ge 1$ for
all values of $m$, with two point of equalities in $m=0$ and $m=\mc$.
The prescription for the choices of the branches of the $\arg\cosh$ function
(that can be changed at the branching point $+1$) explained after 
Eq.~(\ref{eq_varphiprime_mz}) leaves us with two possible real solutions
$\varphi_1$ and $\varphi_2$, which reaches their minimal value $0$ in $m=0$
and $m=\mc$ respectively. These two functions are plotted for $p=3$ in 
Fig.~\ref{fig_varphi_p3_critical}.

This apparent degeneracy of the lowest eigenvalues is however lifted with
an exponentially small correction in $N$.
Let us denote $\alpha_p$ the rate at which this gap closes with $N$, i.e.
\beq
\alpha_p = - \lim_{N \to \infty} \frac{1}{N} 
\ln\left( \underset{s \in [0,1]}{\min} [E_1^{(0)}(s) - E_0^{(0)}(s)]  
\right) \ .
\eeq
At leading order this rate can be computed from the overlap between the
two quasi-eigenvectors $\phi_1(m)=e^{-N\varphi_1(m)}$ and $\phi_2(m)=e^{-N\varphi_2(m)}$,
as explained at the beginning of this section. We thus have
\bea \label{eq_overlap_large_dev} 
\alpha_p &=& 
- \lim_{N \to \infty} \frac{1}{N} \ln | \la \phi_1 | \phi_2 \ra | = 
- \lim_{N \to \infty} \frac{1}{N} 
\ln \left | \int_{-1}^{1} \dd m \, e^{-N \varphi_1(m,s_{\rm c},\ec) 
- N \varphi_2(m,s_{\rm c},\ec)}  \right | \\
&=& \inf_m [ \varphi_1(m,s_{\rm c},\ec) + \varphi_2(m,s_{\rm c},\ec)]  \ .
\eea
The shape of the sum $\varphi_1 + \varphi_2$ is also displayed in 
Fig.~\ref{fig_varphi_p3_critical}. It is minimal and constant on the whole
interval $[0,\mc]$; indeed the two functions are solutions of
Eq.~(\ref{eq_varphiprime_mz}) for the same value of the parameters $e,s$, and
only differ in the opposite choice of the branch of the $\arg\cosh$ function 
for their derivative on $[0,\mc]$. As a consequence $\alpha_p$ can be simply 
computed by integrating the derivative of $\varphi$, i.e.
\beq 
\alpha_p = \frac{1}{2} \int_{0}^{\mc} \dd m  
\ \ach \left(- \frac{\ec +s_{\rm c} \, m^p}{(1-s_{\rm c}) \sqrt{1-m^2}} \right) 
\ .
\label{eq_result_alpha_p}
\eeq
This formula, complemented by the values of $\mc$, $s_{\rm c}$ and $\ec$ as
a function of $p$ given in (\ref{eq_critical_parameters}), is one of the main
results of the statics part of the paper, giving a very explicit analytical
prediction of the exponentially small gap at the first-order transition.

The numerical values of $\alpha_p$ thus obtained are displayed in 
Table~\ref{table_gaps_p}, along with a comparison with the data reported
by J\"org et al in~\cite{jorg10}. The authors of this paper obtained
$\alpha_p$ both by exact diagonalization of the matrices $\hHzero$ for
finite $N$ (with an extrapolation in the limit $N\to \infty$) and by a
semi-classical instantonic computation. Our results agrees very well with 
theirs. One can set up an asymptotic expansion of $\alpha_p$ at large $p$, that
results in
\beq
\alpha_p =  \frac{\ln 2}{2} - \frac{\pi^2}{12p} + O\left(\frac{1}{p^2}\right) 
\label{eq_result_alpha_p_large_p}
\ ,
\eeq
as stated in the Table. The details of this computation are deferred to
Appendix~\ref{app_alpha_p_large_p}. The interpretation of $\alpha_p$ for
even values of $p$ shall be discussed in the next section.

\begin{table}[h]
\center
\begin{tabular}{|c|c|c|c|c|c|c|c|}
\hline
$p$ & $\Gamma_{\rm c}$ & $s_{\rm c}$ & $\mc$ 
& $\frac{\alpha_p}{\ln 2}$ (diagonalization)~\cite{jorg10} 
& $\frac{\alpha_p}{\ln 2}$ (instanton)~\cite{jorg10} 
& $\frac{\alpha_p}{\ln 2}$ from Eq.~(\ref{eq_result_alpha_p}) \\
 \hline
 3 & 1.2991 & 0.4350 & 0.8660  & 0.126(3) & 0.1251 & 0.1252 \\
 4 & 1.1852 & 0.4576 & 0.9428  & - & - & 0.2127\\
 5 & 1.1347 & 0.4685 & 0.9682  & 0.270(3) &0.2686 & 0.2680\\
 6 &  1.1059 & 0.4749 & 0.9798 & - & - & 0.3057\\
 7 & 1.0873 & 0.4791 & 0.9860  & 0.335(3) & 0.3335 & 0.3329\\
 8 & 1.0743 & 0.4821 & 0.9897  & - & - & 0.3535 \\
 9 & 1.0647 & 0.4843 & 0.9922  &0.370(3) & 0.3699 &0.3695\\
 13 & 1.0426& 0.4896 & 0.9965 & 0.410(3) & 0.4105 & 0.4093\\
 17 & 1.0318 & 0.4922 & 0.9980 & 0.431(3) &0.4315 & 0.4306 \\
 21 & 1.0253 & 0.4937 & 0.9987 & 0.445(3) & 0.4445 & 0.4437\\
 31 &1.0168 & 0.4958 & 0.9994 & 0.462(3) & 0.4623 &0.4618\\
 $p \to \infty$ & $1+\frac{1}{2p}$ & $\frac{1}{2}-\frac{1}{8p}$ & 
$1-\frac{1}{2p^2}$ &  $\frac{1}{2} - \frac{1.15}{p}$ & - &
$ \frac{1}{2} - \frac{\pi^2}{12 \log2} \frac{1}{p}$\\
\hline
\end{tabular}
 \caption{Exponential rate of decay of the gap between the groundstate and the 
first excited state at the first-order transition 
for the $p$-spin model, divided by $\ln 2$ to ease the comparison
with the results of~\cite{jorg10}. The last column is our result, computed
from Eq.~(\ref{eq_result_alpha_p}). The 
fifth column is the extrapolation from finite $N$ exact 
diagonalization~\cite{jorg10}, and the sixth one results from an instantonic 
computation~\cite{jorg10}. Thermodynamic parameters 
$\Gamma_{\rm c}=(1-s_{\rm c})/s_{\rm c}$, $s_{\rm c}$ and $\mc$ of the system at 
the critical point are also given. The last line gives an equivalent of these
quantities at the leading order in $1/p$ in the large $p$ limit.}
\label{table_gaps_p}
\end{table}

\subsubsection{The exponentially small gap between the two ferromagnetic
phases for even $p$}
\label{subsubsec_gap_ferro}

For even values of $p$ the classical part of the Hamiltonian, $\hH(s=1)$, is
invariant under the reversal of the longitudinal magnetization, its groundstate
is thus doubly degenerate, with eigenstates fully polarized along the $\pm z$
direction. As soon as the transverse field is switched on, i.e. for $s<1$,
this strict degeneracy is lifted. However in the ferromagnetic phase,
i.e. for $s > s_{\rm c}$, this lifting is weak, and the gap between the
groundstate and the first excited state is exponentially small in $N$, 
of the form $e^{-N \beta_p(s)}$ at the leading order.
We shall now compute this rate $\beta_p(s)$, following essentially the same
lines as in Sec.~\ref{subsubsec_gap_first_order}. A similar study for $p=2$
can be found in~\cite{schulman76}.

\begin{figure}[h]
\centerline{
\includegraphics[width=8.3cm]{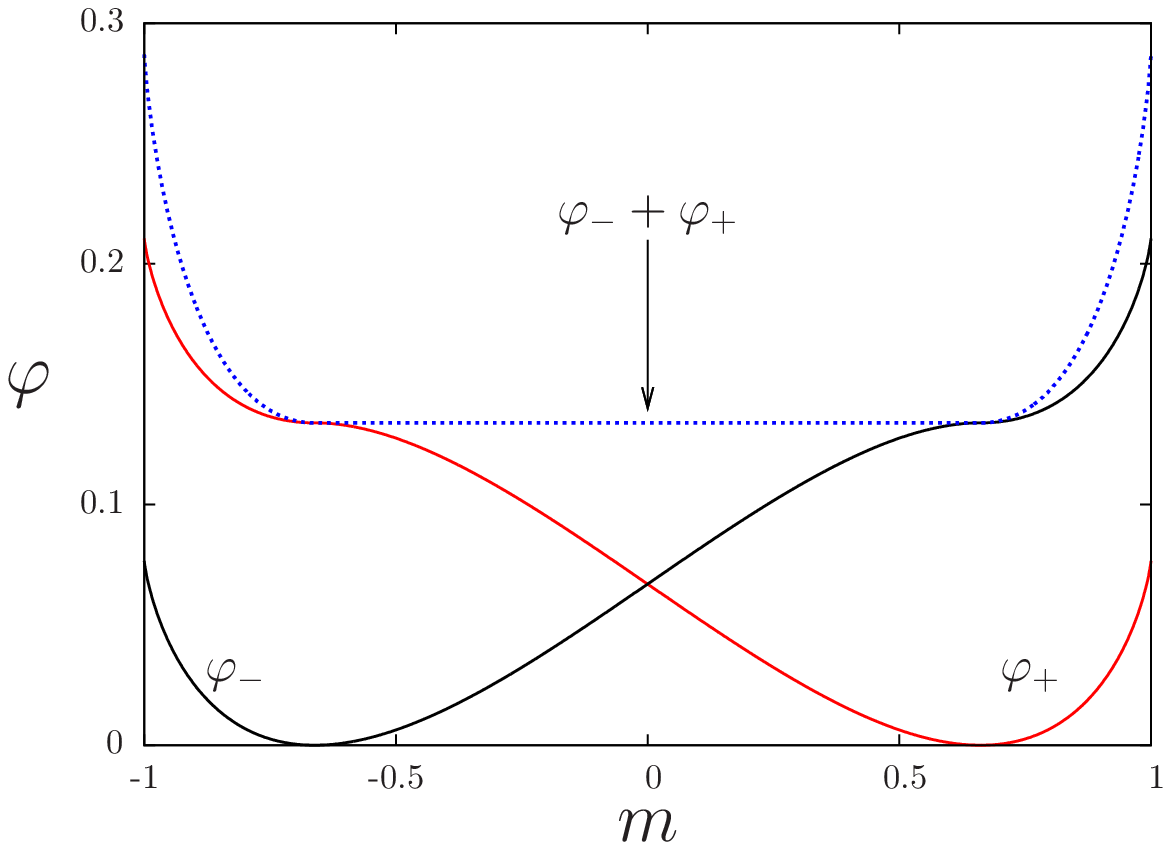} \hspace{6mm}
\includegraphics[width=8.3cm]{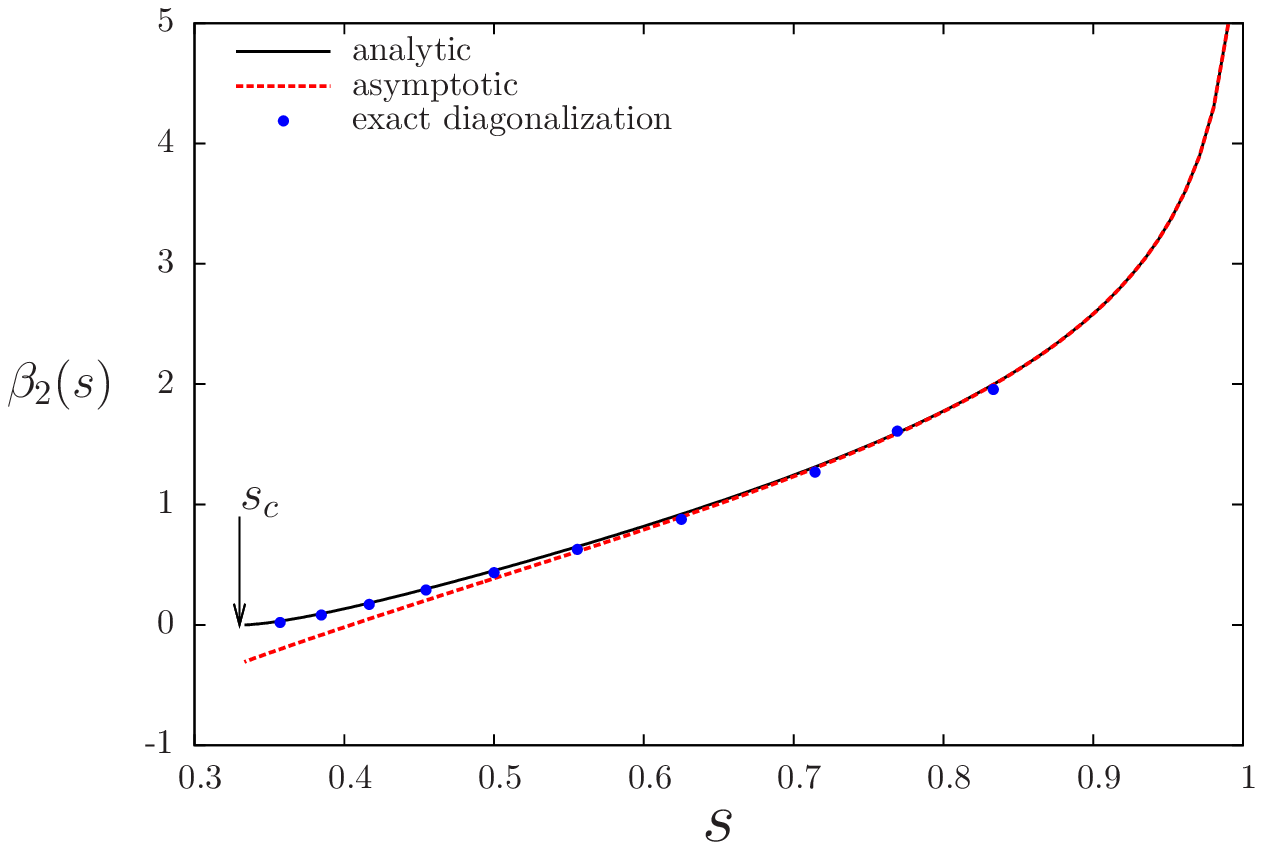}
}
\caption{Left panel: the two eigenstate functions $\varphi_\pm$ for $s=0.4$ 
and $p=2$, at the groundstate energy, and their sum.
Right panel: exponential scaling of the gap $\beta_2(s)$ between the two 
ferromagnetic solutions for $p=2$. The solid black curve has been obtained
from Eq.~(\ref{eq_beta_p}), the symbols are the results of exact 
diagonalization extrapolated in the limit $N\to\infty$~\cite{florent_p2},
the red dashed curve is the leading term in the $s\to 1$ limit, see
Eq.~(\ref{eq_beta_p2_asymptotic}). 
}
\label{fig_gaps_ferro}
\end{figure}

In the ferromagnetic phase of even $p$ models the infimum in the definition 
(\ref{eq_egs_thermo}) of the groundstate energy is reached in $\pm m_*(s)$, 
where the spontaneous longitudinal magnetization $m_*(s)$ is solution of
Eq.~(\ref{eq_mag_thermo}). Hence the argument of $\arg\cosh$ in 
Eq.~(\ref{eq_varphiprime_mz}) is $\ge 1$ for
all values of $m$, touching 1 in $\pm m_*(s)$. One can thus construct
two solutions $\varphi_\pm$ of Eq.~(\ref{eq_varphiprime_mz}), that vanish
in $\pm m_*(s)$. An example for $p=2$ is displayed in 
Fig.~\ref{fig_gaps_ferro}. As above one obtains the rate $\beta_p(s)$
by computing the overlap between these two quasi-eigenstates. This yields
\beq 
\beta_p(s) =  \frac{1}{2} \int_{-m_*(s)}^{m_*(s)} \dd m \,  \ach \left(  
- \frac{\egs(s)+s\, m^p}{(1-s) \sqrt{1-m^2}} \right) \ .
\label{eq_beta_p}
\eeq
This formula compares very well with the results of exact 
diagonalization for $p=2$~\cite{florent_p2}, as shown in the right panel
of Fig.~\ref{fig_gaps_ferro}.

In the classical limit $s \to 1$ the gap vanishes for all values of $N$,
in consequence the rate $\beta_p(s)$ diverges in this limit. One can
study this asymptotic behavior more precisely. One has $m_*(s) \to 1$ and 
$\egs(s) \to -1$ in this limit, and the factor $1/(1-s)$ makes the argument
of the $\ach$ function in (\ref{eq_beta_p}) diverge for all values of $m$
inside the domain of integration. One can then use the asymptotic expansion
$\ach(y) \sim \ln(2y)+o(1)$ for $y$ large to obtain
\beq 
\beta_p(s) \sim  
\frac{1}{2} \int_{-1}^{1} \dd m \,
\ln \left(\frac{2}{1-s}\frac{1-m^p}{\sqrt{1-m^2}} \right)
\sim - \ln(1-s) + \hbeta_p \ ,
\eeq
where the constant $\hbeta_p$ can be expressed in terms of the harmonic
number function $H(x) = \int_{0}^1 \frac{1-t^x}{1-t}\dd t$, as
\beq
\hbeta_p =  
\int_{0}^{1} \dd m \,
\ln \left(\frac{2(1-m^p)}{\sqrt{1-m^2}} \right) = \ln 2 - 
H \left( \frac{1}{p}\right)+ \frac{1}{2}H\left(\frac{1}{2}\right) \ .
\eeq
In particular for $p=2$ we obtain 
\beq
\beta_2(s) \sim - \ln(1-s) -1+2\ln 2 \ ,
\label{eq_beta_p2_asymptotic}
\eeq 
which is also plotted for comparison in the right panel of 
Fig.~\ref{fig_gaps_ferro}.

Another interesting limit case concerns the behaviour of $\beta_2(s)$ around
the threshold $s_{\rm c}$ of the second-order transition of the $p=2$ model. 
The rate of 
exponentially small splitting has to vanish in this limit, since the 
groundstate of the paramagnetic phase is no longer quasi-degenerate. More 
precisely, using the asymptotic behaviors $m_*(s) \sim 3 \sqrt{s-s_{\rm c}}$
when $s \to s_{\rm c}^+$ and $\ach(1+y)\sim \sqrt{2y}$ when $y\to 0^+$, one can 
expand the expression (\ref{eq_beta_p}) of $\beta_2(s)$ and obtain after a 
short computation that 
$\beta_2(s) = 9 (s-s_{\rm c})^{3/2} + O((s-s_{\rm c})^{5/2})$.
This exponent $3/2$ was first predicted in~\cite{botet83} on the basis of an
adaptation of Finite Size Scaling to mean-field systems (and found also
in~\cite{dusuel04} from the scaling of singular finite $N$ corrections). 
The argument of~\cite{botet83} leads
indeed to the value $\numf \dc$, where $\numf$ and $\dc$ are the mean-field 
value of the exponent controlling the divergence of the correlation length and
the upper critical dimension of the universality class to which the studied
model belongs. In the present case $\numf =1/2$, as in the $\phi^4$ theory,
but one has to take $\dc=3$: classical models in this universality class have
an upper critical dimension of $4$, however in the Suzuki-Trotter formulation
a $d$-dimensional quantum model is mapped onto a classical model with an 
additional imaginary time dimension of length $\beta$, and thus correspond to 
a $d+1$-dimensional classical model in the zero temperature limit.
From this value of the exponent and the behavior
of the gap in the paramagnetic phase (see Eq.~(\ref{eq_gap_finite_p2})) the
authors of~\cite{botet83} 
could deduce the scaling with $N$ of the gap in the critical regime 
$s \approx s_{\rm c}$. Let us reproduce here their argument. Suppose that the
gap $E_1(s)-E_0(s)$ satisfies a scaling assumption in the double limit 
$N \to \infty$, $s\to s_{\rm c}$, i.e.
\beq
E_1(s)-E_0(s) \sim N^{-x} \F((s-s_{\rm c}) N^{x'}) \ , 
\label{eq_fss_p2}
\eeq
with $\F$ a scaling function and $x,x'$ two exponents to be determined. This
assumption can agree with the study of the ferromagnetic phase only if 
$x'=2/3$, with $\F(z) \approx \exp[- 9 z^{3/2}]$ as $z \to + \infty$. On
the other hand, approaching the transition from the paramagnetic phase
leads to a closing of the (finite) gap as a square root 
(see Eq.~(\ref{eq_gap_finite_p2})), hence $\F(z) \sim 2 \sqrt{2} (-z)^{1/2}$ as 
$z \to - \infty$ and $x=x'/2=1/3$. The scaling assumption and the behavior
of the gap as $N^{-1/3}$ in the critical regime of the $p=2$ model were
checked numerically in~\cite{botet83}. 

\begin{figure}
\centerline{
\includegraphics[width=8.3cm]{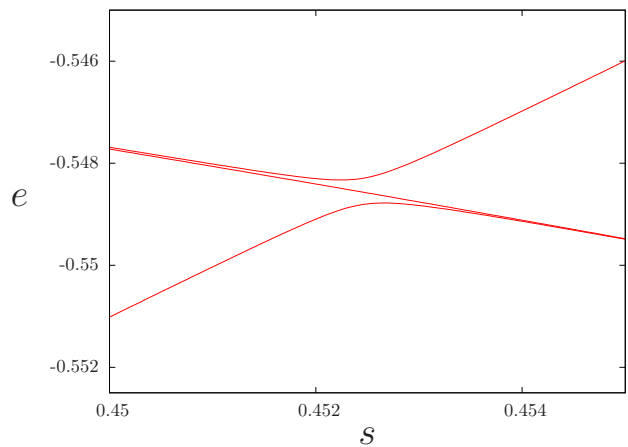}}
\caption{The energy of the three lowest levels of the $p=4$ model in the
neighborhood of the first-order transition, obtained from numerical 
diagonalization with $N=40$.}
\label{fig_p4_sc}
\end{figure}

Let us finally discuss the structure of the gaps between the lowest states
of a model with $p \ge 4$ even, in the neighborhood of its first-order 
transition. For the choice of parameters $(s,e)=(s_{\rm c},\ec)$, the argument
of the $\arg\cosh$ function in Eq.~(\ref{eq_varphiprime_mz}) reaches the 
branching point $1$ in $-\mc,0$ and $+\mc$, one can thus construct three
distinct quasi-eigenvectors with rate $\varphi(m)$ vanishing for these
three magnetizations. One could think that the reasoning presented
at the beginning of this section, that reduces to the diagonalization of
a two by two matrix, is invalidated. This is however not the case, as is
best understood by looking at the three lowest levels of
the $p=4$ model plotted on Fig.~\ref{fig_p4_sc}. On the ferromagnetic side
of the transition the groundstate (resp. the first excited state) is the
symmetric (resp. antisymmetric) combination of the ferromagnetic 
quasi-eigenvectors concentrated on $\pm \mc$, with an exponentially small
splitting of order $\exp[-N \beta_p(s_{\rm c})]$, while the second excited
state is the metastable continuation of the paramagnetic groundstate.
The avoided crossing of order $\exp[-N \alpha_p]$ thus occurs between the
groundstate and the second excited state; as 
$\alpha_p = \frac{1}{2} \beta_p(s_{\rm c})$, this gap is much more opened
than the one between the ferromagnetic states. The fact that the first excited 
state has no level repulsion at the avoided crossing is easily understood
from the additional symmetry of even $p$ models discussed at the end of 
Sec.~\ref{subsubsec_spin_sectors}: the paramagnetic and symmetric
combination of ferromagnetic states belongs to the sector invariant with
respect to the reversal of the longitudinal magnetization, while the
antisymmetric combination is in the other, disconnected, sector.
Note that on the paramagnetic side of the transition the splitting of
order $\exp[-N \beta_p(s_{\rm c})]$ occurs between the first and second
excited states.

\subsubsection{Exponentially small gaps between excited states}
\label{subsubsec_gap_excited}

In the two previous cases we computed the exponentially small splitting
between two quasi-degenerate groundstates, the ferromagnetic and 
paramagnetic ones at the first-order transition in 
Sec.~\ref{subsubsec_gap_first_order} and the two ferromagnetic ones in 
Sec.~\ref{subsubsec_gap_ferro}. It should however be clear 
(see for instance the left panel of Fig.~\ref{fig_salient_p3_ssp}) that
exponentially small gaps occur not only between the two lowest eigenstates, but 
also between excited ones. In this subsection we shall explain how to
adapt the computation in that case, and in the next one we will in particular 
obtain the exponential rate of closing of the gaps encountered by the 
metastable continuation of the ferromagnetic and paramagnetic phases, which 
shall be a crucial ingredient for the analysis of the annealing dynamics in 
Sec.~\ref{sec_dynamics}.

In terms of the leading order eigenvalue equation (\ref{eq_varphiprime_mz}),
an exponentially small splitting between two eigenstates shows up as the
existence of two distinct solutions $\varphi_{1,2}(m,s,e)$ of
Eq.~(\ref{eq_varphiprime_mz})
that both fulfill 
the condition $\underset{m}{\inf} [\Re \, \varphi(m,s,e)]=0$. We shall call 
$\gamma(s,e)$ the rate at which this gap closes, i.e. it is at the leading
order of the form $e^{-N \gamma(s,e)}$. As previously explained this rate
is obtained from the scalar product between the two quasi-eigenvectors.
The two functions $\varphi_{1,2}$ only differ by a choice of branch of the
$\arg\cosh$ function on an interval $[m_1(s,e),m_2(s,e)]$. This implies
that their imaginary part is the same for all $m$ 
(see Eq.~(\ref{eq_def_branches})), hence the scalar product between the two
eigenvectors depends only on the real part of $\varphi_{1,2}$. 
This leads to:
\beq 
\bes
\gamma(s,e)&=- \lim_{N \to \infty} \frac{1}{N} \ln | \la \phi_1 | \phi_2 \ra | 
= - \lim_{N \to \infty} \frac{1}{N} 
\ln \left | \int_{-1}^{1} \dd m \, 
e^{-N \varphi^*_1(m,s,e) -N \varphi_2(m,s,e)}\right | \\
&= \inf_m \Re [\varphi_1(m,s,e) + \varphi_2(m,s,e)] 
=\frac{1}{2} \int_{m_1(s,e)}^{m_2(s,e)} \dd m \, 
\ach \left(\frac{|e+s \, m^p|}{(1-s) \sqrt{1-m^2}} \right)\ .  
\end{split}
\label{eq_gamma_s_e} 
\eeq

Let us now describe the regions in the $(s,e)$ plane where exponentially small
gaps occur. The discussion above shows that their occurence can be traced back
to the number of times the argument of the $\arg\cosh$ function in 
Eq.~(\ref{eq_varphiprime_mz}) reaches the branching points $\pm 1$, 
in other words the number of solutions $m \in [-1,1]$ of the equations
\beq
e=-s \, m^p - (1-s) \sqrt{1-m^2} \qquad \text{or} \ \ 
e=-s \, m^p + (1-s) \sqrt{1-m^2} \ .
\label{eq_nb_sol}
\eeq
A moment of thought reveals that for any value of $(e,s)$ in the authorized
range of eigenvalues (defined in (\ref{eq_e_authorized})) this number is 
either 2, 4 or 6 (counting twice the marginal case of a branching point
touched quadratically and not crossed). The first case corresponds to a
non-degenerate eigenstate, the two others to exponentially small gaps
between eigenstates. The frontiers between these domains correspond to
the disappearance of some solutions of the equations (\ref{eq_nb_sol}), which
define implicitly $m$ as a function of $s,e$. Their boundary can thus be
obtained as the limits of validity of the implicit function theorem. After
a short computation one realizes that they are given by curves $e(s)$ of the
form (\ref{eq_nb_sol}), with $m$ replaced by one of the solutions (stable or
instable) of Eq.~(\ref{eq_mag_thermo}). Let us be more concrete by 
distinguishing between various cases:
\begin{itemize}
\item for $p=2$, when $s \le s_{\rm c}$ the spectrum is made of non-degenerate
eigenvalues with $e\in[-(1-s),(1-s)]$. When $s \ge s_{\rm c}$ the low-energy
part of the spectrum ($e\in [\egs(s),-(1-s)]$) has doubly degenerate 
eigenvalues with an exponentially small gap between them, while the 
high-energy spectrum ($e\in[-(1-s),1-s]$) is non-degenerate. These two
regimes are depicted in the left panel of Fig.~\ref{fig_smallgaps_p2}, and
agree with the qualitative features described in Sec.~\ref{sec_salient}
on the basis of the numerical diagonalization (cf. the left panel of
Fig.~\ref{fig_salient_p2}).
For clarity a zoom of the latter is presented on the right panel of
Fig.~\ref{fig_smallgaps_p2}, in the neighborhood of the line $e=-(1-s)$,
which is indeed the point where the energy splitting of excited ferromagnetic 
states is no more exponentially small.
\begin{figure}[h]
\centerline{
\includegraphics[width=8.3cm]{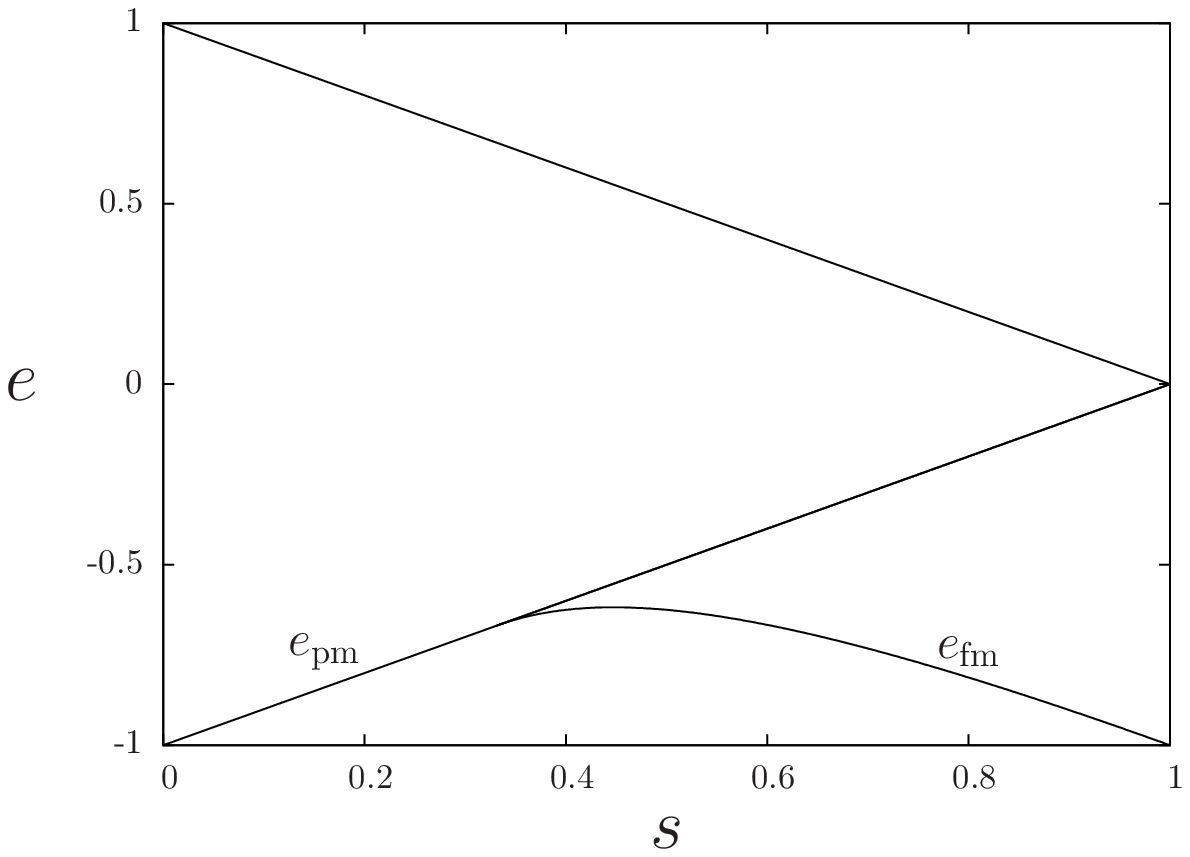} \hspace{6mm}
\includegraphics[width=8.3cm]{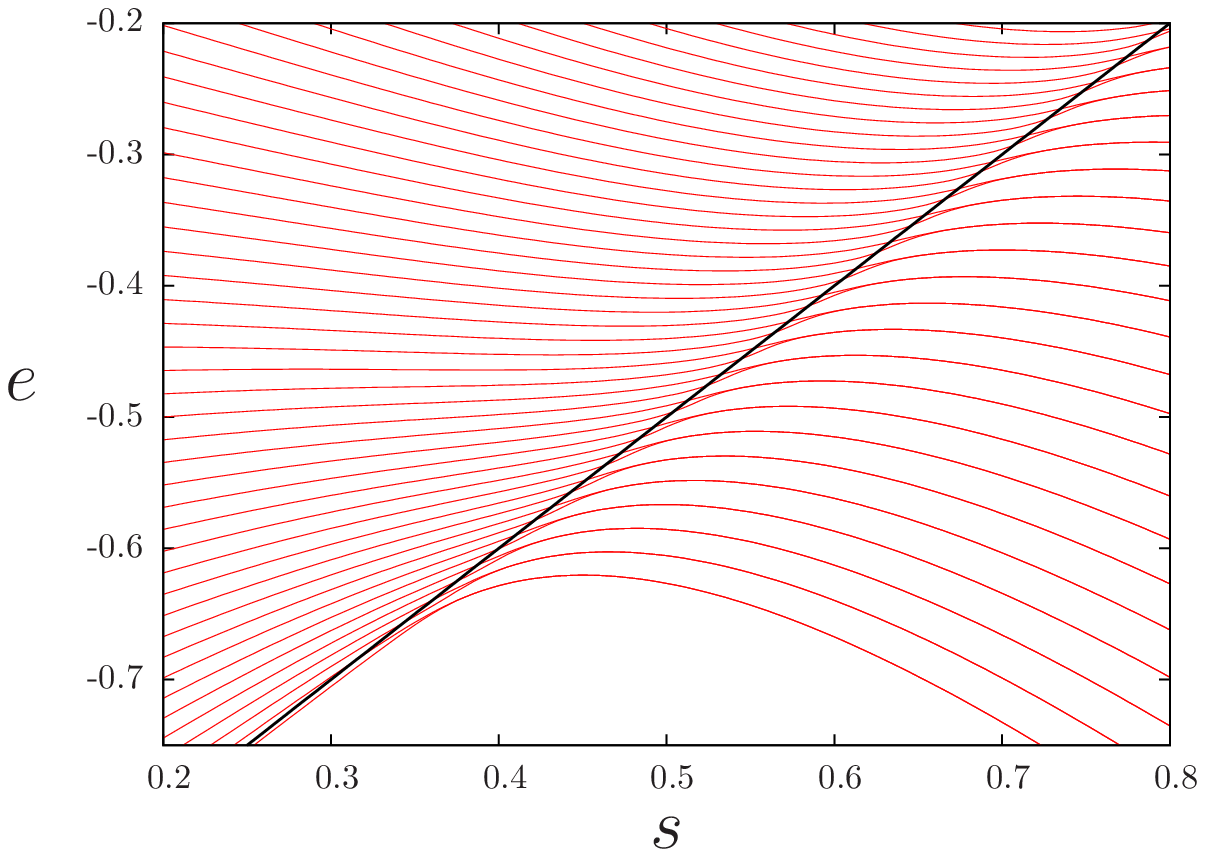}
}
\caption{Left panel: the two regimes in the $(s,e)$ plane for the
$p=2$ model. Right panel: a zoom of the spectrum obtained by numerical
diagonalization for $N=80$ around the line $e=-(1-s)$ where the gaps
between excited ferromagnetic states are no longer exponentially small.}
\label{fig_smallgaps_p2}
\end{figure}
\item for $p\ge 3$ odd, the spectrum is symmetric under $e \to -e$ (as
explained at the end of Sec.~\ref{subsubsec_spin_sectors}), we shall thus
describe only its part with negative $e$. The equation 
(\ref{eq_mag_thermo}) has only $m=0$ as a solution for $s\le \ssp$, with
an associated energy $\epm(s)=-(1-s)$, while for $s\ge \ssp$ there are
three solutions $0<\mi(s) < m_*(s)$ with energies $\epm(s)$, 
$\ei(s)$ and $\efm(s)$. These three energy curves are drawn on the plots
of Fig.~\ref{fig_smallgaps_p3} for $p=3$; the energies corresponding to
doubly-degenerate eigenstates with exponentially small gaps between them
are in the range $[\max[\efm(s),\epm(s)],\ei(s)]$ for $s\ge \ssp$.
On the right panel we have superimposed the spectrum for $N=320$, one
sees indeed that the avoided crossings occur precisely in this regime.
All other authorized values of the energy correspond to non-degenerate
eigenvalues.
\begin{figure}
\centerline{
\includegraphics[width=8.3cm]{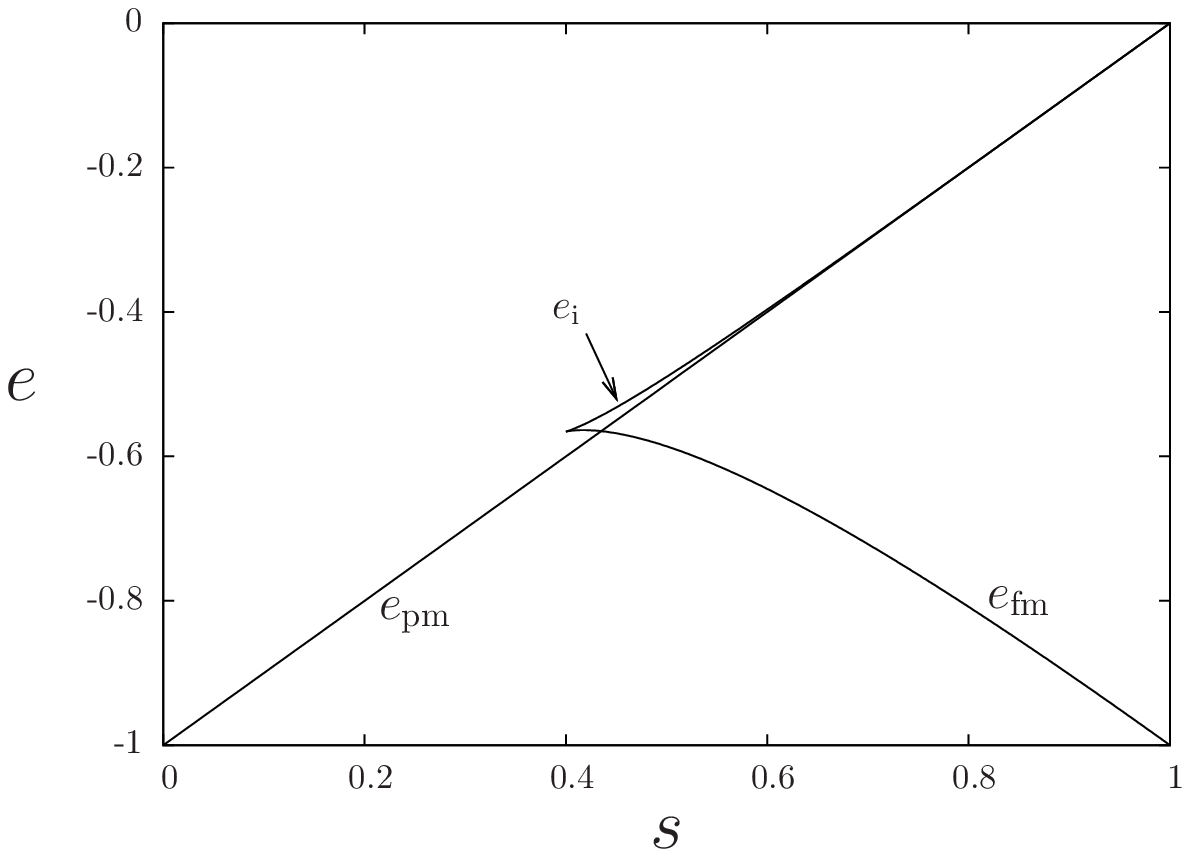} \hspace{6mm}
\includegraphics[width=8.3cm]{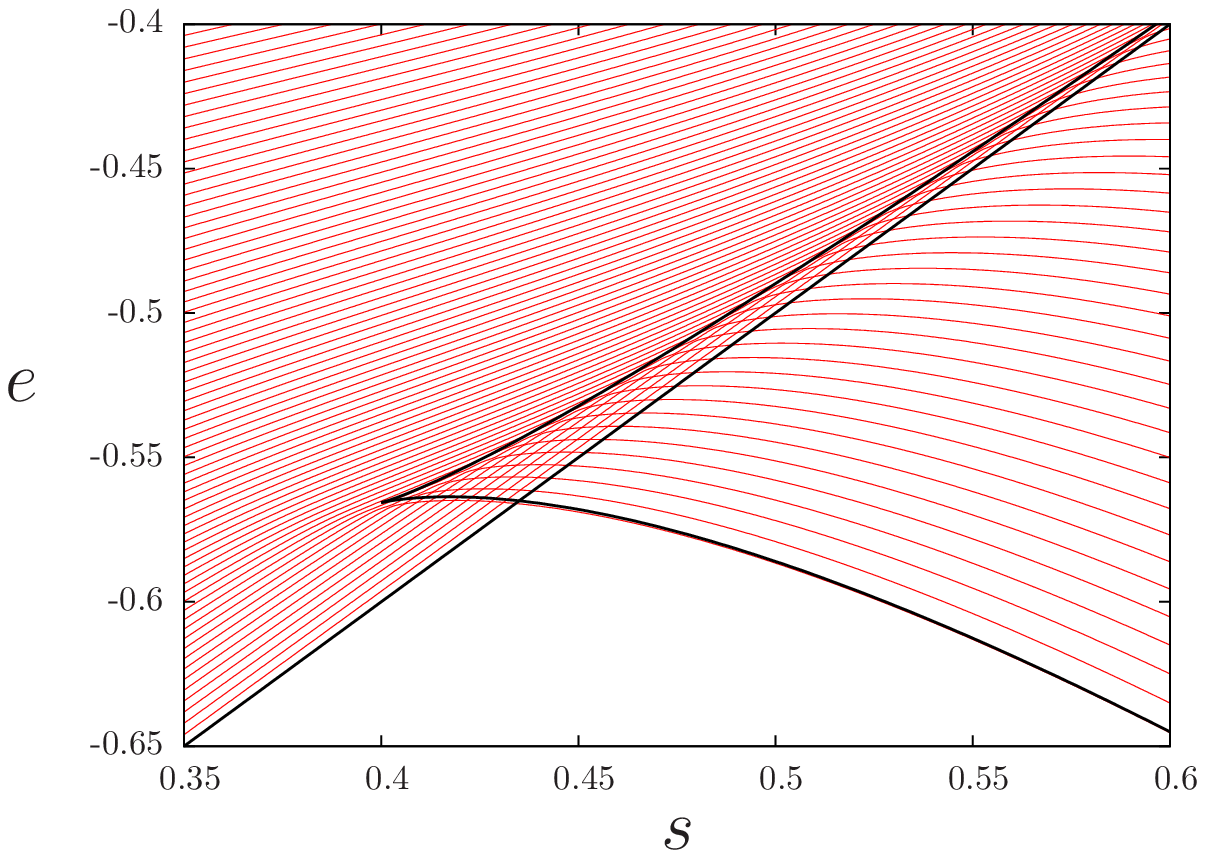}
}
\caption{Left: the three areas in the negative energy part of the spectrum
of $p=3$. 
Right: a blow up of the data from numerical diagonalization in
the region of exponentially small gaps for $N=320$.}
\label{fig_smallgaps_p3}
\end{figure}
\item for $p\ge 4$ even, the phenomenology is mixed between the one of
the $p=2$ and the $p\ge 3$ odd cases. Three zones are to be distinguished
in the $(e,s)$ plane (see Fig.~\ref{fig_smallgaps_p4}). 
One corresponds to the ferromagnetic phase,
with doubly quasi-degenerate eigenstates of opposite magnetizations, for 
$s \ge s_{\rm c}$ and $e\in[\egs(s),\epm(s)]$. In the area $s\ge \ssp$, 
$e\in[\max[\efm(s),\epm(s)],\ei(s)]$ there are three valid solutions
of Eq.~(\ref{eq_varphiprime_mz}). As explained at the end of
Sec.~\ref{subsubsec_gap_ferro} avoided crossings in this area, that
are of order $e^{-N\gamma(s,e)}$, only occur between continuation of levels
coming from the paramagnetic zone and combination of ferromagnetic
quasi-eigenvectors that have the same parity under the reversal of the
longitudinal magnetization. The splitting between symmetric and antisymmetric 
combinations of ferromagnetic states is much smaller, of order
$e^{-2 N\gamma(s,e)}$.
This phenomenon comes from the additional symmetry of
even $p$ models discussed at the end of Sec.~\ref{subsubsec_spin_sectors},
and is illustrated on the right panel of Fig.~\ref{fig_smallgaps_p4}.
The other authorized regime in the $(e,s)$ plane leads to single solutions of
Eq.~(\ref{eq_varphiprime_mz}).
\begin{figure}
\centerline{
\includegraphics[width=8.3cm]{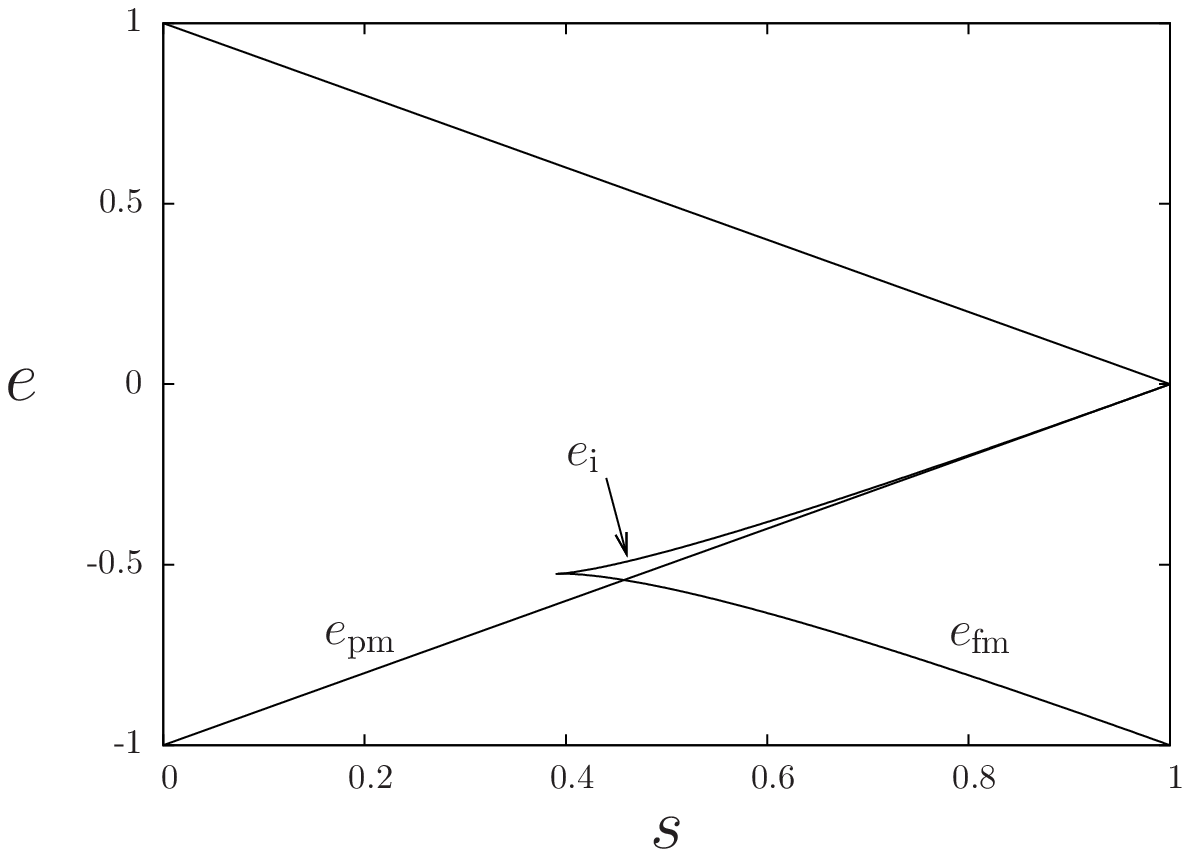} 
\hspace{6mm}
\includegraphics[width=8.3cm]{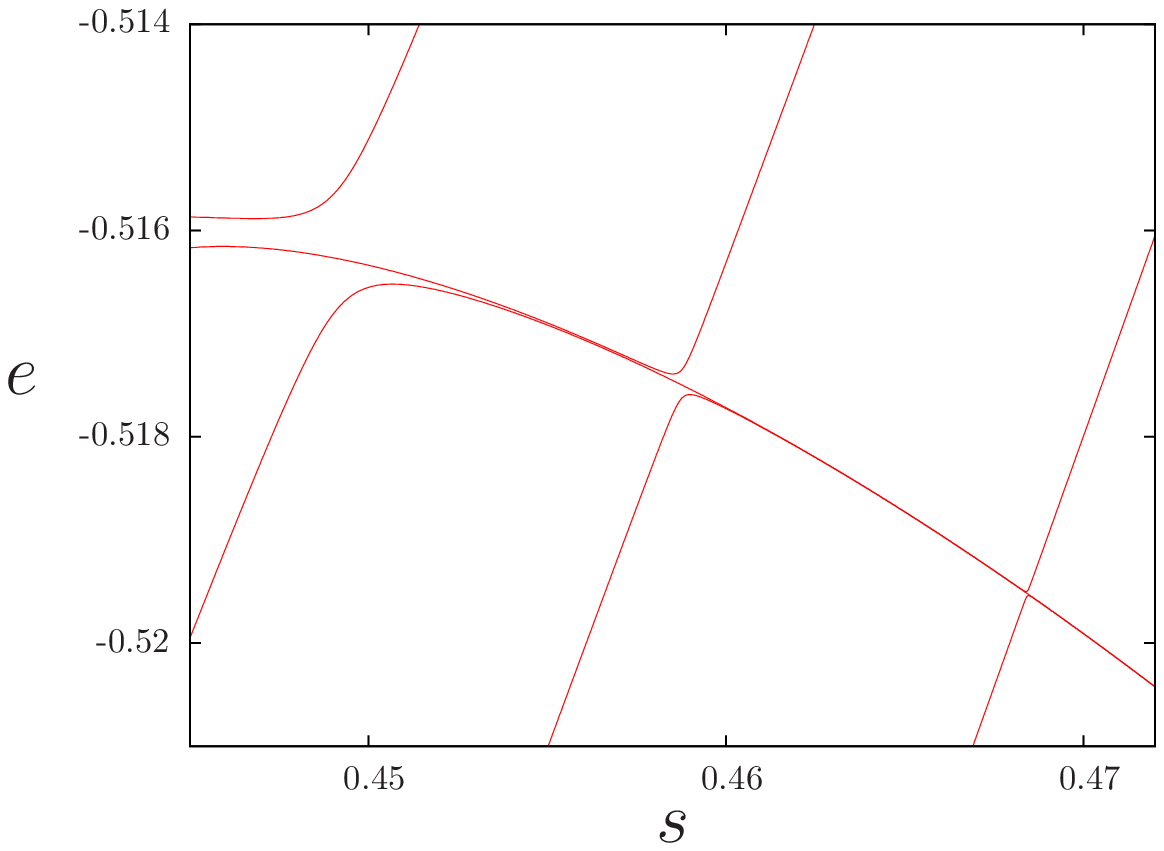}
}
\caption{Left: the three areas in the spectrum
of the $p=4$ model. Right: avoided crossings between paramagnetic excited 
states and quasi-degenerate ferromagnetic excited states, for $p=4$ and
$N=40$.}
\label{fig_smallgaps_p4}
\end{figure}
\end{itemize}

\subsubsection{Exponentially small gaps encountered by the metastable 
states}
\label{subsubsec_gap_metastable}

\begin{figure}
\centerline{
\includegraphics[width=8.3cm]{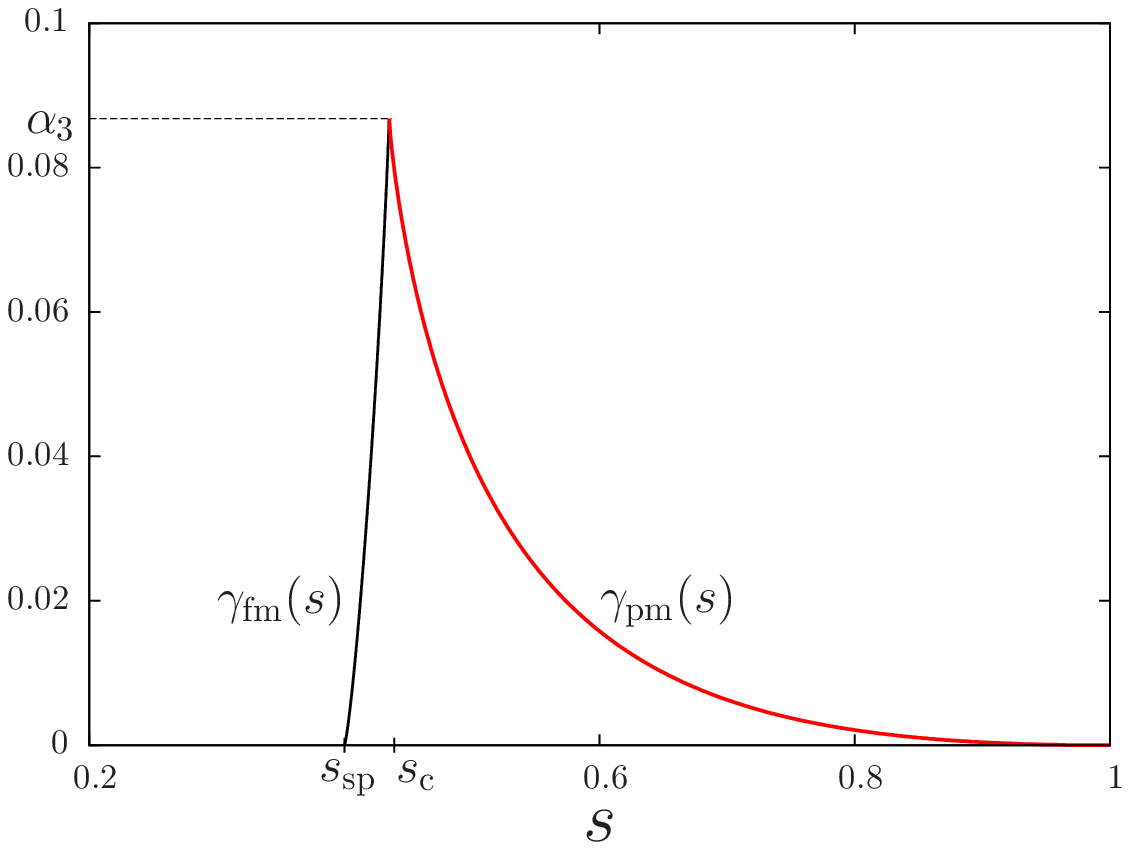} 
\hspace{6mm}
\includegraphics[width=8.3cm]{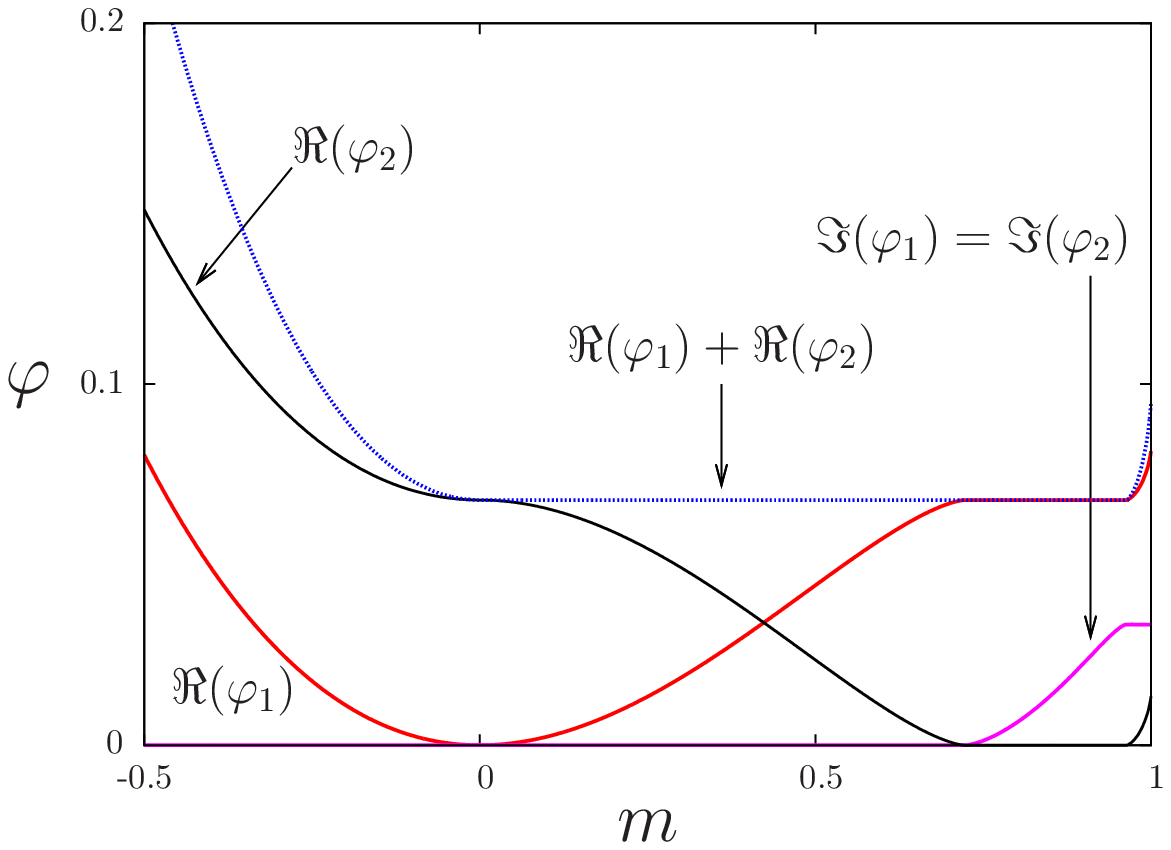}
}
\caption{Left panel: the exponential rates $\gammapm(s)$ and $\gammafm(s)$
of the gaps encountered along the metastable continuation of the paramagnetic
and ferromagnetic groundstates, for $p=3$. 
Right panel: the functions $\varphi_{1,2}$ yielding $\gammapm(s=0.45)$.}
\label{fig_gammas}
\end{figure}

In view of the application of these computations to the annealing dynamics
in Sec.~\ref{sec_dynamics}, the most important case to consider is
the size of the gaps encountered along the metastable continuations of
the paramagnetic and ferromagnetic groundstates. We shall thus define
$\gammapm (s) = \gamma(s,\epm(s))$ for $s \in [s_{\rm c},1]$ and 
$\gammafm(s) = \gamma(s,\efm(s))$ for $s \in [\ssp,s_{\rm c}]$. These
quantities are plotted for $p=3$ in Fig.~\ref{fig_gammas}, along with an 
example of the functions $\varphi_{1,2}$ involved in the computation
of $\gammapm$ for one value of $s$. The numerical evaluation of these
quantities is easy thanks to the explicit expression (\ref{eq_gamma_s_e}).
One can also perform analytically some expansions around special values:
\begin{itemize}
\item 
$\gammafm(s)$ vanishes at $\ssp$ as $\hgammap (s-\ssp)^{5/4}$, with
the prefactor expressed as
\beq
\hgammap = \frac{6\sqrt{2}}{5} 
\frac{(p-1)^{\frac{5(p-3)}{8}}}{p^{\frac{5}{4}} (p-2)^{\frac{5p-12}{8}}}
\left(1 + p \frac{(p-2)^{\frac{p-2}{2}}}{(p-1)^{\frac{p-1}{2}}}
\right)^{\frac{5}{2}} \ .
\label{eq_gammafm_s_to_ssp}
\eeq
The exponent $5/4$ is in agreement with the reasoning of~\cite{botet83} 
recalled in Sec.~\ref{subsubsec_gap_ferro}. Indeed the spinodal transition
is in the universality class of cubic field theories, with the upper critical
dimension (taking into account the imaginary time direction) $\dc=5$, and
the mean-field value of the critical exponent for the divergence of the 
correlation length $\numf=\frac{1}{4}$. One can also adapt the Finite Size 
Scaling argument of~\cite{botet83} to predict that for large but finite values 
of $N$ the gaps encountered in the neighborhood of the spinodal should scale 
as $N^{-4/25}$. This follows from a scaling hypothesis of gaps of the
form given in Eq.~(\ref{eq_fss_p2}), combined with the exponent $x'=4/5$ 
obtained
above from the limit $s\to \ssp^+$, and the closing of the finite gaps
in the limit $s\to \ssp^-$, argued to occur with an exponent 1/5 at the end of
Sec.~\ref{sec_finite_gaps}.

\item
On the other hand the vanishing of $\gammapm$ when $s\to 1$ is non-universal
(i.e. depends on $p$),
one finds indeed 
\beq
\gammapm(s=1-\delta) \sim \tgamma_p \, \delta^{\frac{2}{p-2}} \ , \qquad
\tgamma_p = \frac{1}{2^{\frac{p}{p-2}}} \int_0^1 \dd x \, x\sqrt{1-x^{p-2}} 
\ .
\label{eq_gammapm_s_to_one}
\eeq

\item
In the neighborhood of $s_{\rm c}$ the behavior of $\gammafm$ and $\gammapm$
exhibit a singularity of the form $(s-s_{\rm c}) \ln(s-s_{\rm c})$, more precisely
\beq
\gammapm(s_{\rm c}+\delta) \sim  
\alpha_p + \widetilde{\eta}_p \, \delta \ln(\delta) \ , \qquad
\gammafm(s_{\rm c}-\delta) \sim  
\alpha_p + \widehat{\eta}_p \, \delta \ln(\delta) \ ,
\label{eq_gammas_s_to_sc}
\eeq
where the constants $\widetilde{\eta}_p$ and $\widehat{\eta}_p$ are given by
\beq
\widetilde{\eta}_p = 
\frac{(p-1)^{p-\frac{5}{2}}}{p^\frac{p}{2} (p-2)^\frac{p-3}{2}} 
\left(1+  \frac{ p^\frac{p}{2}(p-2)^\frac{p-2}{2}}{(p-1)^{p-1}} \right)^2 \ ,
\qquad 
\widehat{\eta}_p = \sqrt{(p-1)(p-2)} \, \widetilde{\eta}_p \ .
\eeq
\end{itemize}

\section{Quantum annealing of the models}
\label{sec_dynamics}

This section is devoted to the study of the annealing of the models whose 
static properties were considered above, and is organized as follows. 
We shall first (in Sec.~\ref{sec_dynamics_definitions}) define precisely the 
dynamics and the quantities of interest to be studied in this context. 
Then we will review the phenomenology of the simple, two-level, Landau-Zener 
problem in Sec.~\ref{sec_dynamics_LZ} and discuss on this basis the expected 
features of the annealing dynamics. A further simplified model
is introduced as an aside in Sec.~\ref{subsubsec_def_full_matrix}, that
will be used as a benchmark for the comparison with numerical results.
The actual computations and results will 
then be presented in two sections, divided according to the scaling of the 
annealing time with the system size; in Sec.~\ref{subsec_exptimes} we shall 
consider annealings on exponentially large times, while in 
Sec.~\ref{subsec_csttimes} we will study the behavior of the dynamics when 
the thermodynamic limit is taken for a finite annealing rate.
The main results of Sec.~\ref{subsec_exptimes} and \ref{subsec_csttimes} are
presented for odd values of $p \ge 3$ for which the phase-transition is first 
order and not mixed up with the quasi-degeneracy of
the ferromagnetic states. We briefly comment in 
Sec.~\ref{subsec_otherp} on the behavior for even values of $p$,
in particular $p=2$ (the Curie-Weiss model) whose annealing was already 
studied in~\cite{santoro08,solinas08,itin09_1,itin09_2}.

\subsection{Definitions}
\label{sec_dynamics_definitions}

As sketched in the introduction the quantum annealing procedure, or quantum
adiabatic algorithm, aims at finding the groundstate of some final Hamiltonian
$\hHf$ via an interpolation from an initial Hamiltonian $\hHi$ whose 
groundstate is easy to construct. The system evolves from time $t=0$ to 
$t=T$, the total running time of the algorithm, according to the
Schr\"{o}dinger equation with an Hamiltonian interpolating (for instance 
linearly) between $\hHi$ and $\hHf$. In terms of the reduced time 
$s=t/T \in[0,1]$, this reads
\beq 
\frac{i}{T} \frac{\dd}{\dd s} |\phi_T(s) \ra  = \hH(s) | \phi_T (s) \ra \ ,
\qquad  
\hH(s) = (1-s) \hHi + s\hHf  \ ,
\label{eq_schrodinger_annealing} 
\eeq
with the initial condition that $|\phi_T(0)\ra$ is the (normalized) 
groundstate of $\hHi$ (we set $\hbar=1$ from now on).
The outcome of the algorithm for an annealing time $T$ is thus the final 
state $| \phi_T(1)\ra$, which ideally, if $T$ is much larger than the 
adiabatic time, is close to the groundstate of $\hHf$.

It is however interesting, in particular for the approximability issues 
mentioned in the introduction, to study this procedure also for $T$ smaller
than the adiabatic time. We shall quantify the deviation from adiabaticity
by computing the final energy density defined as
\beq 
\efin(T,N) = \frac{1}{N} \la \phi_T(1) | \hHf | \phi_T(1) \ra \ ,
\eeq 
and comparing it to the groundstate energy density $\egs$ of the final 
Hamiltonian $\hHf$: the residual energy density is thus $\eres=\efin-\egs$.
Another relevant energy density to compare $\efin$ to is the trivial one 
achieved when the interpolation time vanishes, i.e. when one computes the 
average energy of the final Hamiltonian with respect to the groundstate of 
the initial one: $\etriv=\frac{1}{N} \la \phi_T(0) | \hHf | \phi_T(0)\ra$.
Indeed $\egain=\etriv-\efin$ is the gain in energy density that is achieved
by the evolution during the time $T$. Note that with the normalization we
chose for the models one has $\egs=-1$ and $\etriv=0$.

Our analytical results will all be obtained in the thermodynamic limit
$N \to \infty$, but with two different scaling of $T$ with $N$ that shall
be distinguished typographically. If $T$ is kept fixed when $N$ diverges we
shall denote
\beq 
\efin(T) = \lim_{N \to \infty} \efin(T,N) \ ;
\eeq 
this regime will be studied in Sec.~\ref{subsec_csttimes}.
On the other hand if $T$ scales exponentially with $N$ 
(as in Sec.~\ref{subsec_exptimes}) we call $\tau$ this
exponential rate and define
\beq 
\efin(\tau) = \lim_{N \to \infty} \efin(T=e^{N \tau},N) \ .
\eeq 
We shall argue in the following that, as far as intensive quantities like
the energy density are concerned, these two regimes are the only relevant ones
for $p\ge 3$ (see Sec.~\ref{subsec_otherp} for a discussion of the
different case $p=2$), i.e. polynomial scalings of $T$ with $N$ are just
limiting cases of the two regimes above. Note also that in the thermodynamic
limit, for both regimes, the quantum fluctuations of the final energy density
are neglectible, hence a description in terms of the average energy only
is meaningful.

The definitions above are valid for any choice of the inital and final
Hamiltonians. From the point of view of potential applications they are
of course most interesting when $\hHf$ has a groundstate that is a priori hard
to find and when it is easy to prepare the system in the groundstate of 
$\hHi$. In the following we shall consider the dynamics of the annealing of
the models whose statics were studied in the first part, that is use a $p$-spin
interaction and a transverse field as initial and final Hamiltonian. Obviously
neither of these Hamiltonians has a groundstate which is hard to find, hence
they can only be considered as toy models for the application of the quantum 
adiabatic algorithm. However they share some properties (first-order 
transitions, metastability, spinodals) with more realistic random 
combinatorial optimization problems~\cite{qXOR,young10}, while being much 
easier to
study both analytically and numerically. Because of this unrealistic character
both choices of the transverse field as $\hHi$ and the ferromagnetic 
interaction as $\hHf$ or viceversa are equally relevant, and it will be very
instructive to consider these two types of evolution. Let us define them more
precisely:
\begin{itemize}
\item The \textit{annealing towards the ferromagnet} corresponds to the
choice $\hHi = - N\hmx$, $\hHf= - N(\hmz)^p$, i.e.
\hbox{$\hH(s) = -(1-s)N \hmx - s N (\hmz)^p$}
is precisely the Hamiltonian (\ref{eq_def_model}) studied in the first
part of the paper.

\item The \textit{annealing towards the paramagnet} corresponds to
the reverse choice $\hHi= - N(\hmz)^p$, $\hHf = - N\hmx$, in other
words the evolution with the Hamiltonian (\ref{eq_def_model}) is made
with $s$ decreasing from 1 to 0. To avoid confusion we shall denote
$u=1-s$ instead of $s$ the reduced time in this case, i.e. study
the following equation:
\beq 
\frac{i}{T} \frac{\dd}{\dd u} |\phi_T(u) \ra  
= [-N (1-u) (\hmz)^p - N u \, \hmx  ]  | \phi_T (u) \ra \ .
\label{eq_schrodinger_annealing_back} 
\eeq
\end{itemize}

Note that in all these cases the groundstate of the initial Hamiltonian
$\hHi$ belongs to the fully symmetric sector of maximal spin. The instantaneous
Hamiltonian $\hH(s)$ is block diagonal with respect to the spin decomposition
for all values of $s$, in consequence the state $|\phi_T(s) \ra$ remains in the
maximal spin sector $K=0$ all along the evolution.

\subsection{Finite duration Landau-Zener problem and its expected 
consequences}
\label{sec_dynamics_LZ}

The Landau-Zener problem~\cite{landau32, zener32} is the simplest example of 
a quantum evolution with a time-evolving Hamiltonian. It involves two levels
of linearly varying energy with a fixed coupling between them:
\beq
i \frac{\dd}{\dd t} \begin{pmatrix} \psi_1(t) \\ \psi_2(t) \end{pmatrix}
= \begin{pmatrix} a t &  \epsilon \\ \epsilon & - a t \end{pmatrix}
\begin{pmatrix} \psi_1(t) \\ \psi_2(t) \end{pmatrix} \ .
\eeq
The initial condition is given by $\psi_1(t\to -\infty)=1$, i.e. the system
is initially in its groundstate. The probability of transition to the excited
state after an infinite time can be computed exactly 
(see~\cite{rojo10,wittig05,volkov06} for modern derivations) and yields
$P_{\rm exc} = \underset{t \to +\infty}{\lim}|\psi_1(t)|^2 =  
\exp(- \pi a \epsilon^2)$.
It is thus a function of the product between the square of the minimal gap
$\epsilon$ at $t=0$ and the velocity $a$ of variation of the energies of the
levels.

Variations of the Landau-Zener model that account for a finite duration of
the interaction have been studied in great details 
in~\cite{vitanov96,vitanov99}. Consider for instance an evolution with
a reduced time $s\in[0,1]$ corresponding to a total physical time $T$, with
two levels that have an avoided crossing at $s=1/2$:
\beq
i \frac{1}{T}\frac{\dd}{\dd s} 
\begin{pmatrix} \psi_1(s) \\ \psi_2(s) \end{pmatrix}
= \begin{pmatrix} a (s-\frac{1}{2}) &  \epsilon \\ 
\epsilon & - a (s-\frac{1}{2}) \end{pmatrix}
\begin{pmatrix} \psi_1(s) \\ \psi_2(s) \end{pmatrix} \ .
\eeq
The probability $P_{\rm exc}(a,\epsilon,T)$ that the evolution starting from 
the groundstate
at $s=0$ leads to the excited state at $s=1$ can be expressed in terms
of special functions~\cite{vitanov96} and simplified in various asymptotic
limits according to the relative ordering of $a,\epsilon$ and $1/T$. In the
present context the relevant regime corresponds to $a$ fixed, 
$\epsilon \to 0$ and $T \to \infty$. Then $P_{\rm exc}$ has a scaling form if
$T$ diverges as $\epsilon^{-2}$, more precisely
\beq
\lim_{\epsilon \to 0} P_{\rm exc}(a,\epsilon,T = \alpha \epsilon^{-2} ) =
\exp[-\pi \alpha/a] \ . 
\eeq
If one further assumes
that both $T$ and $\epsilon$ scales exponentially with a large parameter
$N$, according to $\epsilon(N) = e^{-\gamma N}$ and $T = e^{\tau N}$, then
the probability of excitation reduces to
\beq 
\label{eq_lz_exp_times} 
\lim_{N \rightarrow \infty} P_{\rm exc} = \theta(2 \gamma - \tau) \ ,
\eeq
with $\theta(x)$ the Heaviside step function,
i.e. on this scale either the evolution is sufficiently slow and the
system follows adiabatically the groundstate or it is too fast and
with probability 1 the system goes into the excited state.

Let us now explain the intuitive picture for the dynamics of the $p$-spin
ferromagnetic model (with an odd value of $p \ge 3$) in the large $N$ limit,
that arises from the
combination of the study of this two-level problem and of the results
of Sec.~\ref{sec_statics} (a similar reasoning can be found for instance
in~\cite{santoro02}). Consider first the annealing towards the 
ferromagnet, for a large evolution time $T$, starting from the groundstate
at $s=0$. As long as $s < s_{\rm c}$ the
gap between the groundstate and the first excited state remains finite,
hence for times sufficiently large (but independent of the system size), 
it is expected that the system will remain in the instantaneous ground state. 
There occurs at $s_{\rm c}$ an avoided crossing with an exponentially small gap 
of order $e^{-N \alpha_p}$. Transposing the results of the two-level problem,
two cases have to be distinguished. If the evolution time is exponentially 
large, $T = e^{\tau N}$, and if $\tau \ge 2 \alpha_p$, then the system follows
adiabatically the groundstate at the avoided crossing, and continues on
the instantaneous groundstate. Otherwise the system is in the first
excited state just after the crossing, i.e. on the metastable continuation
of the paramagnetic groundstate. We have seen that this state encounters a
series of avoided crossings, that lead to gaps of order $e^{-N \gammapm(s)}$.
Let us assume that all these avoided crossings are independent, and can be
treated as in a two-level problem. Then, if the evolution time is 
$T=e^{N \tau}$, one is led to the conclusion that the system will remain in
the metastable groundstate until the value $\sturn$ such that 
$\tau = 2 \gammapm(\sturn)$, and from thereon follows the excited 
ferromagnetic state that crossed the paramagnetic metastable state in
$\sturn$. As there is no spinodal limit for the metastable paramagnet 
$\gammapm(s) > 0$ for all $s<1$. Hence for an evolution on sub-exponential 
times $T$ the system should follow the paramagnet until $s=1$, which leads to
a vanishing energy density gain with respect to the trivial one.

A similar reasoning in the case of the annealing towards the paramagnet 
reveals a richer phenomenology.
For an exponentially large annealing time $e^{N \tau}$ with $\tau > 2 \alpha_p$
the groundstate is followed during the whole evolution. If $\tau < 2 \alpha_p$
the metastable ferromagnetic state will be followed until the turning point
$\uturn$ where $\tau = 2 \gammafm(1-\uturn)$, 
then the system follows the paramagnetic excited
state that rejoins the metastable ferromagnet at the turning point. There is
however an important difference with respect to the reverse direction of
annealing: here the ferromagnet has a spinodal limit of metastability.
Hence an evolution on an exponentially long time $e^{N \tau}$, but for 
arbitrarily small values of $\tau$, yields a non-trivial (negative) energy 
density. In addition
the regime of large but sub-exponential $T$ can be expected to be much richer 
than in the previous case: the ferromagnet will be followed for $u<\usp=1-\ssp$,
but for subsequent times this analysis in terms of level crossings can
give no clue.

In this reasoning we have assumed that the various level crossings can
be treated independently one from the others, and apply to each of them
the results of a simple two-level problem. Arguments in favor of this
assumption can be found from static~\cite{schulman76} and 
dynamical~\cite{vitanov99,volkov06} considerations: as can be seen on the 
drawings of the spectrum (see for instance the right panel of 
Fig.~\ref{fig_salient_p3_sc}), an avoided crossing with an exponentially 
small gap affects notably the two colliding levels on an interval of $s$ 
which is also exponentially small. On the other hand two successive crossings 
are located at values of $s$ which are distant of order $1/N$. Similarly in 
the dynamical case the ``duration'' of a crossing (as defined 
in~\cite{vitanov99}) should go like $\exp((-\gamma+\tau)N)$, and therefore 
the influence of a crossing should spread on a range of $s$ of order at most 
$\exp(-\gamma N)$. 

In the following sections we shall present the explicit results obtained
from this reasoning, and compare them with the results of numerical
integration of Schr\"odinger's equation for finite values of $N$. In 
particular we will test the assumption of independence of the different 
crossings. The numerical results for the annealing of the $p$-spin model
having strong finite-size corrections, we shall first introduce a
simplified model that shares some of the properties of the
$p$-spin model but with smaller finite-size effects.

\subsection{A further simplified model (the $p\to \infty$ limit)}
\label{subsubsec_def_full_matrix}

We shall introduce here a simplified version of the models under study, 
first defining it formally and discussing afterwards its relationship
with the main models of the article and with previous works.

\begin{figure}[h]
\center
\includegraphics[width = 8.3cm]{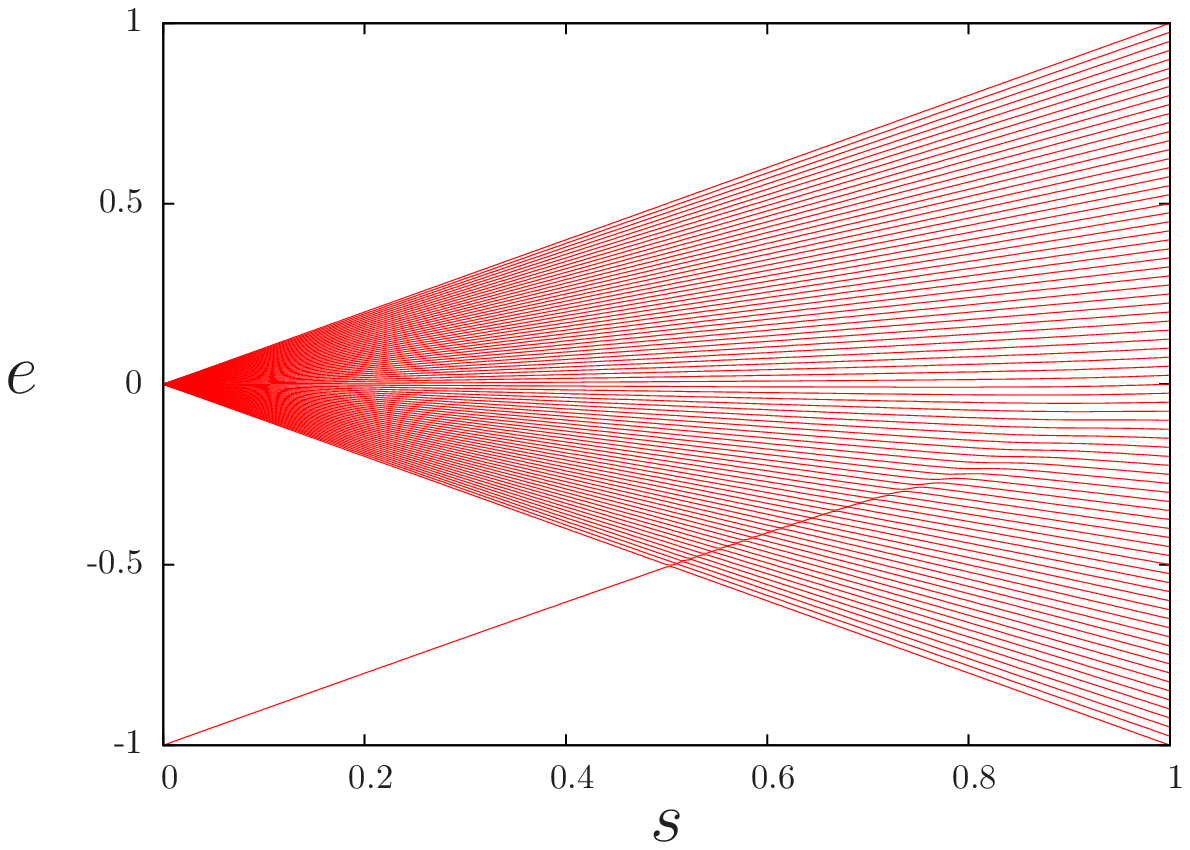}
\includegraphics[width = 8.3cm]{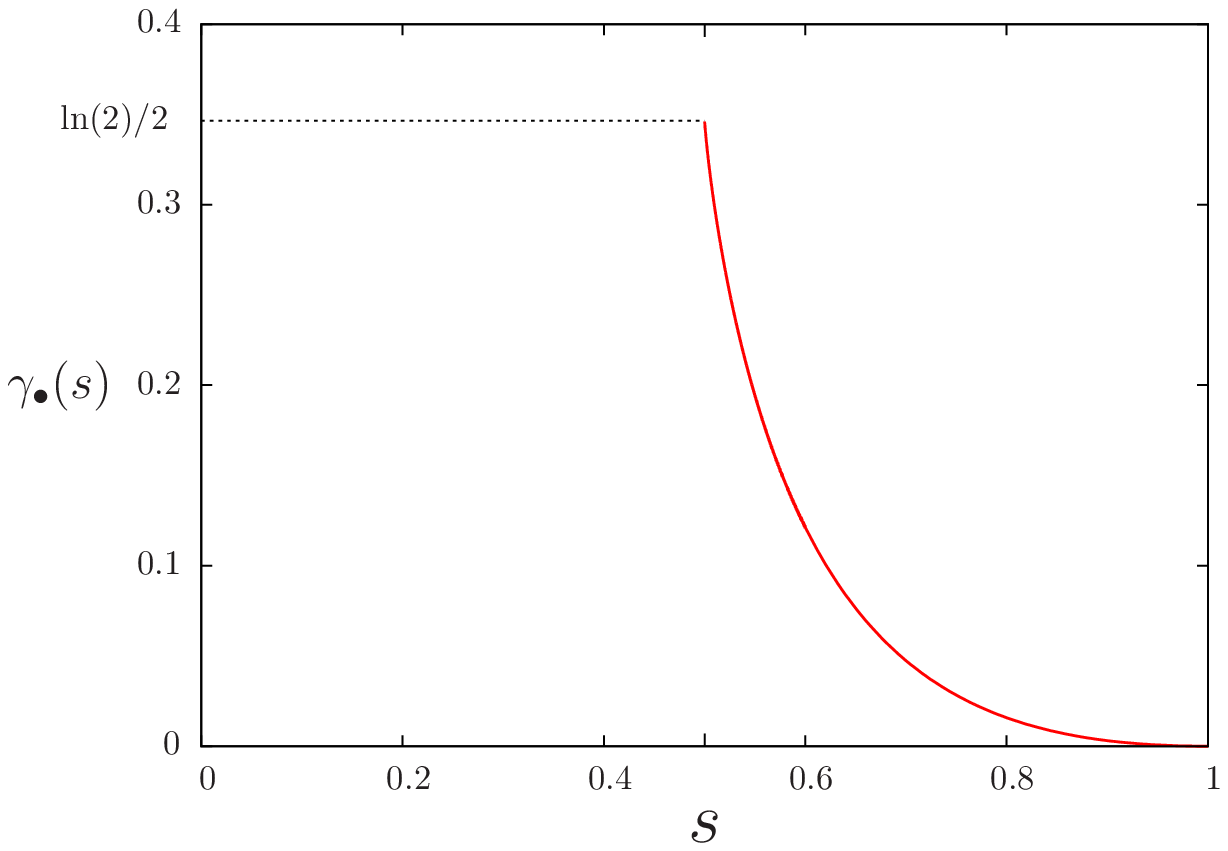}
\caption{Left: spectrum of the operator $\hH_\bullet(s)$ as a function of $s$, 
for $N=80$. In the thermodynamic limit the metastable continuation of the 
groundstate at $s\le 1/2$ exists until $s=1$.
Right: The exponential rate of closing of the gaps at the avoided 
crossings $\gb(s)$, defined in Eq.~(\ref{eq_gb}).
}
\label{fig_spectrum_full_matrix}
\end{figure}

We consider an
interpolating Hamiltonian $\hHb(s)$ acting on the fully symmetric subspace of 
dimension $N+1$. It is given by $\hHb(s)=(1-s) \hJ -N s \hmx$, with the
initial Hamiltonian $\hHi = \hJ$ defined by its matrix elements in the
$x$-diagonal basis,
\beq
\lax m;0 | \hJ | m';0 \rax = - N D_m D_{m'} \ , \qquad 
D_m = \sqrt{\frac{1}{2^N} \binom{N}{N \frac{1+m}{2}}} \ .
\eeq
It is thus a matrix of rank 1, with a single eigenvalue equal to $-N$ and all
other eigenvalues equal to $0$. The spectrum of $\hHb(s)$ is presented on the 
left panel of Fig.~\ref{fig_spectrum_full_matrix}. As $\hJ$ is of rank one
the spectrum of $\hHb(s)$ is essentially equal to the one of $-N s \hmx$
(see for instance~\cite{lowrank} for general results on low rank perturbation 
theory). There is however
a major difference: the isolated eigenvalue of energy density $e=-(1-s)$ 
is continued as a metastable state for $s\in[1/2,1]$, with exponentially 
small avoided crossings of order $e^{-N\gb(s)}$. The computation of $\gb(s)$
is presented in Appendix~\ref{subsubsec_large_dev_full}, and yields
the explicit formula
\beq
\gb(s)=-\frac{1}{2}\ln s + \frac{2 s -1 }{ 4 s} \ln(2 s -1) \ ;
\label{eq_gb}
\eeq
this function is plotted on the right panel of 
Fig.~\ref{fig_spectrum_full_matrix}. For the reasons explained above one
expects that the annealing of the model on sub-exponential time scales
yields a vanishing final energy density, as the metastable continuation
of the groundstate exists until $s=1$; this is confirmed by the analysis
presented in Appendix ~\ref{subsubsec_small_time_full}. 
Exponentially slow annealings with 
$0<\tau < \ln 2$ (i.e. $T \ll 2^N$) should however reach a non-trivial negative 
energy density, larger than the one of the groundstate but smaller than
the trivial one, $\egs<\efin(\tau)<\etriv$ in the notations of 
Sec.~\ref{sec_dynamics_definitions}.

This simplified model is actually (almost) the $p\to \infty$ limit (with
$p$ odd, and the limit on $p$ taken before the limit on $N$) 
of the models studied in the main part of this paper (and was
discussed in these terms in~\cite{jorg10}): 
in the $z$-basis $\hJ$ is
diagonal, with matrix elements $\laz m;0 | \hJ | m;0 \raz = - N \delta_{m,1}$,
to be compared with 
$\underset{p\to \infty}{\lim} \laz m;0 | -N (\hmz)^p | m;0 \raz = 
- N \delta_{m,1} + N \delta_{m,-1}$ if the limit is taken with $p$ odd. If
one focuses on the low-energy part of the spectrum, one can view the evolution
of the simplied model as the evolution towards the paramagnet of the $p$-spin
model in the large odd $p$ limit.

Another justification for the introduction of this simplified model can be
given as follows. Assume that one is given an arbitrary 
Hamiltonian $\hHf$, diagonal
in the computational basis of the $2^N$ classical configurations of spins,
as an optimization problem, and that the problem is to be solved
without using any information about the local structure of these energies
in the configuration space. Then the most natural starting Hamiltonian $\hHi$
for an interpolation is the one connecting any two configurations 
of the Hilbert space with equal probability,
\beq
\hHi=\hJ = - \frac{N}{2^N} \sum_{\us,\us'} 
|\us \rangle \langle \us' | 
= - N |X\rangle\la X | \ , \quad \text{with} \  \
| X \ra = \frac{1}{2^{N/2}} \sum_{\us} |\us\ra \ ,
\eeq
where the normalization chosen is such that $\hJ$ has one eigenvector $|X\ra$ 
with eigenvalue $-N$ and $2^N-1$ eigenvectors with eigenvalue $0$.
Let us denote $\{E_\alpha\}_{\alpha \in [1,M]}$ the distinct energies of
$\hHf$, $d_\alpha$ the number of configurations $\us$ on which $\hHf$ takes
the value $E_\alpha$, and $\widetilde{\mathcal{H}}$ the $M$-dimensional 
Hilbert space generated by the symmetric combinations of the 
states of a given energy: 
\beq 
\widetilde{\mathcal{H}} = \textrm{span} \left\{ | \alpha \ra \right\} \ ,
\qquad  | \alpha \ra = \frac{1}{\sqrt{d_\alpha}} 
\sum_{\us, H_{\rm f}(\us) = E_\alpha}  | \us \ra \ .
\eeq
Then the ground state $| X \ra$ of $\hJ$ is in $\widetilde{\mathcal{H}}$, 
and so is, for any $s$, the vector $|\phi_T(s)\ra$ obtained by the 
evolution according to the Schr\"odinger equation with 
$\hH(s)=(1-s)\hJ + s \hHf$ as interpolating Hamiltonian. The dynamics
can thus be studied in the symmetric subspace $\widetilde{\mathcal{H}}$,
in which the matrix elements of $\hH(s)$ are given by
\beq
\la \alpha | \hH(s) | \beta \ra = s \delta_{\alpha,\beta} E_\alpha
-(1-s) N \frac{\sqrt{d_\alpha d_\beta}}{2^N} \ .
\eeq
The simplified model defined at the beginning of this section is thus
a representative example of this more general construction, in which we
chose $M=N+1$, with equally spaced levels $E_\alpha$ between $-N$ and $+N$,
each with a binomial degeneracy.
The quantum annealing with such an unstructured Hamiltonian $\hJ$ has been 
studied in~\cite{farhi08}. In the context of Grover~\cite{Grover} search 
problem (i.e. with a golf course potential $\hHf$ having a single low energy 
level), it was shown in~\cite{roland01} that a modification of the annealing
procedure could reproduce Grover's quadratic speedup. By slowing down the
interpolation in the neighborhood of the avoided crossing one can indeed 
reduce the adiabatic time to $O(2^{N/2})$.

\subsection{Annealing on exponentially large times}
\label{subsec_exptimes}

\subsubsection{The simplified model}
\label{subsubsec_exp_time_full}

Let us compute the final energy $\efin(\tau)$ after an exponentially
long annealing of duration $T=e^{N\tau}$, for the simplified model of
Sec.~\ref{subsubsec_def_full_matrix}, following the reasoning of
Sec.~\ref{sec_dynamics_LZ}. The turning point $\sturn$ up to which the 
metastable state is followed is given implicitly by 
$2 \gb(\sturn(\tau)) = \tau $,
where the expression of $\gb$ is given in Eq.~(\ref{eq_gb}) 
(if $\tau>2\gb(1/2)=\ln 2$ we set $\sturn(\tau)=1/2$). 
The final energy at the 
end of the annealing is then given by the continuation of the state that 
crosses 
the metastable state at $\sturn$. For this simple model where energy levels 
are at leading order linear functions of $s$ except at the crossings, this 
yields:
\beq \label{eq_eres_full}
\efin(\tau) = - \frac{1 - \sturn(\tau)}{\sturn(\tau)}
= 1 - \frac{1}{\gb^{-1}(\tau/2)} \ ,
\eeq
where $\gb^{-1}$ is the functional inverse of $\gb$, with the convention that
$\gb^{-1}(z)=1/2$ if $z>\frac{\ln 2}{2}$. 

\begin{figure}[h]
\centerline{
\includegraphics[width = 8.3cm]{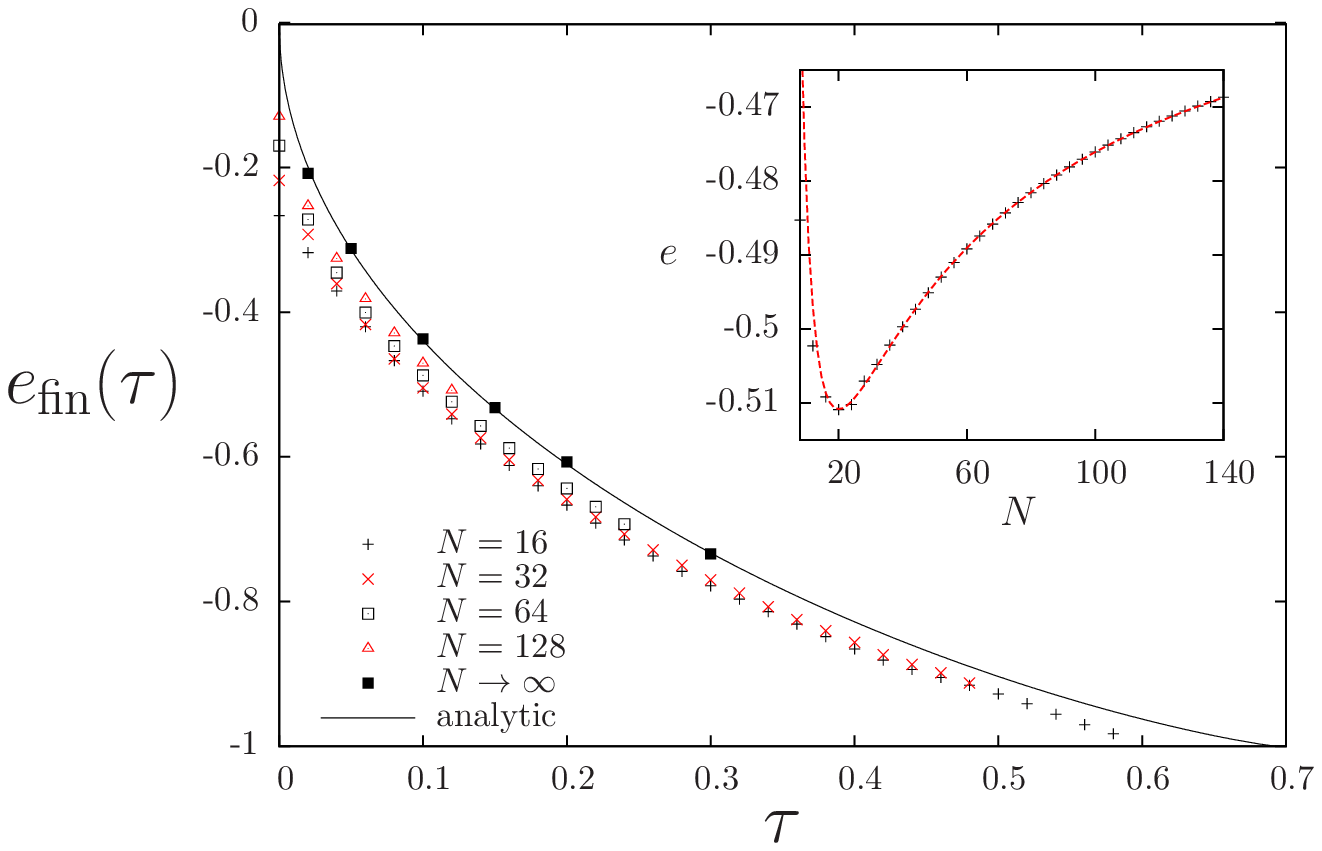}
\includegraphics[width = 8.3cm]{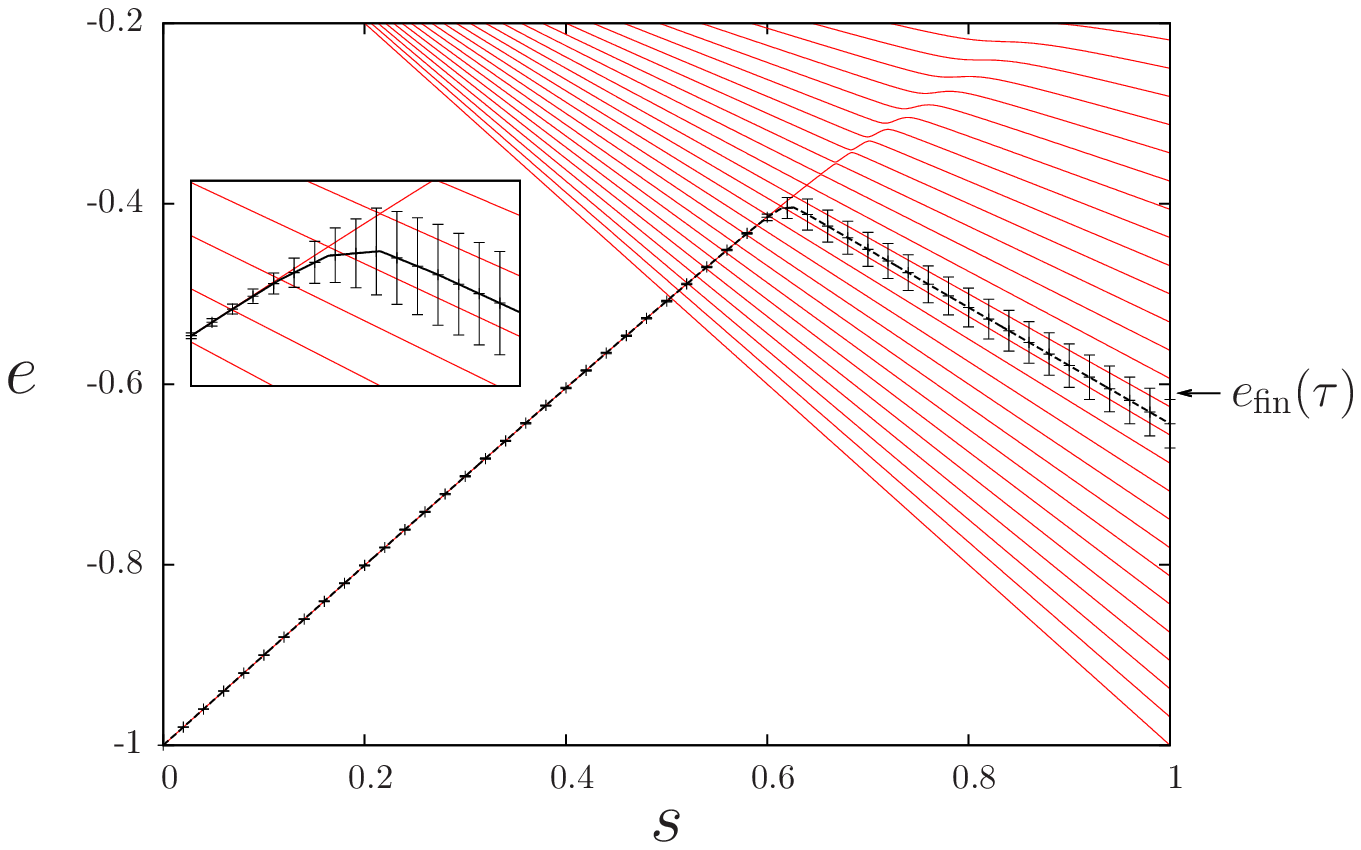}
}
\caption{Left panel: final energy density for the evolution of the
simplified model as a function of $\tau = (\log T)/N$. The solid line
is the analytic prediction (\ref{eq_eres_full}). The symbols are the
results of the integration of Schr\"odinger equation for $N=16,36,64,128$,
and an extrapolation to $N\to \infty$ using corrections in $(\ln N )/N$ 
and $1/N$. The insets shows this fit for $\tau=0.1$, with a fitting function
of the form $f(N) = a + b \ln(N)/N + c/N$, with the results of a best fit
on $N \ge 20$ given by $a = -0.438$, $b = -1.463$,  $c = 2.926$.
Right panel: black symbols and error bars represent the average and standard 
deviation of the instantaneous energy, for the evolution of the simplified 
model with $N=64$, $\tau=0.2$. The red lines correspond to the spectrum
of $\hHb(s)$. The inset is a zoom around the turning point, the arrow
on the right is the prediction of Eq.~(\ref{eq_eres_full}) for the final 
energy density in the thermodynamic limit.
}
\label{fig_eres_full}
\end{figure}

A comparison of this analytical prediction with the results obtained by 
numerical integration of the Schr\"{o}dinger equation (see 
Appendix~\ref{app_schrodinger_evolution} for details on the procedure we used) 
is presented in Fig.~\ref{fig_eres_full}. The left panel displays the
result of Eq.~(\ref{eq_eres_full}) along with curves $\efin(T=e^{N\tau},N)$
obtained numerically for some finite values of $N$. We extrapolated these
results in the $N\to \infty$ limit with finite size corrections of the form
$\efin(T=e^{N\tau},N) = a(\tau) + b(\tau) \frac{\ln N}{N} + 
c(\tau) \frac{1}{N} + o(1/N)$,
a form that can be expected to arise because of polynomial corrections
to the exponentially small gaps; the inset shows the very good quality
of such a fit already for small values of $N$. The extrapolated curve $a(\tau)$
agrees with the analytical prediction (\ref{eq_eres_full}) within $1 \%$.

The right panel of Fig.~\ref{fig_eres_full} provides a further confirmation 
of the analysis in terms of independent two level Landau-Zener problems.
The black symbols with error bars represent the quantum average and standard
deviation of the instantaneous energy,
\beq
e_T(s) = \frac{1}{N}\la \phi_T(s) | \hH(s) | \phi_T(s) \ra \ , \qquad
\s_T(s) = \frac{1}{N}\sqrt{\la \phi_T(s) | \hH(s)^2 | \phi_T(s) \ra 
-\left(\la \phi_T(s) | \hH(s) | \phi_T(s) \ra\right)^2    } \ ,
\eeq
computed numerically during an evolution with $N=64$, $T=e^{N\tau}$ for 
$\tau=0.2$. One observes indeed that the average instantaneous energy
follows the metastable groundstate across several crossings, until
the turning point after which it follows adiabatically the levels crossed
there. The standard deviation is almost constant in time except 
around the turning point where it grows slightly, reflecting the fact that 
for finite $N$ a few levels (those with gaps close to $T^{-1/2}$) get 
populated. The independence of the crossings is even more apparent in the 
inset, which shows that the slope of $e_T(s)$ jumps significantly for three 
values of $s$ that correspond precisely to the locations of avoided crossings.

For completeness let us state the asymptotic expansions of 
$\efin(\tau)$ around $\tau=0$ and $\tau=\ln 2$, that are easily deduced from 
the behavior of $\gb(s)$ in $s=1$ and $s=1/2$, respectively, and read
\beq
\efin(\tau) \underset{\tau \to 0}\sim - \sqrt{2 \tau} \ , \qquad
\efin(\tau=\ln 2 - \delta) \underset{\delta \to 0^+} \sim 
-1 + 2 \frac{\delta}{\ln(1/\delta)} \ .
\label{eq_efin_simplified_limits}
\eeq

\subsubsection{The annealing towards the ferromagnet}
\label{subsubsec_annealing_ferromagnet}

\begin{figure}[h]
\center
\includegraphics[width = 8.3cm]{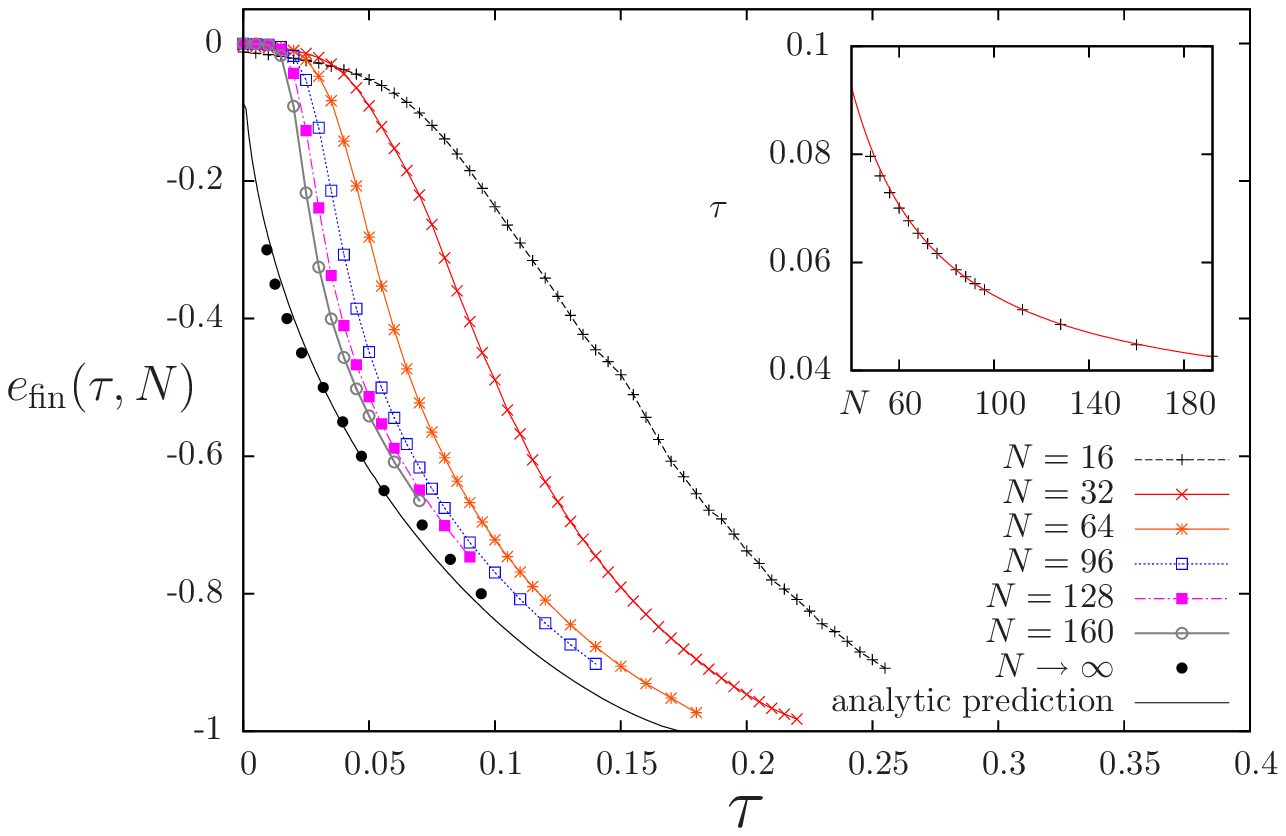}\hspace{6mm}
\includegraphics[width = 8.3cm]{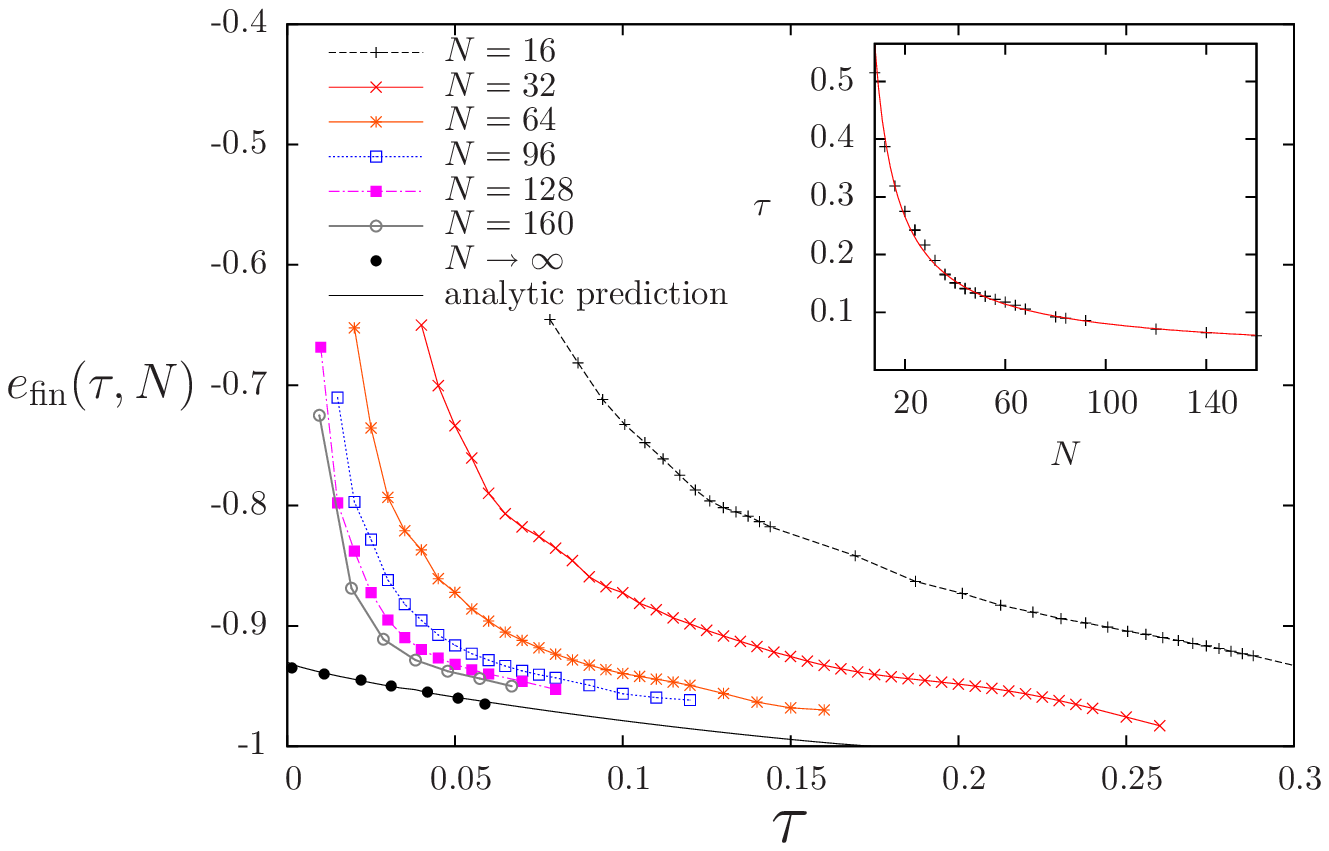}
\caption{The final energy density for the evolution on exponentially
large times, as a function of $\tau = (\log T)/N$. The left (resp. right) 
panel corresponds to the annealing towards the ferromagnet (resp. paramagnet)
for $p=3$. The solid lines at the bottom are the analytic prediction from 
Eq.~(\ref{eq_eres_pspin}) for the left panel, and Eq.~(\ref{eq_eres_pspin_back})
for the right one. The other lines with symbols 
results from the integration of Schr\"odinger equation for various finite
sizes. The black symbols on top of the analytic predictions are the
extrapolation in the $N \to \infty$ limit of the numerical results. 
The latter was performed by fitting, for
various values of $e$, the (exponential) time $\tau(N)$
such that $\efin(T=e^{N\tau(N)},N)$ crossed $e$, with a fitting function
of the form 
$\tau(N)=\tau + \text{cst} \frac{\ln N}{N} + \text{cst} \frac{1}{N}$.
We show in the insets the details of the fit for $e=-0.5$ (left panel) and
$e=-0.945$ (right panel).
}
\label{fig_eres_pspin}
\end{figure}

We now follow the same reasoning for the annealing of the $p$-spin model
towards the ferromagnet. The turning point $\sturn$ is given by
$\sturn(\tau)=\gammapm^{-1}(\tau/2)$, where the function $\gammapm$ was 
computed in
Sec.~\ref{subsubsec_gap_metastable} and plotted on the left panel of
Fig.~\ref{fig_gammas}. We adopt again the convention that 
$\gammapm^{-1}(z)=s_{\rm c}$ if $z \ge 2 \alpha_p$. As already
mentioned in Sec.~\ref{subsubsec_dos} the computation of $\efin(\tau)$ is
then completed by an iso-density argument: for $s\ge \sturn(\tau)$ we assume 
that the evolution follows adiabatically the eigenstate that made an avoided 
crossing with the paramagnetic metastable state at $\sturn$. Because of the
absence of any level crossing in this regime the number of eigenvalues below
the one whose energy we want to follow is constant, by definition. Hence
$\efin$ is fixed by the condition
\beq
\D_0(\sturn(\tau),-(1-\sturn(\tau)))= \D_0(1,\efin(\tau)) = 
\frac{1-(-\efin(\tau))^{1/p}}{2} \ ,
\label{eq_eres_pspin}
\eeq
where the integrated density of states $\D_0(s,e)$ is given in 
Eq.~(\ref{eq_dos}),
and the last equality follows from its explicit expression when $s=1$,
$p$ is odd and $e\le 0$. This prediction is displayed in the left panel
of Fig.~\ref{fig_eres_pspin}, along with results of the
numerical integration of the Schr\"odinger equation for finite $N$.
The finite-size effects in these results are much larger than for the 
simplified model. The extrapolation towards $N\to\infty$ was done searching
the value of $\tau$ that corresponds to a given final energy density
instead of the contrary (see the caption of Fig.~\ref{fig_eres_pspin}
for details), and gives a satisfactory agreement with
the analytic prediction.

The final energy density vanishes in the $\tau \to 0$ limit, as the 
paramagnetic metastable state has no spinodal and can thus be continued until 
$s=1$. 
A more precise asymptotic statement can be obtained by studying the
behavior of the integrated density of states close to $s=1$,
namely
\beq
\D_0(1-\delta,-\delta) \sim \frac{1}{2} - d_p \, \delta^\frac{1}{p} \ , \qquad
d_p = 2^\frac{1-p}{p} - \frac{1}{2\pi}\int_0^2 \dd x \, x^\frac{1-p}{p} 
\acos(1-x) \ ,
\eeq
for odd values of $p$. Combining this expansion with the one of 
$\gammapm(s=1-\delta)$ stated in Eq.~(\ref{eq_gammapm_s_to_one}) yields
\beq
\efin(\tau) \underset{\tau \to 0}{\sim} - e_p \, \tau^\frac{p-2}{2} \ , \qquad
e_p=2^\frac{p+2}{2} d_p^p \, \tgamma_p^\frac{2-p}{2} \ .
\eeq

On the other hand the behavior of $\efin(\tau)$ for (exponential) times 
slightly smaller than the adiabatic time $\tau=2\alpha_p$ corresponds to
the limit where $\sturn \to s_{\rm c}^+$. One can thus invert the expansion
(\ref{eq_gammas_s_to_sc}) for the behavior of $\gammapm$ around $s_{\rm c}$
to obtain the behavior of $\sturn(\tau \to 2 \alpha_p)$. 
Noting that $\D_0(s,-(1-s))$ has a finite derivative with respect to $s$ in 
$s_{\rm c}$, one obtains finally after the simplification of various 
constants:
\beq
\efin(\tau=2 \alpha_p - \delta) \underset{\delta \to 0^+}{\sim} -1 + 2 p
\frac{\delta}{\ln(1/\delta)} \ ,
\eeq
whose form is similar to the one found for the simplified model 
in Eq.~(\ref{eq_efin_simplified_limits}).

\subsubsection{The annealing towards the paramagnet}
\label{subsubsec_exp_ferromagnet}

The annealing towards the paramagnet can be treated along the same lines.
We recall that in this case the interpolation parameter is $u=1-s$.
The evolution follows the metastable ferromagnet until the turning
point $\uturn\in[1-s_{\rm c},1-\ssp]$ such that $\tau = 2\gammafm(1-\uturn)$, 
with
$\gammafm(s)$ the function computed in Sec.~\ref{subsubsec_gap_metastable} and 
plotted on the left panel of Fig.~\ref{fig_gammas}. The isodensity
argument for the continuation of the evolution in the regime
$u \ge \uturn$ reads then
\beq
\D_0(1-\uturn(\tau),\efm(1-\uturn(\tau)))= \D_0(0,\efin(\tau)) = 
\frac{1+\efin(\tau)}{2} \ ,
\label{eq_eres_pspin_back}
\eeq
the last equality being the consequence of the equidistance of the paramagnetic
levels when $u=1$. A comparison of this analytical prediction with numerical 
results is shown on the right panel of Fig.~\ref{fig_eres_pspin}. 
The agreement of the large $N$ extrapolation with the prediction of
Eq.~(\ref{eq_eres_pspin_back}) is again satisfactory.

The small $\tau$ limit of this regime yields a non-trivial energy density,
because the ferromagnetic metastable state has a spinodal limit of existence.
It is given by the continuation of the paramagnetic state that goes to
the spinodal point, and from the formula above reads:
\beq 
\hefin = \lim_{\tau \to 0} \efin(\tau) = -1+2\D_0(\ssp,\esp) \ .
\eeq
We report the value of these energy densities for some values of $p$ in 
Table~\ref{table_eres_pspin}. For large $p$, using the asymptotics of 
(\ref{eq_thermo_spin}), and the fact that the energy of paramagnetic 
levels become linear functions of $s$ in this limit, one gets the asymptotic
behaviour $\hefin \underset{p\to \infty}{\sim} - \frac{2}{\sqrt{p}}$,
a form that agrees very well with the data in Table~\ref{table_eres_pspin}.

The correction of next order in $\tau$ is obtained from the asymptotic
expansion of $\gammafm$ around $\ssp$, given in Eq.~(\ref{eq_gammafm_s_to_ssp}),
which converts into $\uturn(\tau) \sim 1-\ssp - (\tau/ 2 \hgammap)^{4/5} $.
Let us define the positive constant
\beq
M_p = - \left.\frac{\dd}{\dd s} \D_0(s,\efm(s)) \right|_{s = \ssp^+}
= \frac{1}{2 \pi (1-\ssp)} \int_{\mpsp}^{\msp}  \dd m \, 
\frac{\msp^p - m^p }{\sqrt{(1-\ssp)^2 (1-m^2) - (\esp+ \ssp \, m^p)^2}} \ ,
\label{eq_def_Mp}
\eeq
where $\mpsp$ is defined as the negative value of $m$ where the square root
vanishes (one can notice that $M_p=\G(0)$, where $\G(z)$ is the scaling 
function defined in Eq.~(\ref{eq_scaling_D0})). 
Then expanding in Eq.~(\ref{eq_eres_pspin_back}) one obtains
the final energy density behaviour as
\beq
\efin(\tau) \underset{\tau \to 0}{\sim} 
\hefin - 2 M_p  \left(\frac{\tau}{2 \hgammap}\right)^\frac{4}{5}  \ .
\label{eq_efin_tausmall}
\eeq

The opposite limit of quasi-adiabatic times yields exactly the same formula
for the energy density as in the simplified model (which is indeed its 
$p\to \infty$ limit), i.e.
\beq
\efin(\tau=2 \alpha_p - \delta) \underset{\delta \to 0^+}{\sim} -1 + 2
\frac{\delta}{\ln(1/\delta)} \ .
\eeq

\begin{table}[h]
\center
\begin{tabular}{|c|c|c|c|c|}
\hline
$p$ & $\Gammasp$ & $\ssp$ & $\msp$   & $\hefin$  \\
 \hline
3	&1.5	 &0.6 &0.7071 &	-0.9302 \\
4	&1.540 &	0.6062 &	0.8165 &	-0.8259 \\
5	&1.624 &	0.6189 &	0.8660 &	-0.7861 \\
7	&1.812 &	0.6443 &	0.9129 &	-0.6881  \\
9	&1.994 &	0.6660 &	0.9354 &	-0.6187 \\
13	&2.325 &	0.6993 &	0.9574 &	-0.5256 \\
21 & 2.884 & 0.7426 & 0.9747 & -0.4211 \\
31 & 3.462 & 0.7759 & 0.9831 & -0.3500 \\
\hline
\end{tabular}
 \caption{Final energy for an annealing of the $p$-spin model towards the
paramagnet, in the limit of ``small exponential'' times. The thermodynamic 
parameters of the system at the spinodal point are given by 
(\ref{eq_thermo_spin}).}
 \label{table_eres_pspin}
\end{table}

\subsection{Annealing on constant times}
\label{subsec_csttimes}

We now turn to a study of the dynamical properties of the previous models on 
time scales not growing exponentially fast with the size of the system. 
From the analysis of Sec.~\ref{sec_dynamics_LZ} we expect that for the 
simplified model and for the annealing towards the ferromagnet the final
energy density vanishes on such time scales, because the metastable branch 
with exponentially small avoided crossing exists until $s=1$. This is 
confirmed for the simplified model by a technical analysis that is deferred
to Appendix~\ref{subsubsec_small_time_full}. The annealing towards the
ferromagnet can be treated via a semi-classical dynamical 
analysis~\cite{itin09_1,itin09_2,sciolla11}, and this
will confirm its triviality on constant time-scales. The same
semi-classical analysis will on the other hand reveal a rich structure
for the annealing towards the paramagnet on finite time scales.

\subsubsection{Semi-classical dynamics for the annealing towards 
the ferromagnet}
\label{sec_sc_dyn_toferro}

Let us decompose the vector $|\phi_T(s)\ra$ on the $z$-diagonal basis as
\beq
|\phi_T(s)\ra=\underset{m\in\MNzero}{\sum} \phi_T(m,s) |m;0 \raz \ .
\eeq
The Schr\"odinger equation 
(\ref{eq_schrodinger_annealing}) is equivalent to a set of coupled equations 
for these coefficients,
\bea
\frac{i}{N T} \frac{\partial \phi_T(m,s)}{\partial s}  
= - s  \, m^p \, \phi_T(m,s) 
&- & \frac{(1-s)}{2} \sqrt{1-m^2 + \frac{2}{N}(1-m)} \
\phi_T\left(m+\frac{2}{N},s\right) \nonumber \\ 
&-& \frac{(1-s)}{2} \sqrt{1-m^2 + \frac{2}{N}(1+m)} \
\phi_T\left(m-\frac{2}{N},s\right) \ ,
\eea
which is the analog of (\ref{eq_eigenvalue_sigmaz}) in the stationary case.
The semi-classical dynamic Ansatz is $\phi_T(m,s)=e^{-N \varphi_T(m,s)}$,
which yields in the large $N$ limit, with $T$ fixed, the evolution equation
\beq 
\label{eq_schro_large_dev_para} 
- \frac{i}{T} 
\frac{\partial \varphi_T(m,s)}{\partial s} = - s \, m^p - 
(1-s) \sqrt{1-m^2} \cosh \left (2 \varphi_T'(m,s) \right) \ ,  
\eeq
where the prime denotes the derivation with respect to $m$. This corresponds
to (\ref{eq_largedev_sigmaz}) with the replacement 
$e \to - \frac{i}{T} \frac{\partial \varphi_T}{\partial s}$. 
The initial condition is
the groundstate of the pure transverse field, it is thus given by
$\varphi_T(m,0) = \varphi_0(m)$, with $\varphi_0$ defined in 
Eq.~(\ref{eq_varphi_check}). This partial differential equation is
rather difficult to solve numerically. One can however make a further
analytical simplification.

The computation of physical observables that are diagonal in the $\hmz$ basis
only requires the knowledge of the location of the minimum of the real part of 
the large deviation function $\varphi_T$, that we shall denote 
$q_T(s) = \arg \min_{m} \Re \varphi_T(m,s)$. In particular at the end of the
evolution the final energy is given by $\efin(T) = - q_T(1)^p$.
It turns out, as explained in~\cite{sciolla11}, 
that it is 
possible to write a closed system of two differential equations on $q_T(s)$ and
its conjugate momentum, $\tq_T(s) = - \partial_m \Im \varphi_T(q_T(s),s)$.
The evolution equation (\ref{eq_schro_large_dev_para}) implies indeed
 \begin{equation}\label{eq_hamilton_para} \begin{split}
 \frac{1}{T} \frac{\dd}{\dd s}  q_T(s) & = 
\frac{\partial}{\partial \tq} \mathcal{H}(q_T(s),\tq_T(s),s) 
=  2 (1-s) \sqrt{1-q_T(s)^2} \sin(2\tq_T(s)) \ , \\
\frac{1}{T} \frac{\dd}{\dd s} \tq_T(s) & = -  
\frac{\partial}{\partial q} \mathcal{H}(q_T(s),\tq_T(s),s) 
= s \, p \, q_T(s)^{p-1} -
(1-s) \frac{q_T(s)}{\sqrt{1-q_T(s)^2}} \cos(2 \tq_T(s)) \ .
\end{split} \end{equation}
These are Hamilton equations of classical mechanics, with an Hamiltonian
\beq 
\mathcal{H}(q,\tq,s) = -s \, q^p -(1-s) \sqrt{1-q^2} \cos(2 \tq)
\eeq 
obtained from the differential operator on the r.h.s. of 
(\ref{eq_schro_large_dev_para}) by the canonical substitution 
$ m \rightarrow q, i \frac{\partial \varphi_T}{\partial m} \rightarrow \tq$. 
For the sake of completness we explain in Appendix~\ref{appendix_hamilton} the 
derivation of (\ref{eq_hamilton_para}) from (\ref{eq_schro_large_dev_para}), 
along the same lines as in \cite{sciolla11}; note also that similar 
semi-classical equations can be obtained for fermionic models within the
time-dependent Gutzwiller approximation~\cite{ScFa10}.

One can in addition show that
the average instantaneous energy of the evolution according to
the Schr\"odinger equation is precisely
equal to the classical Hamiltonian, namely
\beq
\lim_{N \to \infty} \frac{1}{N} \la \phi_T(s) | \hH(s) | \phi_T(s) \ra
= \mathcal{H}(q_T(s),\tq_T(s),s) \ . 
\eeq

Let us now conclude on the validity of the analysis of 
Sec.~\ref{sec_dynamics_LZ}, i.e. that for annealing times $T$ that are
constant in the thermodynamic limit the final energy $\efin(T)$
vanishes. The initial condition $\varphi_T(m,s=0)=\varphi_0(m)$ implies 
$q_T(s=0)=\tq_T(s=0)=0$. The point $(q,\tq)=(0,0)$ is a 
stationary point of $\mathcal{H}$ for all values of $s$, hence for all 
(finite when $N \to \infty$) values of the annealing time $T$ the solution of
(\ref{eq_hamilton_para}) is $q_T(s)=\tq_T(s)=0$. In particular when $s=1$
the final energy is $\efin(T) = -q_T(1)^p=0$.

\subsubsection{Semi-classical dynamics for the annealing towards the 
paramagnet}

\label{sec_sc_dyn_topara}

The semi-classical analysis of the annealing towards the paramagnet is more
conveniently performed in the $x$-diagonal basis. We write
$|\phi_T(u)\ra=\underset{m\in\MNzero}{\sum} \phi_T(m,u) |m;0 \rax$ with
$\phi_T(m,u)=e^{-N \varphi_T(m,u)}$, and obtain from the Schr\"odinger
equation (\ref{eq_schrodinger_annealing_back}) that $\varphi_T$ evolves
according to
\beq 
\label{eq_schro_large_dev_ferro} 
-\frac{i}{T} \frac{\partial \varphi_T(m,u)}{\partial u} = - u \, m - (1-u) 
(1-m^2)^{p/2} \left( \cosh( 2 \varphi'_T(m,u)) \right)^p  \ ,
\eeq
the dynamical analog of Eq.~(\ref{eq_largedev_sigmax}). The initial condition
corresponds to the groundstate of the $-(\hmz)^p$ term, and is thus given in
this basis by $\varphi_T(m,u=0) = \varphi_0(m)$. The reduction of the partial
differential equation (\ref{eq_schro_large_dev_ferro})
to an Hamiltonian system on $\{q_T(u), \tq_T(u)\}$ follows the same lines
as in the annealing towards the ferromagnet, and yields (see 
Appendix~\ref{appendix_hamilton} for the derivation):
\begin{equation}
\label{eq_hamilton_ferro} 
\begin{split}
 \frac{1}{T} \frac{\dd}{\dd u}  q_T(u) & = 
\frac{\partial}{\partial \tq} \mathcal{H}(q_T(u),\tq_T(u),u) 
= 2 p (1-u)  \left( 1-q_T(u)^2\right) ^{p/2} 
\sin(2\tq_T(u)) \left(\cos(2\tq_T(u))\right)^{p-1} \ , \\
 \frac{1}{T} \frac{\dd}{\dd u} \tq_T(u) & 
= -  \frac{\partial}{\partial q} \mathcal{H}(q_T(u),\tq_T(u),u) 
= u -p (1-u)  q_T(u) \left( 1-q_T(u)^2\right)^{p/2-1} 
\left( \cos 2 \tq_T(u) \right)^p  \ .
\end{split} 
\end{equation}
where the classical Hamiltonian is
\beq 
\mathcal{H}(q,\tq,u) = -u \, q -(1-u) (1-q^2)^{p/2} \cos(2 \tq)^p \ .
\label{eq_cH_ferro}
\eeq
The initial condition is $q_T(0)=\tq_T(0)=0$, and the
final energy is computed at the end of the evolution as 
\hbox{$\efin(T)=-q_T(1)$}.

\begin{figure}[h]
\center
\includegraphics[width = 8.3cm]{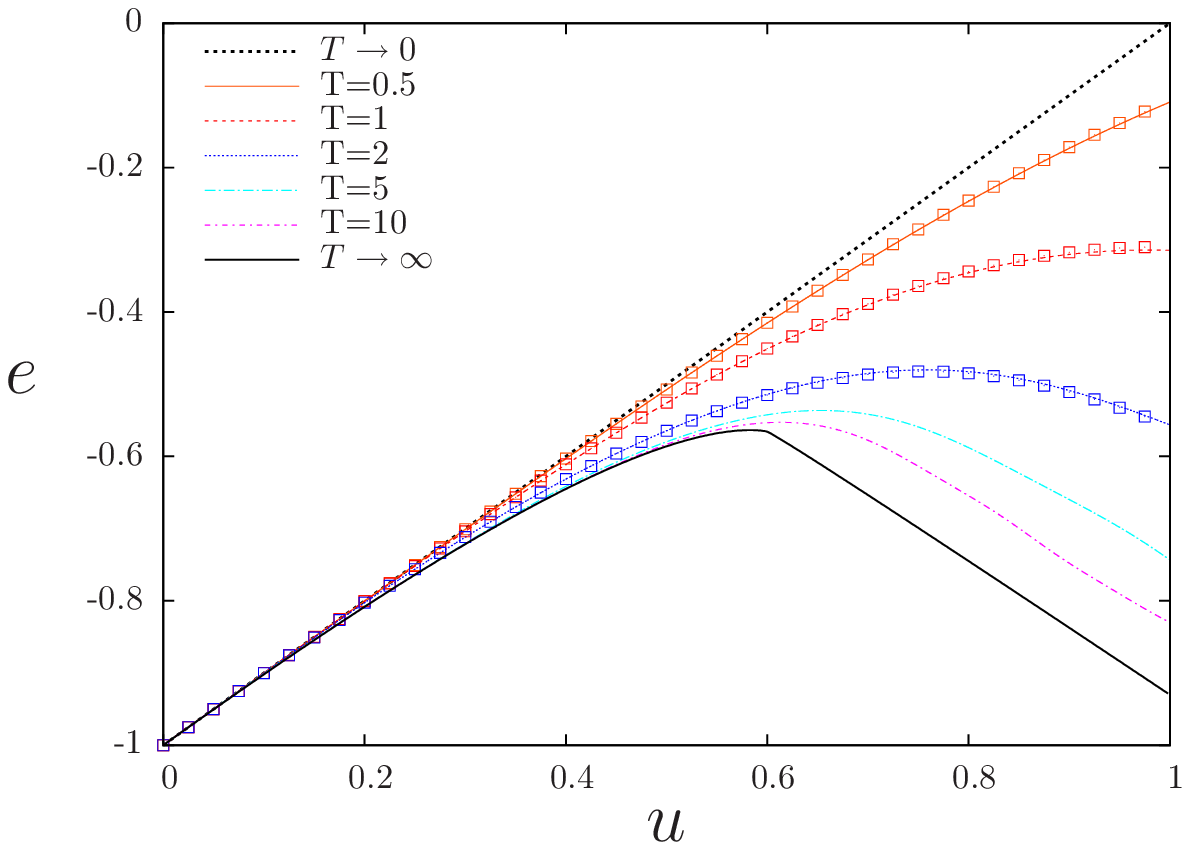}
\includegraphics[width = 8.3cm]{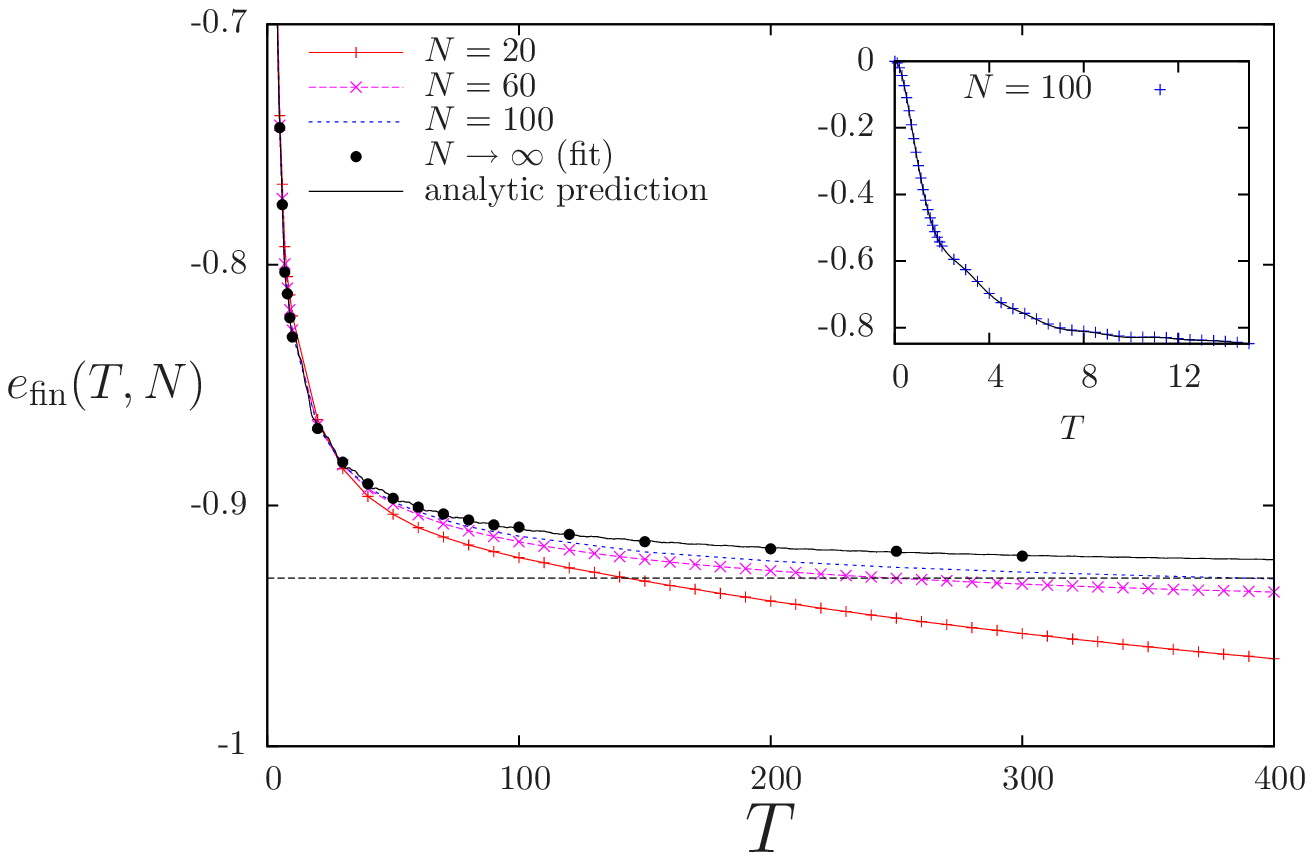}
\caption{Annealing towards the paramagnet of the $p=3$ model, in the
regime of constant times.
Left: evolution of the instantaneous energy as a function of $u$, in
the $N\to \infty$ limit with various values of $T$ (independent of $N$).
The lines are the results of the integration of Hamilton equations
of motion. The symbols are obtained via the integration of Schr\"odinger
equation with $N=80$, for the $T$ values on which they fall on.
The $T\to \infty$ line is the ferromagnetic energy for $u\le\usp=1-\ssp$, and
its continuation with the iso-density argument for larger values of $u$.
Right: the final energy at $u=1$, as a function of $T$. The solid
line has been obtained via the integration of Hamilton equations
of motion, the other lines are the results of the Schr\"odinger evolution
for various values of $N$. The $N\to \infty$ extrapolation was made
with fits of the form $\efin(T,N) = \efin(T) + x(T)/N$. 
The horizontal dashed line is the asymptotic value $\hefin$ for the 
$T\to \infty$ limit (taken after $N\to \infty$), discussed in more details in 
Sec.~\ref{sec_long_T}. The inset shows
a zoom on the small $T$ regime, for which the finite size effects are
very small: the data for $N=100$ are indistinguishable from the results
of the Hamiltonian formalism. 
}
\label{fig_Tfinite_ferro}
\end{figure}

It is easy to integrate numerically the two coupled ordinary differential
equations (\ref{eq_hamilton_ferro}), and we present on 
Fig.~\ref{fig_Tfinite_ferro} some results obtained in this way.
The plot on the left panel shows the instantaneous energy density 
$\mathcal{H}(q_T(u),\tq_T(u),u)$ as
a function of the interpolation parameter $u$, for several (rather small)
values of the annealing time $T$; the agreement
with the integration of Schr\"odinger equation with $N=80$ is already 
excellent. On the right panel we concentrate on the final energy density,
computed from the value of the solution of Hamilton equations in $u=1$,
as a function of $T$. The finite size effects on the results of Schr\"odinger 
equation get stronger for larger values of $T$, yet their extrapolation with
a correction term of order $1/N$ is in very good agreement with the classical
dynamics prediction.

\subsubsection{The long time limit of the annealing towards the paramagnet}
\label{sec_long_T}

Let us now discuss the behavior of the final energy density for large values
of $T$ (yet finite with respect to $N$). We expect that this large $T$ limit
matches the small $\tau$ limit of the exponentially large time regime studied
in Sec.~\ref{subsubsec_exp_ferromagnet}, namely that
\beq
\lim_{T\to \infty} \efin(T) = \lim_{\tau \to 0} \efin(\tau) = \hefin \ .
\label{eq_efin_largeT}
\eeq
In other words we do not foresee an intermediate scaling regime, as far
as the energy density is concerned, between the constant times and
the exponentially large times regimes. 

The intuitive explanation of this statement, in terms of the gap structure
in the spectrum of the quantum Hamiltonian, is the following. 
The gaps encountered on the metastable continuation
of the ferromagnetic groundstate are exponentially small until the spinodal
$\usp=1-\ssp$ is reached, thus for any finite $T$ no turning on the crossing 
paramagnetic states can be performed before $\usp$. Around $\usp$ there are
some polynomially small gaps that would need a polynomially growing time
$T$ to be resolved. However these gaps do not extend to values of $u$ strictly
greater than $\usp$, in the thermodynamic limit. Hence in the limit of large
$T$, taken after the thermodynamic limit, the evolution should follow
the paramagnetic energy levels that join the spinodal point, and hence
lead to a final energy density $\hefin$.

\begin{figure}
\centerline{
\includegraphics[width=4.5cm]{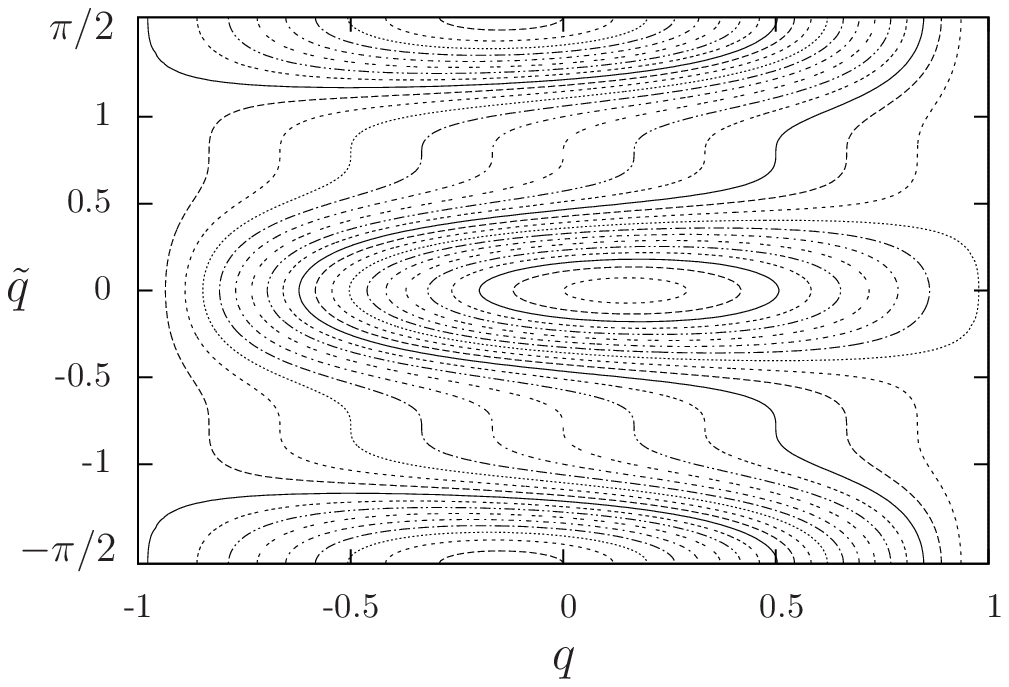}
\includegraphics[width=4.5cm]{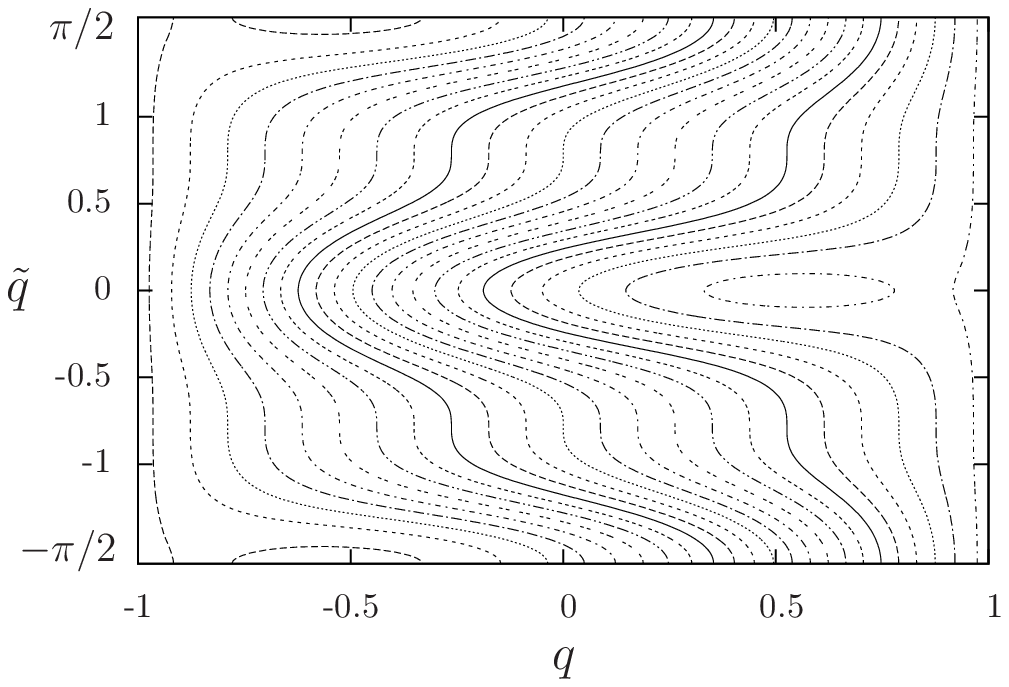}
\includegraphics[width=4.5cm]{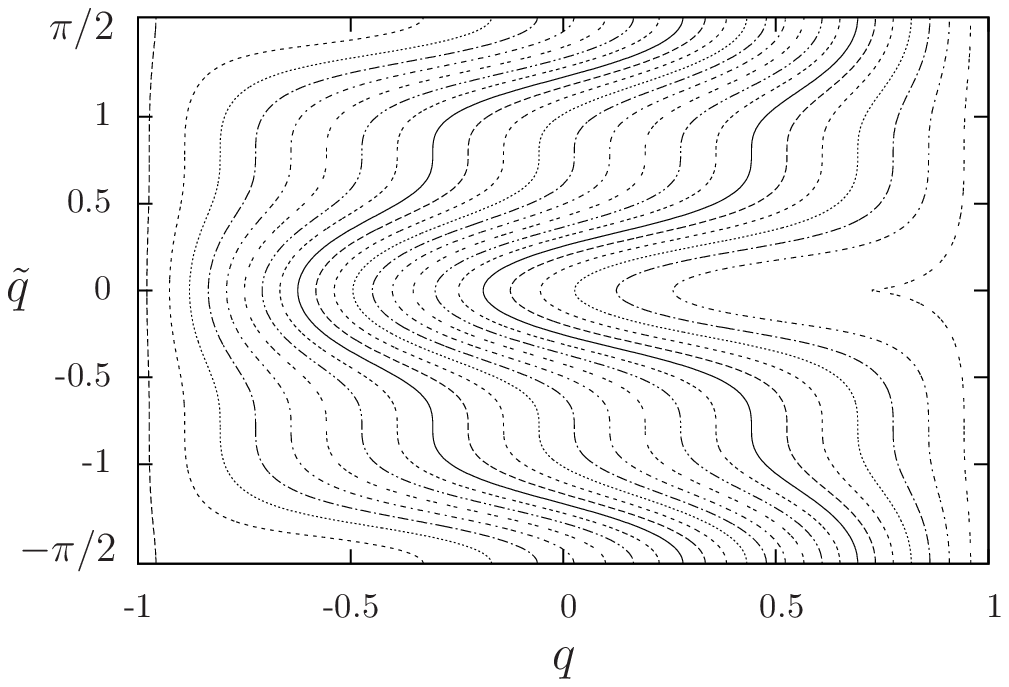}
\includegraphics[width=4.5cm]{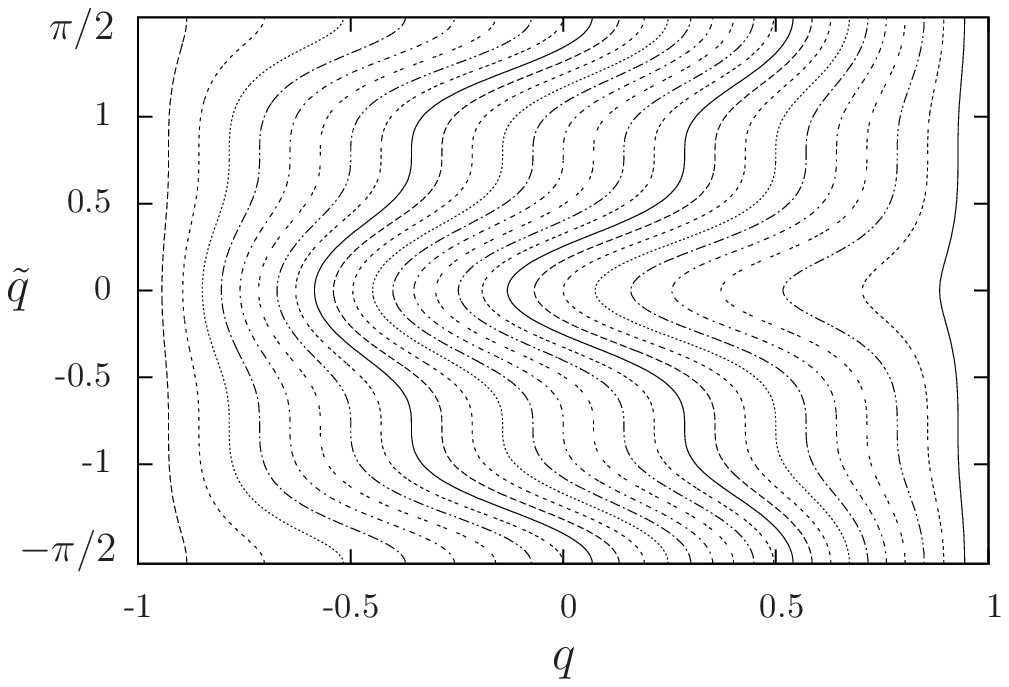}
}
\caption{Phase portraits of the classical Hamiltonian (\ref{eq_cH_ferro})
for $p=3$;
from left to right $u=0.3$, $u=0.57$, $u=\usp=0.6$, $u=0.7$.
}
\label{fig_phase_portrait}
\end{figure}

We shall give now a more quantitative justification of the statement 
(\ref{eq_efin_largeT}), and characterize the asymptotic corrections
$\efin(T)-\hefin$ as $T\to \infty$, by analyzing the classical mechanics 
problem defined in Eqs.~(\ref{eq_hamilton_ferro},\ref{eq_cH_ferro}). 
The large $T$ limit of these equations corresponds to an adiabatic classical 
mechanics evolution, and we shall thus use the tools from the theory of 
classical adiabatic invariants~\cite{arnold06,kerkovian87}. 
Consider first the phase portraits of the classical 
Hamiltonian~(\ref{eq_cH_ferro}), plotted on Fig.~\ref{fig_phase_portrait}.
For $u\le \usp$ the classical Hamiltonian has a local minimum
in $(q,\tq)=(q_*(u),0)$, where $q_*(u)$ is given in terms of the
longitudinal magnetization $m_*$ of the ferromagnetic state by 
$q_*(u)=\sqrt{1-m_*(s=1-u)^2}$ ($q_*$ is in fact the associated transverse
magnetization). The corresponding value of $\mathcal{H}$
is $\efm(s=1-u)$. In consequence the classical mechanics evolution has closed 
trajectories around this minimum, as can be seen on the first two panels
of Fig.~\ref{fig_phase_portrait}. The initial condition $q_T(0)=\tq_T(0)=0$
corresponds to this minimum in $u=0$, hence for $T \to \infty$ the evolution
follows this moving minimum (with corrections of order $1/T$ that will be
discussed below), and reaches the point of coordinates $(\qsp,0)$ at $\usp$ 
(we denote $\qsp=\sqrt{1-\msp^2}=1/\sqrt{p-1}$).
At the spinodal reached in $\usp$ the ferromagnetic metastable state
disappears; in this context this translates into the absence of such closed
trajectories for $u\ge\usp$ (note however that the Hamiltonian is 
$\pi$-periodic in $\tq$), see the two last panels of 
Fig.~\ref{fig_phase_portrait}. 
The $T\to \infty$ evolution for $u \ge \usp$ can be understood in terms of 
classical adiabatic invariants. Let us recall that these are quantities that 
depend on $(q,\tq,u)$ and that have small variations along a trajectory
solution of Hamilton equations, in the limit where the Hamiltonian of the
system has a slow explicit time-dependence with respect to the instantaneous
motion of the system, i.e. here in the large $T$ limit. The simplest adiabatic
invariant (conserved with corrections of order $T^{-1}$) corresponds to a 
passage to action-angle variables, and reads
\beq
\mathcal{I}(q,\tq,u) = 
\underset{\mathcal{H}(q',\tq',u)=\mathcal{H}(q,\tq,u)}{\oint} 
\tq' \, \dd q' \ , 
\eeq
where the integral is performed over a trajectory starting in $(q,\tq)$ that
corresponds to Hamiltonian conservative evolution for a fixed value of $u$. 
Note that in this case the adiabatic invariant depends on $(q,\tq,u)$ only
through $(e,u)$ where $e=\mathcal{H}(q,\tq,u)$ is the fixed energy on the
trajectory.
For the lines of the phase portraits that reach the points $\tq = \pm \pi/2$,
this quantity can be computed by expressing $\tq'$ as a function of
$q'$ and $e=\mathcal{H}(q',\tq')$. Inverting the relation (\ref{eq_cH_ferro})
one obtains
\beq
\mathcal{I}(e,u) = 2 \int_{\qmin(e,u)}^{\qmax(e,u)} \frac{1}{2} 
\acos\left(\left( - \frac{e + u q}{(1-u) (1-q^2)^{p/2}} \right)^\frac{1}{p}
\right) \dd q' \ ,
\eeq
where $\qmin$ and $\qmax$ denote the extremal points of the trajectory.
A moment of thought reveals that for $u\ge \usp$ 
this quantity is proportional to the 
integrated density of states $\D_0$: compare it with the expression of
$\D_0$ in (\ref{eq_dos_first}), and the semi-classical solution of the 
eigenvalue equation expressed in the $x$-basis given in 
Eq.~(\ref{eq_varphiprime_mx}).
This shows that the conservation of adiabatic invariants in the 
$T\to \infty$ limit is strictly equivalent to the iso-density argument
used in the analysis of exponentially large timescales. Hence the classical 
mechanics adiabatic evolution between $u=\usp$ and $u=1$ brings the system to 
the final energy $\hefin$, defined by $\D_0(\ssp,\esp)=\D_0(0,\hefin)$.

\begin{figure}
\centerline{
\includegraphics[width = 8.3cm]{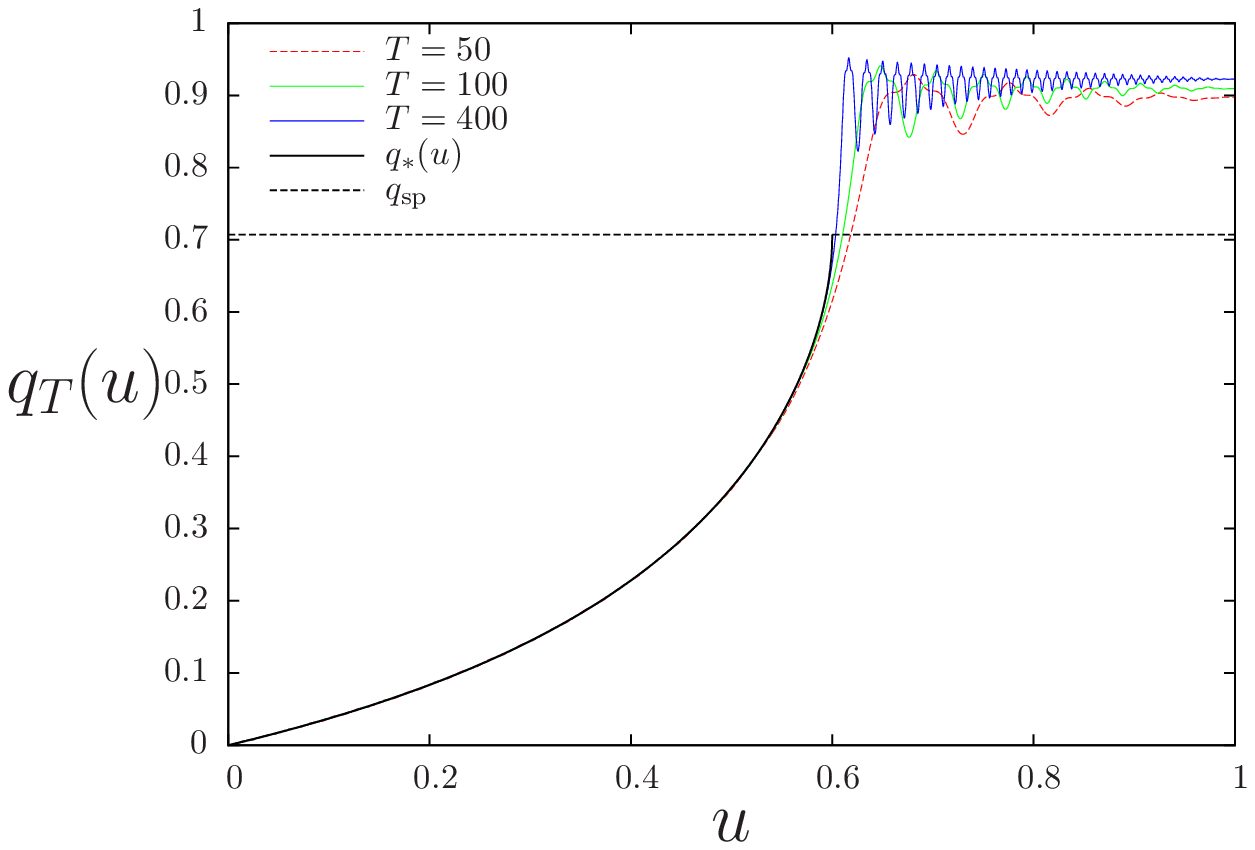} \hspace{7mm}
\includegraphics[width = 8.3cm]{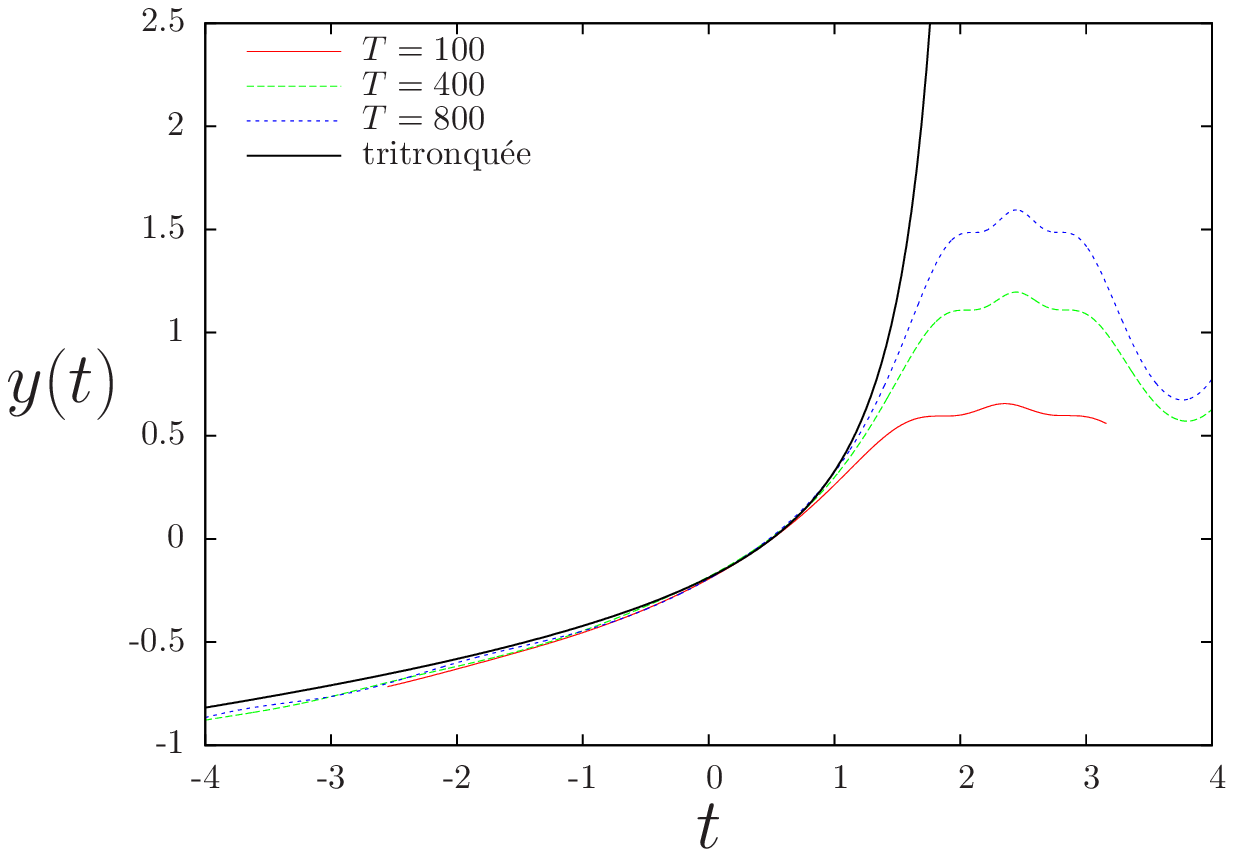}
}
\caption{Left: the solutions $q_T(u)$ of Eq.(\ref{eq_hamilton_ferro})
for $p=3$, $T=50,100,400$, with their adiabatic limit
$q_*(u)$ when $u \le \usp$. Right: similar datas for $T=100,400,800$ 
plotted with the rescaling defined in Eq.~(\ref{eq_rescaling}), 
together with the 
\emph{tritronqu\'ee} solution of the Painlev\'e equation.
}
\label{fig_qT_u}
\end{figure}

We shall now discuss the behavior of $\efin(T) - \hefin$ in the large $T$
limit. One can expect some generic corrections of order $T^{-1}$ to arise because
of the imperfect conservation of the adiabatic invariant for large but
finite $T$. However these effects are subdominant with respect to the
singular corrections due to the bifurcation transition of the classical
mechanics system at $\usp$~\cite{haberman00}.
The left panel of Fig.~\ref{fig_qT_u} displays the functions
$q_T(u)$ for several values of $T$. For large enough $T$ they indeed follow
with a good approximation $q_*(u)$ for $u<\usp$, while they have an 
oscillating behavior for $u>\usp$, in agreement with the shape of the phase
portraits. The critical regime around $\usp$ 
plays however a crucial role, as will be explained now by reconsidering in a 
quantitative way the reasoning above.

For $u<\usp$, the expansion of the Hamiltonian (\ref{eq_cH_ferro}) around its 
minimum in $(q,\tq)=(q_*(u),0)$ yields
\beq
\mathcal{H}(q,\tq,u) = \efm(s=1-u) + \frac{1}{2} g(u) \omega(u)^2 (q-q_*(u))^2 +
\frac{1}{2 g(u)} \tq^2 + O((q-q_*(u))^3,\tq^4,(q-q_*(u))\tq^2 ) \ ,
\eeq
which corresponds to an harmonic oscillator centered in $q_*(u)$, with
mass and pulsation given by
\beq
g(u) = \frac{1}{4p(1-u)(1-q_*(u)^2)^{p/2}} \ , \quad \omega(u) = 2 p (1-u)
(1-q_*(u)^2)^{\frac{p}{2}-1} \sqrt{1 - (p-1) q_*(u)^2} \ .
\eeq
In the large $T$ limit one can set up an expansion for the trajectory of an
harmonic oscillator with slowly varing parameters (here $q_*,g,\omega$),
under the form of oscillating terms of pulsation $T \omega(u)$ (in the slow 
time $u$) multiplied by slowly varying terms. At the leading order, and taking
into account the initial condition $q_T(0)=\tq_T(0)=0$, one obtains
\bea
q_T(u) &=& q_*(u) - \frac{1}{T}  q_*'(0) 
\sqrt{\frac{g(0)}{\omega(0) g(u) \omega(u) }} 
\sin\left( T \int_0^u \dd u' \, \omega(u') \right)  + O(T^{-2}) \ ,
\label{eq_qT_before_spinodal}
\\
\tq_T(u) &=& \frac{1}{T} g(u) q_*'(u) -  \frac{1}{T}  q_*'(0) 
\sqrt{\frac{g(0) g(u) \omega(u)}{\omega(0) }} 
\cos\left( T \int_0^u \dd u' \, \omega(u') \right) + O(T^{-2}) \ .
\eea
This expansion is only valid for $u < \usp$, because $\omega(u)$ vanishes
as $u \to \usp$. In this limit the harmonic potential is not confining
anymore. To continue the description of the evolution towards larger values 
of $u$ we shall now expand the Hamiltonian around its bifurcation, and look
for a scaling function that will describe the neighborhood of the singularity.
We write:
\beq
\mathcal{H}(q,\tq,u) = \esp - \epfm(\ssp) (u-\usp) + \frac{1}{2 g(\usp)} \tq^2 
- \frac{1}{3!} a_p (q-\qsp)^3 - b_p  (q-\qsp) (u-\usp) + \dots \ ,
\label{eq_H_dev_sp}
\eeq
where $a_p$ and $b_p$ are two positive constants depending on $p$ that can
be obtained as partial derivatives of $\mathcal{H}$ in $(\qsp,0,\usp)$.
The mechanical interpretation of the last three terms is a particle of mass
$g(\usp)$, evolving in an energy potential that has a constant cubic term and 
a linear term whose sign changes as $u$ crosses $\usp$, thus provoking the
disappearance of its stable minimum. Eliminating
$\tq$ from the Hamilton equations of motion that follows from this 
truncated expansion leads to
\beq
\frac{1}{T^2} \frac{\dd^2}{\dd u^2} (q_T(u) - \qsp) = \frac{a_p}{2 g(\usp)}
(q_T(u) - \qsp)^2 + \frac{b_p}{g(\usp)} (u-\usp) \ .
\label{eq_pain_before_scal}
\eeq
We define a scaling function $y(t)$ with
\beq
\begin{cases}
y = T^{2/5} \, C_p^{(y)} \, (q-\qsp)  \ , \qquad 
C_p^{(y)} = \frac{g(\usp)}{b_p} \left(\frac{a_p b_p}{12 g(\usp)^2}\right)^{3/5} 
\ , \\
t = T^{4/5} \, C_p^{(t)} \, (u-\usp)   \ , \qquad
C_p^{(t)} = \left(\frac{a_p b_p}{12 g(\usp)^2} \right)^{1/5}= 
\frac{\left(\frac{8}{3} \right)^\frac{1}{5} p^\frac{3}{5}
\frac{(p-2)^\frac{3p-4}{10}}{(p-1)^\frac{3p-5}{10}}}
{\left(1+ p \frac{ (p-2)^\frac{p-2}{2}}{(p-1)^\frac{p-1}{2}}\right)^\frac{2}{5}
} 
\ .
\end{cases}
\label{eq_rescaling}
\eeq
The scalings with $T$ of these changes of variables are chosen in such a way 
that the three terms of Eq.~(\ref{eq_pain_before_scal}) are of the same order.
The constants $C_p^{(y)}$ and $C_p^{(t)}$ are more arbitrary, and have been 
chosen here in order for the scaling function $y(t)$ to be solution of
the canonical form of the first Painlev\'e equation, 
$y''(t) = 6 \, y(t)^2 + t$. We have only given $C_p^{(t)}$ explicitly above as
$C_p^{(y)}$ will not appear in the final result.
There exists of course an infinite family of solutions of the Painlev\'e 
equation, selected for instance by the value of $(y,y')$
at a given $t$. In our case the solution will be selected by a matching 
argument between the $u \to \usp$ limit of the first regime $u<\usp$ described
by Eq.~(\ref{eq_qT_before_spinodal}), and the $t\to - \infty$ limit of the
regime described by the Painlev\'e equation (in a neighborhood of $\usp$
of order $T^{-4/5}$). To expand Eq.~(\ref{eq_qT_before_spinodal}) we note that 
in the limit $u \to \usp$ one has
$q_*(u)=\qsp-\sqrt{2 b_p (\usp -u)/a_p}+O((\usp -u)^{3/2} )$ and
$\omega(u) \sim \text{cst} \, (\usp-u)^{1/4}$, where here and in the following
we denote $\text{cst}$ positive constants whose precise values are not 
necessary for the reasoning. These expansions yields, in terms of the
rescaled variables $y$ and $t$,
\beq
y(t) \underset{t \to -\infty}\sim - \sqrt{\frac{-t}{6}} - 
\text{cst} \ T^{-1/2} (-t)^{-1/8} 
\sin(T (\text{cst} - \text{cst} \, (-t)^{5/4}) ) \ .
\eeq
The leading term $-\sqrt{-t/6}$ is common to several solutions of the
first Painlev\'e equation; however we note here that the amplitude of the 
oscillating term vanishes as $T\to \infty$ (for a fixed large $t$), hence the 
scaling function should be given by a monotonous solution with the 
$-\sqrt{-t/6}$ asymptotic behaviour. This was shown in~\cite{Joshi2001}
to imply that $y(t)$ is Boutroux~\cite{boutroux} \emph{tritronqu\'ee} solution. 
This solution was studied in great details in~\cite{Joshi2001}, in particular
the location of its smallest real pole $t_0$ was determined numerically with 
great accuracy, and found to be $t_0=2.3841687\dots$.
On the right panel of Fig.~\ref{fig_qT_u} we compare the \emph{tritronqu\'ee} 
solution of the Painlev\'e equation (determined numerically with its values
$(y(0),y'(0))$ given in~\cite{Joshi2001}) with the curves $q_T(u)$, rescaled
according to (\ref{eq_rescaling}). Their agreement improves as $T$ increases,
as expected for a scaling function. One can compute the instantaneous
energy in the regime described by the Painlev\'e equation, namely for 
$u \sim \usp + T^{-4/5} t/C_p^{(t)}$ with $t<t_0$, and find from
(\ref{eq_H_dev_sp}) that it is given by $\esp - \epfm(\ssp) (u-\usp) + 
O(T^{-6/5})$. Let us also compute the integrated density of states
associated to such energies,
\beq
\D_0\left(s=\ssp - T^{-4/5} \frac{t}{C_p^{(t)}}, 
e = \esp - \epfm(\ssp) T^{-4/5} \frac{t}{C_p^{(t)}} \right) \sim
\D_0(\ssp,\esp) + M_p T^{-4/5} \frac{t}{C_p^{(t)}} \ ,
\eeq
where $M_p$ was defined explicitly in Eq.~(\ref{eq_def_Mp}). As explained
above the conservation of mechanical classical invariants corresponds to
the conservation of the integrated density of states for $u \ge \usp$.
Our prediction for the large $T$ behaviour of the final energy density 
thus reads
\beq
\efin(T) \sim \hefin  +2 M_p \frac{t_0}{C_p^{(t)}} T^{-4/5} \ .
\label{eq_efin_largeT_correc}
\eeq
Indeed the largest violation of the conservation of the adiabatic invariant
is obtained by taking $t\to t_0$, the limit of existence of the scaling
regime described by the Painlev\'e equation.

It is rather peculiar that a scaling function matching two different regimes
is defined only on a part of the real axis (here $t < t_0$). In fact at the
end of the Painlev\'e regime the values of $u$ are still close to the 
singularity ($u-\usp= O(T^{-4/5})$), hence the periods of the orbits encountered
at those times are divergent. It has been shown in~\cite{haberman00} how
to deal with this third regime of time, that matches the $t\to t_0$ limit
with $u=\usp + \varepsilon$, where $\varepsilon$ is arbitrary small but
independent on $T$. In particular it was found that the additional corrections
to the adiabatic invariant
due to this regime are of order $T^{-5/6}$, i.e. asymptotically neglectible 
with respect to those we have computed, yet larger than the regular $T^{-1}$
corrections to the action adiabatic invariant.

\begin{figure}
\centerline{\includegraphics[width=9cm]{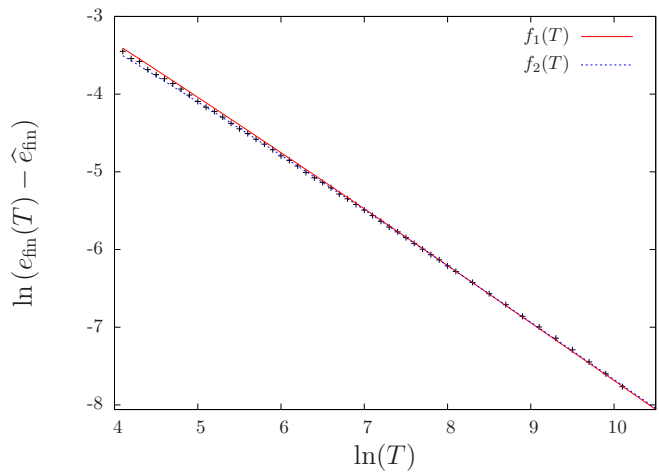}}
\caption{The large $T$ limit of the final energy density for the annealing
towards the paramagnet of the $p=3$ model. The symbols have been obtained
via the integration of Hamilton equation of motion, for $T$ as large as
24000. The two lines are of the form $\efin(T)-\hefin = a\, T^{-4/5} + 
b \, T^{-5/6} + c \, T^{-1}$. In both cases $a$ was fixed by the analytical 
prediction from Eq.~(\ref{eq_efin_largeT_correc}), the function $f_1$
was obtained with $c=0$ and using $b$ as a fitting parameter, while for $f_2$
we fitted the data with both $b$ and $c$.}
\label{fig_efin_largeT}
\end{figure}

We have checked the analytical prediction (\ref{eq_efin_largeT_correc}) against
numerical integrations of the Hamilton equations of motion, and present these
results on Fig.~\ref{fig_efin_largeT}. We could not achieve a good
agreement with the data using only the form (\ref{eq_efin_largeT_correc});
indeed, even for the largest times $T=24000$ we could reach, the
subdominant correction term of order $T^{-5/6}$~\cite{haberman00} is
comparable to the leading one (the difference between the two exponents
$4/5$ and $5/6$ is tiny). Including this correction term as a fitting
parameter yields a very good agreement with the data, that is further
improved with the inclusion of the regular $T^{-1}$ corrections.
We have also checked for other values of $p$ a similar
agreement with the prediction of Eq.~(\ref{eq_efin_largeT_correc}).

\subsection{Even values of $p$}
\label{subsec_otherp} 

Let us finally discuss the annealing for even $p$ models that was left aside
in the previous discussion. As explained at the end of 
Sec.~\ref{subsubsec_spin_sectors} the models with even $p$ enjoy an additional 
symmetry, the conservation of the parity of the magnetization in the $x$ basis.
This implies that the dynamics of the even $p\ge 4$ models has exactly the
same properties as the odd $p \ge 3$ cases. Indeed the dynamics is confined
to the subspace of parity equal to the one of the groundstate. In that subspace
the ferromagnetic levels are unique, and all the structure of the gaps in the
spectrum is qualitatively the same as for odd $p\ge 3$ models.

The case $p=2$, studied in~\cite{santoro08,solinas08,itin09_1,itin09_2}, is on the
contrary very different. Consider first the annealing towards the paramagnet.
The only relevant timescale, as far as the energy density is concerned, is
the one of finite $T$ when $N\to \infty$. Indeed, as explained at the beginning
of Sec.~\ref{sec_long_T}, resolving the gaps of order $N^{-1/3}$ encountered 
around $u_{\rm c}=2/3$ (hence considering interpolation times of order $N^{2/3}$)
is necessary only to end up the evolution in the groundstate, not to reach
energy densities equal to the one of the groundstate. On the finite $T$ 
timescale the semi-classical analysis of Sec.~\ref{sec_sc_dyn_topara} is
thus relevant, and allows to cover the full range of energy densities
between $\efin(T=0)=0$ and $\efin(T \to \infty)=-1$. Moreover the large $T$
corrections to the energy density are much less singular than for $p\ge 3$,
because the bifurcation at $u_{\rm c}$ is of a different type. The scaling
regime is described by the second Painlev\'e equation~\cite{itin09_1,itin09_2} instead
of the first one, and this should lead to corrections of the form 
$\efin(T)\sim -1 + \text{cst} /T$.

The annealing towards the $p=2$ ferromagnet has a much richer structure.
For all finite $T$, in the thermodynamic limit, the semi-classical analysis 
of Sec.~\ref{sec_sc_dyn_toferro} predicts that $\efin(T)=0$. 
Indeed the initial condition $(q_T(s=0),\tq(s=0))=(0,0)$
is a stationary point of $\mathcal{H}$ for all values of $s$, even though
it becomes unstable for $s\ge s_{\rm c}$. Reaching non-trivial final energy
densities thus requires interpolation time-scales that grows with $N$; their
precise scaling is a delicate problem that we leave for future work.
Indeed a preliminary treatment, within the formalism of this paper, 
reveals that the gaps that close along the line
$e=-(1-s)$ (corresponding thermodynamically to the unstable solution $m=0$
of Eq.~(\ref{eq_mag_thermo})) do so as $1/\ln N$, and not polynomially in $N$
as happens for the groundstate. Such a logarithmic behaviour was already 
discussed in~\cite{HeScGe05,RiPa09}.

\section{Conclusions}
\label{sec_conclu}

Let us give a partial summary of this work and propose a few directions
for future research. One of our main results in the statics part of the paper 
is the formula (\ref{eq_result_alpha_p}) that gives the exponential rate of 
closing of the gap at a first-order phase transition (i.e. for $p \ge 3$ in 
this class of models), under the form of a semi-classical tunneling amplitude
between the paramagnetic and ferromagnetic states that cross at the 
transition. At a second-order phase transition (here for $p=2$) we recover 
the results of~\cite{botet83,dusuel04} on the polynomial 
scaling $N^{-1/3}$ of the gap, from a matching between the square-root closing 
of the finite gap in the paramgnetic phase (see Sec.~\ref{sec_finite_gaps}) 
and the behavior of the exponential splitting of the two ferromagnetic 
groundstates around the transition (studied in Sec.~\ref{subsubsec_gap_ferro}).

The detailed description of the spectral properties
of the models studied here, in particular the density of states and the
rate of closing of exponentially small gaps, relies on the analysis
of the solutions of the semi-classical eigenvalue equation 
(\ref{eq_largedev_sigmaz}). It is actually straightforward to write
its generalization, and thus to perform the same subsequent steps of analysis,
for any model of spins whose Hamiltonian only depend on the total
magnetizations $\hmx,\hmy,\hmz$. For concreteness let us give these
generalizations for three examples:
\begin{itemize}
\item In the LMG model the Hamiltonian reads
$\hH/N = -  \Gamma \hmz -  \gamma_x (\hmx)^2 -  \gamma_y (\hmy)^2$,
and the generalization of (\ref{eq_largedev_sigmaz}) is
\beq
e = - \Gamma \, m - \gamma_x (1-m^2) \cosh^2(2 \varphi'(m)) + \gamma_y (1-m^2)
\sinh^2(2 \varphi'(m)) \ .
\eeq
For the density of states this should yield formulas equivalent to those
obtained in~\cite{ribeiro08}.

\item For the models of~\cite{FiDuVi11}, where the interactions along two
axis are raised to arbitrary powers, one can write the Hamiltonian as
$\hH/N=-\gamma_z (\hmz)^p - \gamma_x (\hmx)^{p'}$ and the eigenvalue equation,
in the thermodynamic limit, as
\beq
e = - \gamma_z \, m^p - \gamma_x (1-m^2)^\frac{p'}{2} \cosh(2 \varphi'(m))^{p'}
\ .
\eeq

\item The authors of~\cite{SeNi12} introduced an antiferromagnetic coupling
in the interpolating Hamiltonian of the annealing,
under the form $\hH/N=-s[\lambda (\hmz)^p - (1-\lambda) (\hmx)^2] - (1-s) \hmx$.
This yields
\beq
e = - s \lambda m^p + s (1-\lambda) (1-m^2) \cosh(2 \varphi'(m))^2 - (1-s)
\sqrt{1-m^2}  \cosh(2 \varphi'(m)) \ .
\eeq

\end{itemize}

In the Section~\ref{sec_dynamics} devoted to the annealing dynamics of
the fully-connected $p$-spin models we have analyzed the final energy density
$\efin$ after an evolution on a time $T$. 
Our analytical results have been obtained
in the thermodynamic limit; let us emphasize the necessity, in this limit,
to define precisely the scaling of $T$ with the system size $N$. The results,
and the methods employed to derive them, are indeed very different according
to the timescale investigated. In Sec.~\ref{subsec_exptimes} we studied
annealing times $T$ growing exponentially with $N$, in terms of the 
Landau-Zener mechanism controlled by the exponentially small gaps encountered
by metastable states. The regime where $T$ is kept fixed while the limit
$N\to \infty$ is performed first was analyzed in Sec.~\ref{subsec_csttimes},
via a reduction to a classical mechanics problem~\cite{sciolla11}. We have 
argued that these two regimes are the only relevant ones for $p\ge 3$, 
and as far as the energy density is concerned; resolving finite (extensive) 
energy differences would require in some cases the study of an intermediate
timescale, with $T$ growing polynomially or logarithmically with $N$.
An outcome of our analysis is the crucial role played by spinodals in
the annealing of mean-field models encountering a first-order transition:
in the limit where $T$ is large but finite with respect to $N$, or 
exponential with $N$ but with an infinitesimal growth rate $\tau$, an 
annealing follows the metastable groundstate until its disappearance at
the spinodal, and reaches at the end of the evolution an excited energy
density $\hefin$ corresponding to the state that crosses the metastable state 
at the spinodal. This energy separates what can be achieved on sub-exponential
times ($e \ge \hefin$), from the range of energies $\egs \le \efin \le \hefin$
that require an exponentially large annealing time to be reached.
In the models studied here the paramagnetic state is always metastable and
has no spinodal, hence the annealing from the paramagnet has a trivial
finite-time regime ($\hefin=0$); this motivated the complementary study
of the annealing in the reverse direction (from the ferromagnet to the 
paramagnet) which exhibits a non-trivial boundary $\hefin$ between the two
timescales.

As we already emphasized the models studied in this paper are only toy-models
as far as the difficulty of finding their groundstates is concerned; from
this point of view both directions of the annealing (from the paramagnet
to the ferromagnet or viceversa) are equally relevant.
We conjecture that some of the results we obtained may remain true
for the quantum annealing of more difficult combinatorial optimization 
problems as those of~\cite{qXOR,young10}. In particular the scalings of
the final energy density for the small $\tau$ limit of exponentially
large time-scales (see Eq.~(\ref{eq_efin_tausmall})) and for the large $T$ 
limit of constant timescales (see Eq.~(\ref{eq_efin_largeT_correc})) with the 
exponent $4/5$ could be generic for all mean-field models encountering a 
first-order transition followed by a separate spinodal along their quantum 
annealing (the location of the spinodals for the XORSAT problem were
determined in~\cite{qXOR}), be there fully-connected or diluted, as long
as they are mean-field. Indeed in these combinatorial optimization
problems the paramagnetic state from which one starts the annealing procedure
has a spinodal limit of existence (exactly as the ferromagnetic state of
the toy models studied in the present paper).
One of the several open questions in this
context would be the determination of $\hefin$, in other words the
generalization of the iso-integrated density argument that applies only to
the fully-connected models whose Hilbert space can be decomposed in 
disconnected spin sectors. One possible road for this calculation in
the context of diluted mean-field models would be the quantum extension
of the ``state following method''~\cite{ZdKr10} (related to the Franz-Parisi 
potential~\cite{FrPa95}), that answers a similar question for classical 
annealing dynamics.

In the design of a quantum annealing algorithm there is some freedom in the
choice of the initial Hamiltonian $\hHi$ (it should however have a groundstate
that is easy to prepare, and its construction should not assume a detailed
knowledge of the sought-for groundstate of the final Hamiltonian $\hHf$).
To avoid the phase transitions that appear when $\hHi$ is a transverse field
it was for instance proposed in~\cite{farhi_random} to randomize the direction 
of the transverse fields on each spin. Very recently another proposal was to
include antiferromagnetic couplings in the interpolating 
Hamiltonian~\cite{SeNi12}; in this way it is possible to avoid the first-order
phase transition by making a detour in the $(s,\lambda)$ plane (see 
also~\cite{ribeiro06} for a similar phenomenon). The annealing towards the 
ferromagnet for $p\ge 3$ studied in the paper was particularly inefficient 
because the groundstate of the initial Hamiltonian (the transverse field 
$- \hmx$) remained metastable all the way to $s=1$. One can thus wonder 
whether taking a ferromagnetic coupling $-(\hmx)^{p'}$ with $p'\ge 2$ 
(this corresponds to the models of~\cite{FiDuVi11}) would help. The answer 
is no, the metastability until $s=1$ persists for all values of $p'$, as long 
as $p\ge 3$. Instead the antiferromagnetic coupling $(\hmx)^2$ introduced 
in~\cite{SeNi12} helps the annealing because their groundstate $|0;0\rax$ 
has a much larger overlap with the groundstate $|1;0\raz$ of the target
Hamiltonian $-(\hmz)^p$ than has the groundstate $|1;0\rax$ of the transverse 
field.

In the fully-connected ferromagnetic models studied in this paper the 
condition for a thermodynamic first-order transition (i.e. $p\ge 3$) 
coincided with the existence of exponentially small gaps at the transition. 
There can however be exceptions to this rule: the case $p=p'=2$ 
of~\cite{FiDuVi11} exhibits a discontinuity in the derivative of the 
groundstate energy but no exponentially small gaps. This peculiarity is due 
to the coincidence of the spinodals with the first-order transition, the 
states that cross become unstable right after the transition. A similar 
situation was shown to happen in antiferromagnetic chains of odd lenghts 
with periodic boundary conditions~\cite{LaMoScSo12}.

\acknowledgments

We warmly thank 
Fabrizio Altarelli,
Laura Foini, 
Florent Krzakala, 
Marc M\'ezard,
R\'emi Monasson, 
Alberto Rosso and
Francesco Zamponi
for useful discussions related to this work, and in particular 
LF with whom the first steps of this work were taken, and
FK for giving us the extrapolated datas for the rate of closing
of the exponential ferromagnetic gap for $p=2$ plotted in 
Fig.~\ref{fig_gaps_ferro}.

\appendix

\section{Large $p$ expansion of the closing rate of the gap}
\label{app_alpha_p_large_p}

This appendix is devoted to the derivation of the asymptotic expansion 
(\ref{eq_result_alpha_p_large_p}) for the rate of the exponential closing
of the gap at the first-order transition, in the large $p$ limit.
Let us first compute the limit of $\alpha_p$. Simplifying the expression 
(\ref{eq_result_alpha_p}) with $\mc = 1$, $s_{\rm c}=1/2$, $\ec = -1/2$, one
obtains:
\beq  
\alpha_p \to  
\frac{1}{2}\int_0^1 \dd m \, \ach \left ( \frac{1}{\sqrt{1-m^2}} \right) 
=  \frac{1}{2} \int_0^1 \dd m \, \arg \tanh (m) =  \frac{\ln 2}{2} \ ,
\eeq
as argued for in~\cite{jorg10}.
For the computation of $\alpha_p$ at order $1/p$ the corrections 
to $\mc$ and $s_{\rm c}$ given in Eq.~(\ref{eq_critical_parameters_largep})
are actually irrelevant and one has:
\beq  
\bes \alpha_p &=  \frac{1}{2} 
\int_0^1 \dd m \, \ach \left( \frac{1}{\sqrt{1-m^2}} \left(1-m^p\right) \right) 
+ O \left( \frac{1}{p^2} \right) 
\\ & =  \frac{1}{2} \sum_{k=0}^\infty \frac{1}{k!} \int_0^1 
\dd m \, \left( \frac{-m^p}{\sqrt{1-m^2}}\right)^k 
\left( \frac{\dd}{\dd^k} \ach \right) \left( \frac{1}{\sqrt{1-m^2}} \right)
+ O \left( \frac{1}{p^2} \right) \ .
\end{split} \eeq
In general one has $ \frac{\dd}{\dd^k} \ach x = \sum_{i=\frac{k+1}{2}}^{k-1} c_{i,k} 
\frac{x^{2i-(k+1)}}{(x^2-1)^{i-1/2}}$, and therefore the $k-$th integral above is 
found to be: 
\beq
(-1)^k \sum_i c_{i,k} \int_0^1 \dd m \, m^{kp+1-2i} = 
\frac{(-1)^k}{kp} \sum_i c_{i,k}+ O\left(\frac{1}{p^2}\right) =  
\frac{-(k-1)!}{kp}+ O \left( \frac{1}{p^2} \right) \ ,
\eeq
where we used that $\sum_i c_{i,k} = \lim_{x \rightarrow \infty} x^k 
\frac{\dd^k}{\dd x^k} 
\ach (x)= \lim_{x \rightarrow \infty} x^k \frac{d^k}{d x^k} \ln x = 
(-1)^{k-1} (k-1)! x^{-k}$. 
We obtain finally the expansion of Eq.~(\ref{eq_result_alpha_p_large_p}):
\beq \bes  
\alpha_p &=  \frac{\ln 2}{2}- \frac{1}{2p} 
\sum_{k=1}^{\infty} \frac{(k-1)!}{k.k!} +O\left(\frac{1}{p^2}\right) \\& 
= \frac{\ln 2}{2}- \frac{1}{2p} \sum_{k=1}^\infty \frac{1}{k^2} 
+ O\left(\frac{1}{p^2}\right) =  \frac{\ln 2}{2} - \frac{\pi^2}{12p} 
+ O\left(\frac{1}{p^2}\right)  \ .
\end{split}\eeq

\section{Technical details on the simplified model}

\subsection{Statics}

\label{subsubsec_large_dev_full}

We justify in this appendix the formula (\ref{eq_gb}) for the rate of
closing of the gaps of the simplified model, along the metastable
continuation of its groundstate for $s \ge 1/2$. We look for
an eigenvector of $\hHb(s)$, with an eigenvalue $N e$, under the form
$\underset{m\in\MNzero}{\sum} \phi(m,s,e) |m;0 \rax$. The coefficients of this 
decomposition are solutions of
\begin{equation*}
\label{eq_vp_full_matrix} 
e \, \phi(m,s,e) = -s m \phi(m,s,e) - (1-s) 
D_m \sum_{m' \in \MNzero} D_{m'}  \phi(m',s,e) \ . 
\end{equation*}
For $s = 0$ the lowest eigenstate is given exactly by 
$\phi(m,0,-1) = D_m$, which can be written at the leading order in the
thermodynamic limit $e^{-N \varphi_0(m)}$, with $\varphi_0$ given in 
Eq.~(\ref{eq_varphi_check}). For $0 \leq s < 1/2$ we construct an
approximation $\phi(m,s)$ of the groundstate eigenvector as
\beq 
\phi(m,s) = \frac{\widetilde{\phi}(m,s)}{\| \widetilde{\phi}(s) \|} \ , 
\qquad \text{with} \ \
\widetilde{\phi}(m,s) = \frac{\phi(m,0,-1)}{1-\frac{s}{1-s}m} \ .
\eeq 
Indeed one has :
\begin{equation}
\label{eq_vp_full_matrix_2}
\bes -\left(1-\frac{s}{1-s}m \right) \phi(m,s) =
- \frac{\phi(m,0,-1)}{ \| \widetilde{\phi}(s) \|} &=  
- D_m \sum_{m'\in \MNzero} D_{m'} \frac{ \phi(m',0,-1)}{\| \widetilde{\phi}(s) \|} 
\\ & =  - D_m \sum_{m' \in \MNzero} D_{m'}  \phi(m',s) 
+ O \left(\frac{\phi(m,s)}{\sqrt{N}}\right) \ .
\end{split} 
\end{equation}
Thus :  
\begin{equation} 
-(1-s) \phi(m,s) =  -s m \phi(m,s) - (1-s) D_m \sum_{m'\in \MNzero} 
D_{m'} \phi(m',s) + O \left(\frac{\phi(m,s)}{\sqrt{N}}\right)  \ .
\end{equation}
This shows that, for $s<1/2$, the groundstate has energy close to 
$-(1-s)$ and takes the form 
\begin{equation}
\label{eq_large_dev_gs_full} 
\phi(m,s,-(1-s)) = e^{- N \varphi_0(m) + O(\sqrt{N})}  \ .
\end{equation}

For $s \geq 1/2$, there appears a divergence in the definition of 
$\phi(m,s)$ at $m= \frac{1-s}{s}$, a sign of the avoided crossing with an 
eigenvector localized near $m = \frac{1-s}{s}$ in the $x$-basis. 
This divergence is lifted by constructing the symmetric and antisymmetric 
combinations of these two quasi-eigenvectors. Let 
$\phi_{\pm}(m,s) = \frac{1}{\sqrt{2}} \left( (1-\delta_{m,\frac{1-s}{s}}) 
\phi(m,s) \pm \delta_{m,\frac{1-s}{s}} \right)$. 
Then one can see that $\phi_\pm$ still satisfies 
(\ref{eq_vp_full_matrix_2}), and thus correspond to two quasi-eigenvectors at 
the location of the avoided crossing. At the leading exponential order the
gap between the two eigenstates that crosses for some value of $s$ is given by 
the overlap between the metastable state of eigenvector close to 
$e^{-N\varphi_0(m)}$ and the localized state in $m_0(s)=\frac{1-s}{s}$. As a 
consequence $\gb(s) = \varphi_0((1-s)/s)$, which explains the origin of 
Eq.~(\ref{eq_gb}).

\subsection{Annealing with a sub-exponential interpolation time}
\label{subsubsec_small_time_full}

In this appendix we present an analysis of the annealing of the simplified
model with an interpolation time growing sub-exponentially with $N$.
The Schr\"odinger equation on the vector 
$|\phi_T(s) \ra = \underset{m\in \MNzero}{\sum} \phi_T(m,s) | m;0 \rax$ reads :
\beq 
\label{eq_schrodinger_full} 
\frac{i}{T} \frac{\dd \phi_T(m,s)}{\dd s} = 
- s N m \phi_T(m,s) - N (1-s) D_m \sum_{m' \in \MNzero} D_{m'} \phi_T(m',s) 
= - s N m \phi_T(m,s) - N (1-s) f_T(s) \ ,
\eeq
where we introduced $f_T(s) = \underset{m \in \MNzero}{\sum} D_m \phi_T(m,s)$. 
By summing (\ref{eq_schrodinger_full}) over $m$ we obtain:
\beq 
\label{eq_schrodinger_full_f} 
\frac{i}{T} \frac{\dd f_T(s)}{\dd s} = - s N \la m \ra_{s,T} - N(1-s)f_T(s) 
\eeq
with $\la m \ra_{s,T} = \underset{m\in\MNzero}{\sum} m D_m \phi_T(m,s)$. 
Note that $\la m \ra_{s=0,T}=0$. Assume first that one can neglect 
$\la m \ra_{s,T}$ in (\ref{eq_schrodinger_full_f}). Then using the initial value 
$f_T(0)=1$, one obtains $f_T(s) =e^{  i NT (s-s^2/2) }$. 
Replacing into (\ref{eq_schrodinger_full}) gives:
\beq 
\frac{i}{T} \frac{\dd \phi_T(m,s)}{\dd s}  = 
- s N m \phi_T(m,s) - N (1-s)  e^{  i NT (s-s^2/2) } \ .
\label{eq_full_simplified}
\eeq
A solution of the associated homogenous equation is 
$\phi_T^{({\rm h})}(m,s) =  e^{iT s^2Nm/2}$. Writing the solution 
of the complete equation (\ref{eq_full_simplified}) as 
$\phi_T(m,s) = \lambda_T(m,s) \phi_T^{({\rm h})}(m,s)$ leads to:
\beq 
\frac{i}{T} \frac{\dd \lambda_T(m,s)}{\dd s} = 
- N(1-s) e^{\frac{-iTs^2Nm}{2}} e^{iNT\left(s - s^2/2\right)} \ ,
\eeq
with $\lambda_T(m,s=0)=D_m$. Let us now write 
$\lambda_T(m,s) = h_{N,T}(m,s) e^{N g_T(m,s)}$, and assume that 
$\underset{N \to \infty}{\lim} \frac{1}{N} \ln \frac{\dd g_T(m,s)}{\dd s} = 0$ 
and $\underset{N \to \infty}{\lim} \frac{1}{N} \ln h_{N,T}(m,s) = 0$. 
Then it is easily found 
that $g_T(m,s) = - \varphi_0(m) - iT s^2m/2 + iT (s- s^2/2)$, and thus 
$\phi_T(m,s) = \phi_T(m,s=0) h_{N,T}(m,s) f_T(s)$. 
The two conditions above then reduces to 
$\lim_{N \to \infty} \frac{1}{N} \ln T = 0$, that is, that one considers 
sub-exponential times. Finally, it is easy to check that in this regime 
$\la m \ra_{s,T} = 0$, and thus that our derivation is indeed self-consistent.

To summarize, in this case, $\phi_T(m,s)$ is up to subdominant corrections 
equal to $\phi_T(m,0)$ times a phase independent on $m$, and the final 
energy $\efin(T)$ is thus identically zero. Therefore we showed that for the 
simplified model:
\beq 
\sup_{a} \lim_{T, N \to \infty} \efin(T=N^a,N) = 
\lim_{T \rightarrow \infty} \efin(T) = 
\lim_{\tau \rightarrow 0} \efin(\tau) = 0 \ ,
\eeq
where the last term comes from the analysis of (small) exponential times of
Sec.~\ref{subsubsec_exp_time_full}.

\section{Numerical integration of the Schr\"{o}dinger equation}
\label{app_schrodinger_evolution}

In this appendix we explain the details of the procedure we used for the
numerical treatment of the finite $N$ dynamics, inspired 
by~\cite{huyg90,poulin11}. Solving the time-dependent Schr\"{o}dinger equation 
(\ref{eq_schrodinger_annealing}) amounts to compute the evolution operator 
$\U(0,1) = \mathcal{T} \left(e^{-i T \int_0^1 \hH(s) \dd s}\right)$ where 
$\mathcal{T}$ denotes the time-ordering operation. It is convenient numerically
to break the time interval $s\in[0,1]$ in $n$ intervals of length
$\Delta s = 1/n$, with equidistant discrete times $s_i = (i-1)/n$. 
This allows to write:
\beq 
\label{eq_apB_break1} 
\U(0,1) =  \mathcal{T} \left(e^{-i T \int_0^1 \hH(s) \dd s}\right) 
= \prod_{i=1}^n \mathcal{T} 
\left(e^{-i T \int_{s_i}^{s_{i+1}} \hH(s) \dd s}\right) 
= \prod_{i=1}^n \U(s_i,s_{i+1})  \ .
\eeq
We are interested in the particular case of a linear dependency of $\hH(s)$ on
$s$: $\hH(s) = (1-s)\hHi + s \hHf$. The approximation
\beq 
\label{eq_apB_break2} 
\begin{split} \U(s,s+\Delta s) = 
\mathcal{T} \left(e^{-i T \int_{s}^{s+\Delta s} \hH(s') \dd s'}\right) 
\rightarrow&  \left(e^{-i T \int_{s}^{s+\Delta s} s' \hHf \dd s'}\right)  
\left(e^{-i T \int_{s}^{s+\Delta s} (1-s') \hHi \dd s'}\right) \\ &=
\left(e^{-i T \frac{2 s \Delta s + \Delta s^2}{2} \hHf}\right)  
\left(e^{-i T \frac{2 (1-s) \Delta s - \Delta s^2}{2} \hHi}\right) \\ & 
\equiv \tU_{\Delta s}(s) \end{split} 
\eeq
gives rise to an error in operator norm 
$\| A \| \equiv \sup_{X, \| X = 1\|} \| AX \|$ bounded by~\cite{huyg90}: 
\begin{equation} 
\| \mathcal{U}(s,s+\Delta s) - \widetilde{\mathcal{U}}_{\Delta s}(s) \| \leq 
\| [\hHi,\hHf] \| \frac{T (\Delta s)^2}{2} + O(\Delta s^3) 
= O(NT\Delta s^2) \ . 
\end{equation}
Indeed in all the cases of interest here the commutator of the initial and
final Hamiltonian has a norm of order $N$.
We define the approximate evolution operator 
$\tU(0,s_i) \equiv \prod_{j=0}^{i-1} 
\tU_{\Delta s}(s_j)$. The triangle inequality
\begin{equation}\begin{split} 
\| \mathcal{U}(0,s_{i+1}) - \tU(0,s_{i+1}) \| &= 
\| \mathcal{U}(0,s_i) (\mathcal{U}(s_i,s_{i+1}) - \tU_{\Delta s}(s_i)) 
+(\mathcal{U}(0,s_i)-\tU(0,s_i) ) \tU_{\Delta s}(s_i)
\| \\ &\leq 
\|\mathcal{U}(s_i,s_i+\Delta s) - \tU_{\Delta s}(s_i) \| 
+\|\mathcal{U}(0,s_i)-\tU(0,s_i)  \| 
\end{split}
\end{equation}
leads by recurrence to
\begin{equation} 
\| \mathcal{U}(0,1) - \widetilde{\mathcal{U}}(0,1) \| \leq O(nNT\Delta s^2)=
O(NT/n) \ .
\end{equation}
One can thus replace the exact evolution operator $\U(0,1)$ by
its approximation $\tU(0,1)$ with a precision of order $\epsilon$
in the evaluation of intensive observables if the number of discretization 
steps $n$ is of order $N T/\epsilon$.

Let us evaluate the total complexity of the procedure.
The dynamical evolution occurs in the fully symmetric sector of the Hilbert
space, hence all operators are actually matrices of size $N+1$. For the
evolution towards the ferromagnet $\hHi = - N \hmx$, $\hHf = -N (\hmz)^p$,
and we work in the basis where $\hmz$ is diagonal. We do not compute all
the matrix elements of $\tU(0,1)$, but rather its product with the initial
state $|\phi_T(0)\ra$, a column vector of size $N+1$. For each time increment
$s_i \to s_{i+1}$ we have to multiply the (approximation) of $|\phi_T(s_i)\ra$
by the two matrices in (\ref{eq_apB_break2}). The first multiplication in 
(\ref{eq_apB_break2}) is computed in a time proportional to $N$ as $\hHf$ is 
a diagonal matrix. The multiplication with the second term is performed
with $O(N^2)$ operations, provided $\hmx$ (whose expression in this basis is 
given in Eq.~\ref{eq_mx_on_mz}) is diagonalized as an initialization step 
(this costs $O(N^3)$ operations). The total cost of the computation is
thus $O(\epsilon^{-1} N^3 T) + O(N^3)$. In the exponentially large times 
regime the second term becomes irrelevant; the limitations of this
numerical method arise from the large times investigated
rather than from the sizes of the matrices themselves. 
The evolution towards the paramagnet is treated similarly, the role of
$\hHi$ and $\hHf$ being simply exchanged with respect to the previous case.
The integration of the dynamics of the simplified model defined in 
Sec.~\ref{subsubsec_def_full_matrix} is slightly easier. One can
indeed exploit the fact that $\hHi=\hJ$ is a matrix of rank one with its
non-zero eigenvalue equal to $-N$, thus
\beq
e^{a \hJ} = \hun -  \frac{(e^{- a N} -1 )}{N} \hJ \ , 
\eeq
with $\hun$ the identity matrix. This avoids the diagonalization of the matrix 
$\hH_i$, and in this case the total complexity of the integration is 
$O(\epsilon^{-1} N^2 T)$, the multiplication of a rank one matrix with a vector
being computable with $O(N)$ operations.

\section{Derivation of Hamilton's equations of motion}
\label{appendix_hamilton}

In this Appendix we give some details of the derivation of 
(\ref{eq_hamilton_para}) from (\ref{eq_schro_large_dev_para}). A similar
computation can be found in~\cite{sciolla11}.

The equation (\ref{eq_schro_large_dev_para}) on $\varphi_T(m,s)$
is of the form
\beq 
- \frac{i}{T} \frac{\partial \varphi_T(m,s)}{\partial s} = 
\mathcal{H}\left(m, i \frac{\partial \varphi_T(m,s)}{\partial m},s\right) 
\eeq where $\mathcal{H}(q,\tq,s)$ is a smooth real function of its parameters. 
We write $\varphi_T (m,s) = g(m,s) - i \theta(m,s)$, with $g$ and $\theta$ 
real-valued functions (from now on we keep implicit the dependence on $T$). 
We assume that there exists a continuous function $q(s)$ such that 
$\frac{\partial g(q(s),s)}{\partial m} =0$ and 
$\frac{\partial^2 g(q(s),s)}{\partial m^2} \neq 0$ for all times $s$ and we 
let $\tq(s) = \frac{\partial}{\partial m} \theta(q(s),s)$.
Then we get:
\beq 
\frac{1}{T} \frac{\partial^2 g(q(s),s)}{\partial m \partial s} = 
- \left. \Im \frac{\dd}{\dd m} \mathcal{H} 
\left(m, i\frac{\partial g}{\partial m} +\frac{\partial \theta}{\partial m}
,s \right) \right |_{m = q(s)} 
= - \frac{\partial \mathcal{H}(q(s),\tq(s),s)}{\partial \tq}  
\frac{\partial ^2 g(q(s),s)}{ \partial m^2} \ .
\eeq
The derivation of the condition 
$\frac{\partial g(q(s),s)}{\partial m} =0$ with respect to $s$ yields:
\begin{equation}
\frac{1}{T}\frac{\dd q(s)}{\dd s} = 
- \frac{1}{T}\frac{ \frac{\partial^2 g(q(s),s)}{\partial m \partial s}}
{ \frac{\partial^2 g(q(s),s)}{\partial m^2}} = 
\frac{\partial \mathcal{H}(q(s),\tq(s),s)}{\partial \tq} \ . 
\eeq
In a similar way we obtain:
\beq \begin{split} 
\frac{1}{T} \frac{\partial^2 \theta(q(s),s)}{\partial m \partial s} &
= - \left. \Re \frac{\dd}{\dd m} 
\mathcal{H} \left(m, i\frac{\partial g}{\partial m}+
\frac{\partial \theta}{\partial m},s \right) \right |_{m = q(s)} \\ &
= - \frac{\partial \mathcal{H}(q(s),\tq(s),s)}{\partial q}  - 
\frac{\partial \mathcal{H}(q(s),\tq(s),s)}{\partial \tq} 
\frac{\partial^2 \theta(q(s),s)}{\partial m^2}  \ .
\end{split} 
\eeq
This leads to 
\begin{equation}
\frac{1}{T}\frac{\dd \tq(s)}{\dd s} = 
\frac{1}{T}\frac{\partial^2 \theta(q(s),s)}{\partial^2 m} 
\frac{\dd q(s)}{\dd s} + \frac{1}{T} 
\frac{\partial^2 \theta(q(s),s)}{\partial m \partial s}
= -\frac{\partial \mathcal{H}(q(s),\tq(s),s)}{\partial q} 
\eeq
Therefore $q(s)$ and $\tq(s)$ obey indeed Hamilton's equations of motion 
for the Hamiltonian $\mathcal{H}(q,\tq,s)$. Note that the reality condition 
on $\mathcal{H}$ corresponds to the Hermitianity of the quantum operator $\hH$,
and that the derivation shows also how Eq.~(\ref{eq_hamilton_ferro}) is a
consequence of Eq.~(\ref{eq_schro_large_dev_ferro}).

\bibliographystyle{h-physrev}
\bibliography{biblio} 

\end{document}